\documentclass[acmtog,nonacm]{acmart}

\usepackage{booktabs} 
\usepackage{multirow}
\usepackage{amsmath}
\usepackage{mathrsfs}
\usepackage{subfigure}
\usepackage{graphicx}
\usepackage[normalem]{ulem}
\usepackage{diagbox}
\usepackage{multicol}

\usepackage[ruled]{algorithm2e} 

\SetAlFnt{\small}
\SetAlCapFnt{\small}
\SetAlCapNameFnt{\small}
\SetAlCapHSkip{0pt}

\acmJournal{TOG}

\begin{document}

\newcommand{\tabincell}[2]{\begin{tabular}{@{}#1@{}}#2\end{tabular}}

\newcommand{\sysName}{DeepFaceDrawing}
\newcommand{\moduleOneFull}{\emph{Component Embedding}}
\newcommand{\moduleTwoFull}{\emph{Feature Mapping}}
\newcommand{\moduleThreeFull}{\emph{Image Synthesis}}
\newcommand{\moduleOne}{\emph{CE}}
\newcommand{\moduleTwo}{\emph{FM}}
\newcommand{\moduleThree}{\emph{IS}}

\title{{\sysName:  Deep Generation of Face Images from Sketches}}

\author{Shu-Yu Chen}

\authornotemark[2]
\affiliation{%
 \institution{Institute of Computing Technology, CAS and University of Chinese Academy of Sciences}}

\email{chenshuyu@ict.ac.cn}

\author{Wanchao Su}
\authornotemark[2]
\affiliation{%
\institution{School of Creative Media, City University of Hong Kong}}

\email{wanchao.su@my.cityu.edu.hk}

\author{Lin Gao}
\authornotemark[1]
\affiliation{%
\institution{Institute of Computing Technology, CAS and University of Chinese Academy of Sciences}}
\email{gaolin@ict.ac.cn}

\author{Shihong Xia}
\affiliation{%
\institution{Institute of Computing Technology, CAS and University of Chinese Academy of Sciences }}
\email{xsh@ict.ac.cn}

\author{Hongbo Fu}
\affiliation{%
\institution{School of Creative Media, City University of Hong Kong}}
\email{hongbofu@cityu.edu.hk}


\authorsaddresses{
$\dag$ Authors contributed equally.\\
$ \ast$ Corresponding author.\\
Webpage: \url{http://geometrylearning.com/DeepFaceDrawing/} \\This is the author's version of the work. It is posted here for your personal use. Not for redistribution. }


\begin{abstract}
Recent deep image-to-image translation techniques allow fast generation of face images from freehand sketches. However, existing solutions tend to overfit to sketches, thus requiring professional sketches or even edge maps as input.
To address this issue,
our key idea is to implicitly model the shape space of plausible face images and synthesize a face image in this space to approximate an input sketch. We take a local-to-global approach.
We first learn feature embeddings of key face components, and push corresponding parts of input sketches towards underlying component manifolds defined by the feature vectors of face component samples.
We also propose another deep neural network to learn the mapping from the embedded component features to realistic images with multi-channel feature maps as intermediate results to improve the information flow.
Our method essentially uses input sketches as soft constraints and is thus able to produce high-quality face images even from rough and/or incomplete sketches. Our tool is easy to use even for non-artists, while still supporting fine-grained control of shape details.
Both qualitative and quantitative evaluations show the superior generation ability of our system to existing and alternative solutions. The usability and expressiveness of our system are confirmed by a user study.

\end{abstract}



\begin{teaserfigure}
    \centering
    \setlength{\fboxrule}{0.5pt}
    \setlength{\fboxsep}{-0.01cm}
    \begin{minipage}{\linewidth}
    \framebox{\includegraphics[width=0.1965\linewidth]{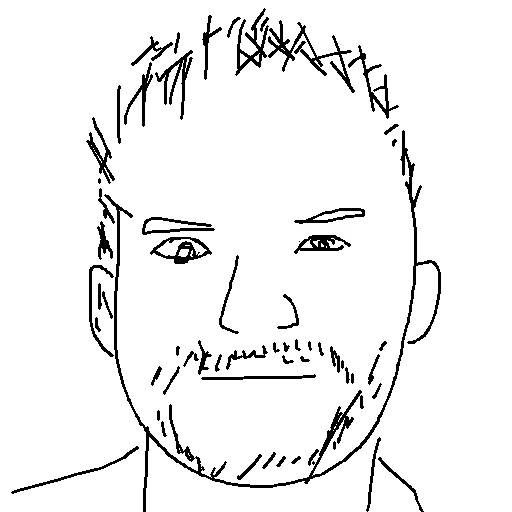}}
    \framebox{\includegraphics[width=0.1965\linewidth]{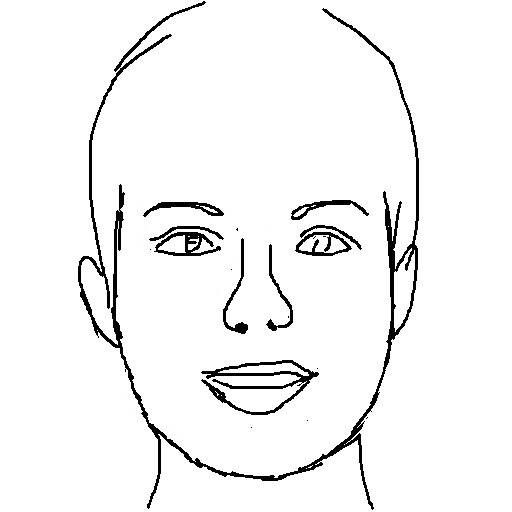}}
    \framebox{\includegraphics[width=0.1965\linewidth]{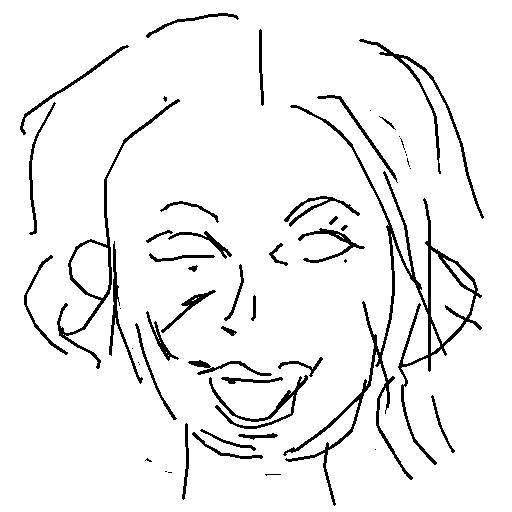}}
    \framebox{\includegraphics[width=0.1965\linewidth]{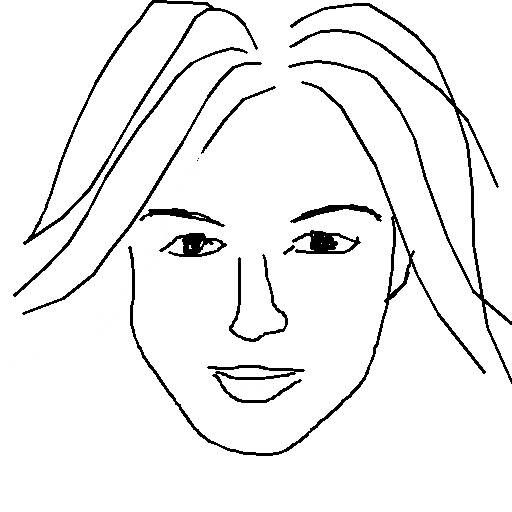}}
    \framebox{\includegraphics[width=0.1965\linewidth]{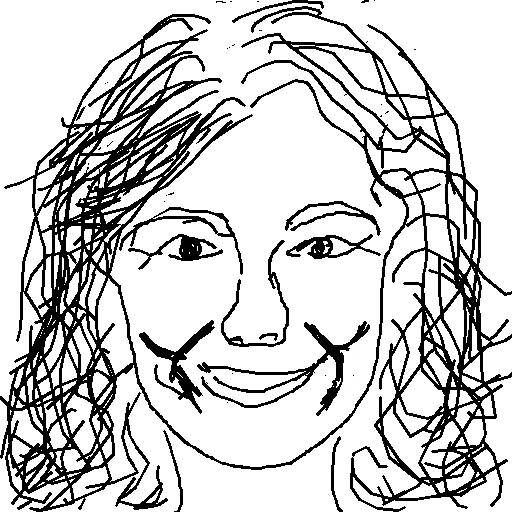}}
    \end{minipage}
    \begin{minipage}{\linewidth}
    \vspace*{-1.5mm }
    \subfigure[]{\framebox{\includegraphics[width=0.1965\linewidth]{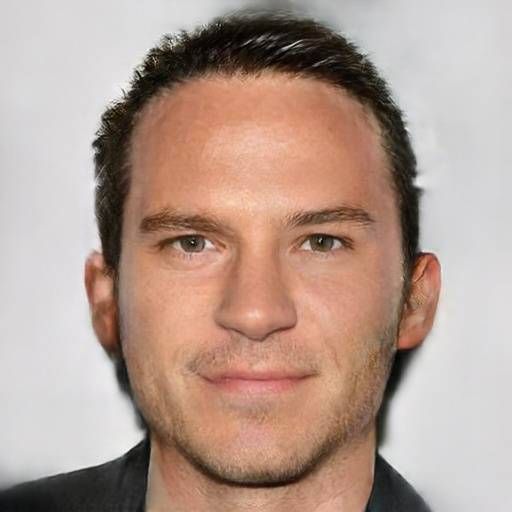}}}
    \subfigure[]{\framebox{\includegraphics[width=0.1965\linewidth]{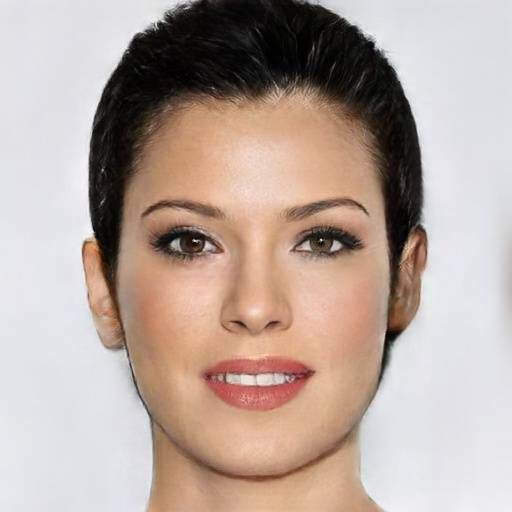}}}
    \subfigure[]{\framebox{\includegraphics[width=0.1965\linewidth]{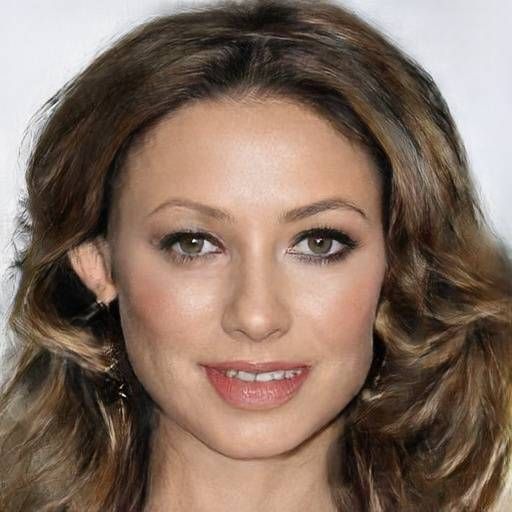}}}
    \subfigure[]{\framebox{\includegraphics[width=0.1965\linewidth]{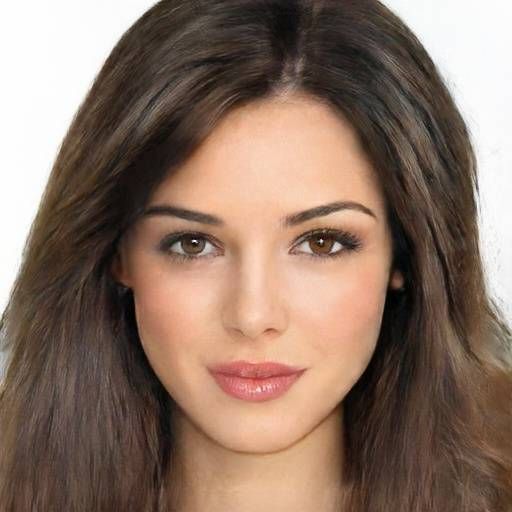}}}
    \subfigure[]{\framebox{\includegraphics[width=0.1965\linewidth]{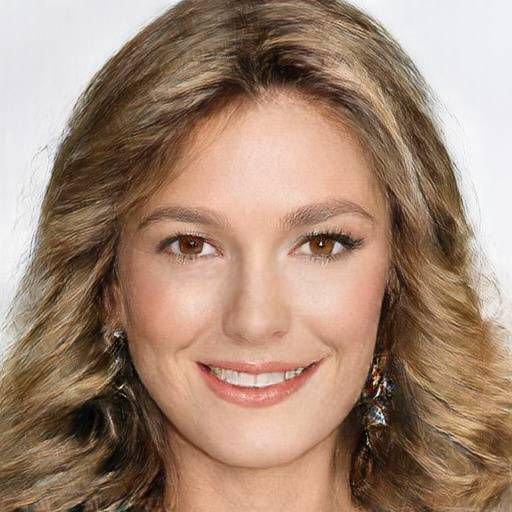}}}
    \end{minipage}
    \caption{
    Our \sysName~system allows users with little training in drawing to produce high-quality face images (Bottom) from rough or even incomplete freehand sketches (Top). Note that our method faithfully respects user intentions in input strokes, which serve more like soft constraints to guide image synthesis.
    }
    \label{fig:teaser}
\end{teaserfigure}



\maketitle

\section{Introduction}
Creating realistic human face images from scratch benefits various applications including criminal investigation, character design, educational training, etc. Due to their simplicity, conciseness and ease of use, sketches are often used to depict desired faces. The recently proposed deep learning based image-to-image translation techniques (e.g.,~\cite{isola2017image, wang2018high}) allow automatic generation of photo images from sketches for various object categories including human faces, and lead to impressive results. 

Most of such deep learning based solutions (e.g., \cite{isola2017image, wang2018high,li2019linestofacephoto,dekel2018sparse}) for sketch-to-image translation often take input sketches almost fixed and attempt to infer the missing texture or shading information between strokes. To some extent, their problems are formulated more like \emph{reconstruction} problems with input sketches as \emph{hard} constraints. Since they often train their networks from pairs of real images and their corresponding edge maps, due to the data-driven nature, they thus require test sketches with quality similar to edge maps of real images to synthesize realistic face images. However, such sketches are difficult to make especially for users with little training in drawing. 

To address this issue, our key idea is to implicitly learn a space of plausible face sketches from real face sketch images and find the closest point in this space to \emph{approximate} an input sketch. In this way, sketches can be used more like \emph{soft} constraints to guide image synthesis. Thus we can increase the plausibility of synthesized images even for rough and/or incomplete input sketches while respecting the characteristics represented in the sketches (e.g., Figure \ref{fig:teaser} (a-d)). Learning such a space globally (if exists) is not very feasible due to the limited training data against an expected high-dimensional feature space. 
This motivates us to implicitly model component-level manifolds, which makes a better sense to assume each component manifold is low-dimensional and locally linear \cite{roweis2000nonlinear}. 
This decision not only helps locally span such manifolds using a limited amount of face data, but also enables finer-grained control of shape details (Figure \ref{fig:teaser} (e)). 

To this end we present a novel deep learning framework for sketch-based face image synthesis, as illustrated in Figure \ref{fig:pipeline}. Our system consists of three main modules, namely, \moduleOne~(\moduleOneFull), \moduleTwo~(\moduleTwoFull), and \moduleThree~(\moduleThreeFull). 
The \moduleOne~module adopts an auto-encoder architecture and separately learns five feature descriptors from the face sketch data, namely, for ``left-eye'', ``right-eye'', ``nose'', ``mouth'', and ``remainder'' for locally spanning the component manifolds. 
The \moduleTwo~and \moduleThree~modules together form another deep learning sub-network for conditional image generation, and map component feature vectors to realistic images. Although \moduleTwo~looks similar to the decoding part of \moduleOne, by mapping the feature vectors to 32-channel feature maps instead of 1-channel sketches, it improves the information flow and thus provides more flexibility to fuse individual face components for higher-quality synthesis results. 

Inspired by~\cite{lee2011shadowdraw}, we provide a shadow-guided interface (implemented based on \moduleOne) for users to input face sketches with proper structures more easily (Figure \ref{fig:interface}). Corresponding parts of input sketches are projected to the underlying facial component manifolds and then mapped to the corresponding feature maps for conditions for image synthesis. 
Our system produces high-quality realistic face images (with resolution of $512 \times 512$), which faithfully respect input sketches. We evaluate our system by {comparing with the existing and alternative solutions, both quantitatively and qualitatively. The results show that our method produces visually more pleasing face images. The usability and expressiveness of our system are confirmed by a user study}.  
We also propose several interesting applications using our method.

\section{Related Work}

Our work is related to existing works for drawing assistance and conditional face generation. We focus on the works closely related to ours. A full review on such topics is beyond the scope of our paper.

\subsection{Drawing Assistance}

Multiple guidance or suggestive interfaces (e.g., \cite{iarussi2013drawing}) have been proposed to assist users in creating drawings of better quality. For example, Dixon et al.~\shortcite{dixon2010icandraw} proposed \emph{iCanDraw}, which provides corrective feedbacks based on an input sketch and facial features extracted from a reference image.
\emph{ShadowDraw} by Lee et al. ~\shortcite{lee2011shadowdraw} retrieves real images from an image repository involving many object categories for an input sketch as query and then blends the retrieved images as shadow for drawing guidance. Our shadow-guided interface for inputting sketches is based on the concept of \emph{ShadowDraw} but specially designed for assisting in face drawing.
Matsui et al. ~\shortcite{matsui2016drawfromdrawings} proposed \emph{DrawFromDrawings}, which allows the retrieval of reference sketches and their interpolation with an input sketch. Our solution for projecting an input sketch to underlying component manifolds follows a similar retrieval-and-interpolation idea but we perform this in the learned feature spaces, without explicit correspondence detection, as needed by \emph{DrawFromDrawings}.
Unlike the above works, which aim to produce quality sketches as output, our work treats such sketches as possible inputs and we are more interested in producing realistic face images even from rough and/or incomplete sketches.

Another group of methods (e.g., \cite{arvo2000fluid,igarashi2006interactive}) take a more aggressive way and aim to automatically correct input sketches. For example, Limpaecher et al.~\shortcite{limpaecher2013real} learn a correction vector field from a crowdsourced set of face drawings to correct a face sketch, with the assumption that such face drawings and the input sketch is for a same subject.
Xie et al.~\shortcite{xie2014portraitsketch} and Su et al.~\shortcite{su2014ez} propose optimization-based approaches for refining sketches roughly drawn on a reference image.
We refine an input sketch by projecting individual face components of the input sketch to the corresponding component manifolds. However, as shown in {Figure \ref{fig:attempt_}}, directly using such refined component sketches as input to conditional image generation might cause artifacts across facial components. Since our goal is sketch-based image synthesis, we thus perform sketch refinement only implicitly.

\begin{figure}[t]
    \centering
    \subfigure[]{\includegraphics[width=0.3\linewidth]{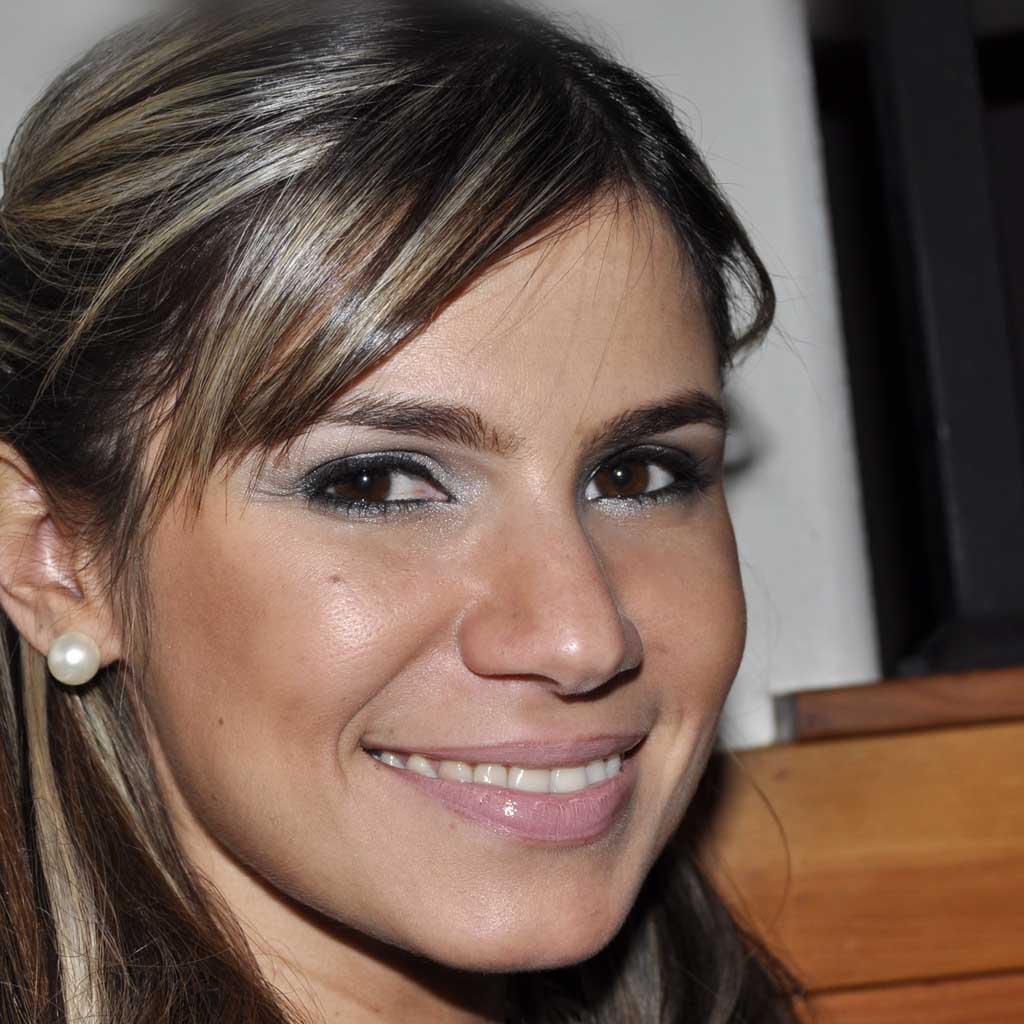}}
    \subfigure[]{\includegraphics[width=0.3\linewidth]{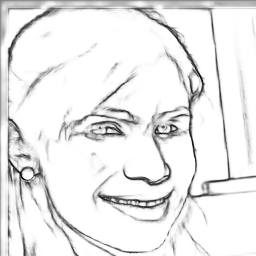}}
    \subfigure[]{\includegraphics[width=0.3\linewidth]{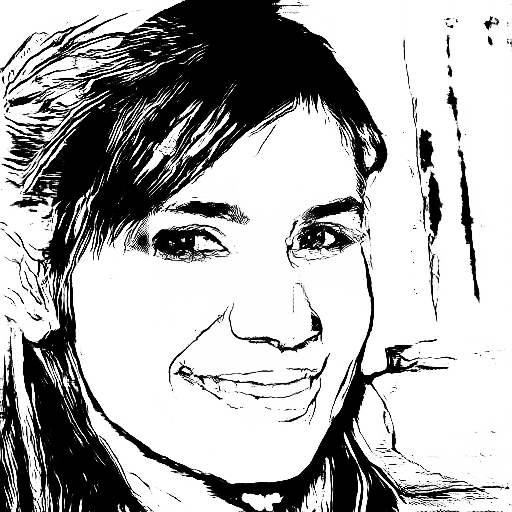}}\\
    \vspace*{-2mm}
    \subfigure[]{\includegraphics[width=0.3\linewidth]{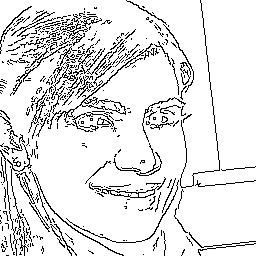}}
    \subfigure[]{\includegraphics[width=0.3\linewidth]{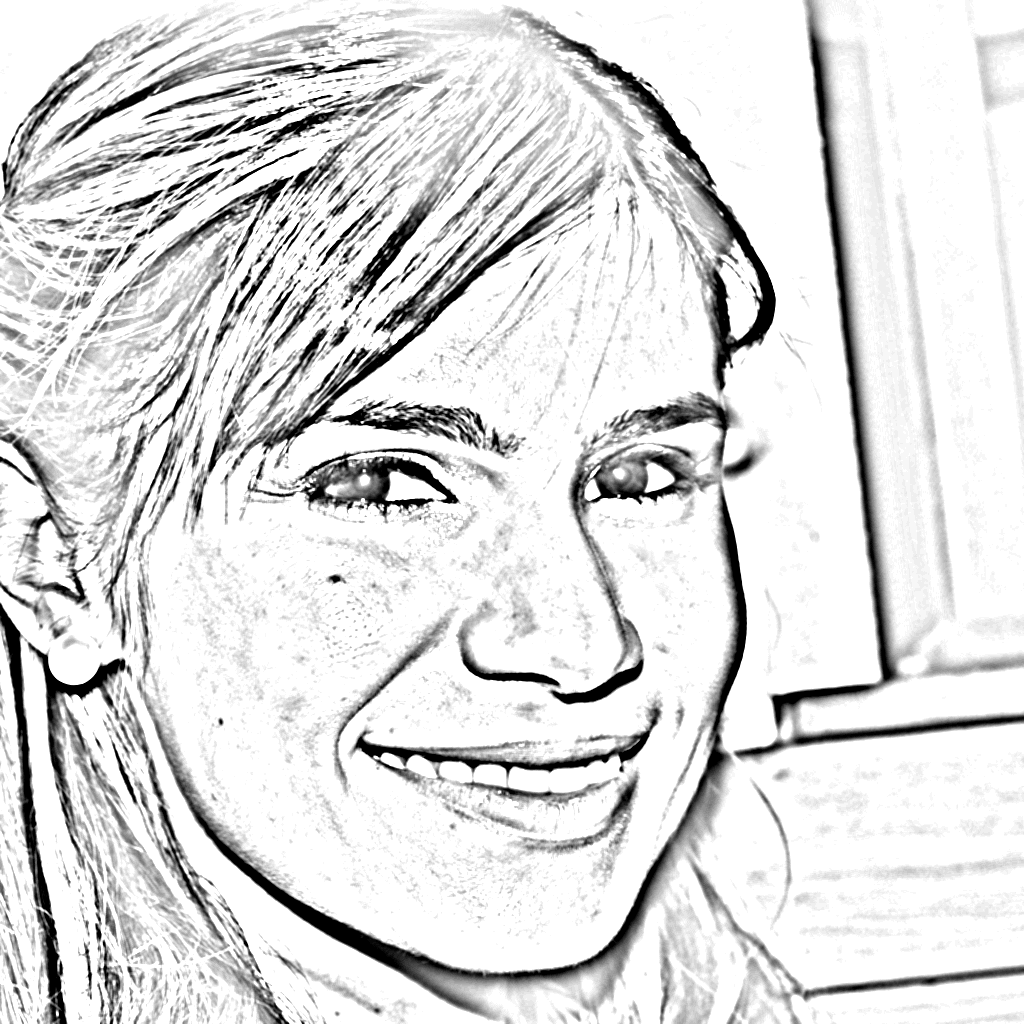}}
    \subfigure[]{\includegraphics[width=0.3\linewidth]{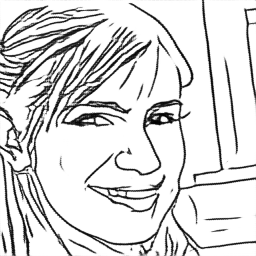}}
    \caption{The comparisons of different edge extraction methods. (a): Input real image. (b): Result by HED \cite{xie15hed}. (c): Result by APDrawingGAN \cite{YiLLR19}. (d): Canny edges \cite{canny1986a}. (e): the result by the Photocopy filter in Photoshop. (f): Simplification of (e) by \cite{SimoSerraSIGGRAPH2016}. Photo (a) courtesy of \copyright~ LanaLucia.}
    \label{fig:edge_compare}
\end{figure}

\begin{figure*}
    \centering
    {\includegraphics[width=0.98\linewidth]{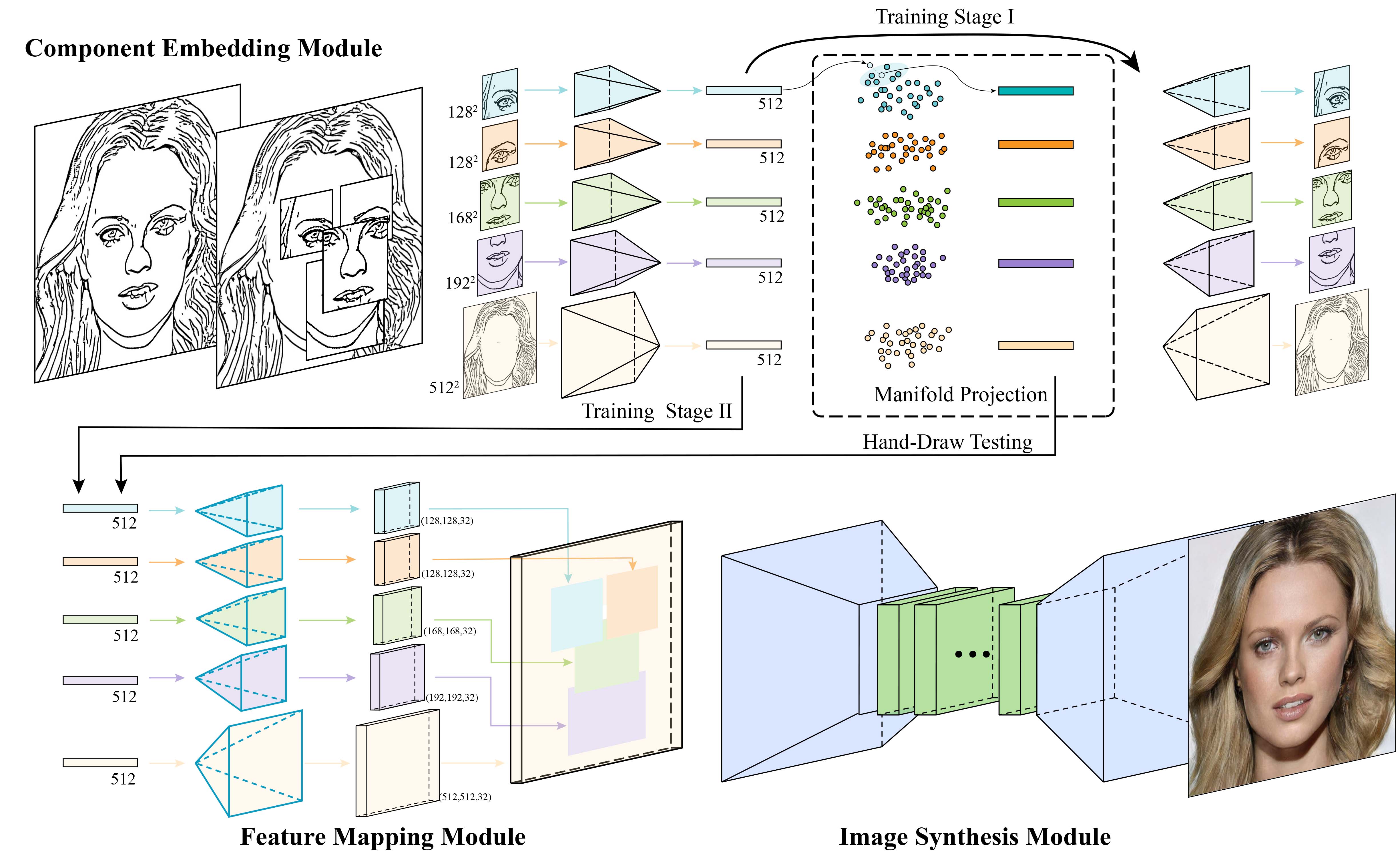}}
    \caption{Illustration of our network architecture. The upper half is the \moduleOneFull~module. We learn feature embeddings of face components using individual auto-encoders. The feature vectors of component samples are considered as the point samples of the underlying component manifolds and are used to refine an input hand-drawn sketch by projecting its individual parts to the corresponding component manifolds.
    The lower half illustrates a sub-network consisting of the \moduleTwoFull~(\moduleTwo) and the \moduleThreeFull~(\moduleThree) modules. The \moduleTwo~module decodes the component feature vectors to the corresponding multi-channel feature maps ($H \times W \times 32$), which are combined according to the spatial locations of the corresponding facial components before passing them to the \moduleThree~module.
    }
    \label{fig:pipeline}
\end{figure*}

\subsection{Conditional Face Generation}
In recent years, conditional generative models, in particular, conditional Generative Adversarial Networks \cite{Goodfellow2014generative} (GANs), have been popular for image generation conditioned on various input types.
Karras et al. ~\shortcite{karras2019style} propose an alternative for the generator in GAN that separates the high level face attributes and stochastic variations in generating high quality face images.
Based on conditional GANs \cite{mirza2014conditional}, Isola et al.~\shortcite{isola2017image} present the \emph{pix2pix} framework for various image-and-image translation problems like image colorization, semantic segmentation, sketch-to-image synthesis, etc.
Wang et al.~\shortcite{wang2018high} introduce \emph{pix2pixHD}, an improved version of \emph{pix2pix} to generate higher-resolution images, and demonstrate its application to image synthesis from semantic label maps.
Wang et al. ~\shortcite{wang2019example} generate an image given a semantic label map as well as an image exemplar.
Sangkloy et al. \shortcite{sangkloy2017scribbler} take hand-drawn sketches as input and colorize them under the guidance of user-specified sparse color strokes.
These systems tend to overfit to conditions seen during training, and thus when sketches being used as conditions, they achieve quality results only given edge maps as input.
To address this issue, instead of training an end-to-end network for sketch-to-image synthesis, we exploit the domain knowledge and condition GAN on feature maps derived from the component feature vectors.

Considering the known structure of human faces, researchers have explored component-based methods (e.g., \cite{huang2017beyond}) for face image generation. For example, given an input sketch, Wu and Dai~\shortcite{wu2009sketch} first retrieve best-fit face components from a database of face images, then compose the retrieved components together, and finally deform the composed image to approximate a sketch. Due to their synthesis-and-deforming strategy, their solution requires a well-drawn sketch as input.
To enable component-level controllability, Gu et al.~\shortcite{gu2019mask} use auto-encoders to learn feature embeddings for individual face components, and fuse component feature tensors in a mask-guided generative network. Our \moduleOne~module is inspired by their work. However, their local embeddings learned from real images are mainly used to generate portrait images with high diversity while ours learned from sketch images are mainly for implicitly refining and completing input sketches.

Conditional GANs have also been adopted for local editing of face images, via interfaces either based on semantic label masks \cite{gu2019mask,CelebAMask-HQ,wang2019example} or sketches \cite{portenier2018faceshop,jo2019sc}. While the former is more flexible for applications such as component transfer and style transfer, the latter provides a more direct and finer control of details, even within face components. Deep sketch-based face editing is essentially a sketch-guided image completion problem, which requires the completion of missing parts such that the completed content faithfully reflects an input sketch and seamlessly connects to the known context. It thus requires different networks from ours. The \emph{SN-patchGAN} proposed by Jo and Park ~\shortcite{jo2019sc} is able to produce impressive details for example for a sketched earring. However, this also implies that their solution requires high-quality sketches as input. To tolerate the errors in hand-drawn sketches, Portenier et al. \shortcite{portenier2018faceshop} propose to use smoothed edge maps as part of the input to their conditional completion network. Our work takes a step further to implicitly model face component manifolds and perform manifold projection.

Several attempts have also been made to generate images from incomplete sketches. To synthesize face images from line maps possibly with some missing face components, Li et al. \shortcite{li2019linestofacephoto} proposed a conditional self-attention GAN with a multi-scale discriminator, where a large-scale discriminator enforces the completeness of global structures. Although their method leads to visually better results than \emph{pix2pix} \cite{isola2017image} and \emph{SkethyGAN} \cite{chen2018sketchygan}, due to the direct condition on edge maps, their solution has poor ability to handle hand-drawn sketches.
Ghosh et al.~\shortcite{ghosh2019interactive} present a shape generator to complete a partial sketch before image generation, and present interesting auto-completion results. However, their synthesized images still exhibit noticeable artifacts, since the performance of their image generation step (i.e., \emph{pix2pixHD} \cite{wang2018high} for single-class generation and \emph{SkinnyResNet}~\shortcite{ghosh2019interactive} for multi-class generation) heavily depends on the quality of input sketches.
A similar problem exists with the progressive image reconstruction network proposed by You et al.~\shortcite{you2019pi}, which is able to reconstruct images from extremely sparse inputs but still requires relatively accurate inputs.

To alleviate the heterogeneity of input sketches and real face images, some researchers resort to the unpaired image-to-image methods (e.g., \cite{zhu2017unpaired,yi2017dualgan,huang2018multimodal}).
These methods adopt self-consistent constraints to solve the lack of paired data. While the self-consistent mechanism ensures the correspondence between the input and the reconstructed input, there is no guarantee for the correspondence between the input and the transformed representation. Since our goal is to transform sketches to the corresponding face images, these frameworks are not suitable for our task.
In addition, there are some works leveraging the image manifolds. For example, Lu et al. ~\shortcite{Lu2017Face} learn a fused representation from shape and texture features to construct a face retrieval system. In contrast, our method not only retrieves but also interpolates the face representations in generation.
Zhu et al. ~\shortcite{zhu2016generative} first construct a manifold with the real image dataset, then predict a dense correspondence between a projected source image and an edit-oriented ``feasible'' target in the manifold, and finally apply the dense correspondence back to the original source image to complete the visual manipulation. In contrast, our method directly interpolates the nearest neighbors of the query and feeds the interpolation result to the subsequent generation process.
Compared to Zhu et al. ~\shortcite{zhu2016generative}, our method is more direct and efficient for the sketch-based image generation task.

\section{Methodology}

The 3D shape space of human faces has been well studied (see the classic morphable face model ~\cite{blanz1999morphable}).
A possible approach to synthesize realistic faces from hand-drawn sketches is to first project an input sketch to such a 3D face space \cite{han2017deepsketch2face} and then synthesize a face image from a generated 3D face. However, such a global parametric model is not flexible enough to accommodate rich image details or support local editing.
Inspired by \cite{gaosdmnet2019}, which shows the effectiveness of a local-global structure for faithful local detail synthesis, our method aims for modeling the shape spaces of face components in the image domain.

To achieve this, we first learn the feature embeddings of face components (Section \ref{para: module1}). For each component type, the points corresponding to component samples implicitly define a manifold.
However, we do not explicitly learn this manifold, since we are more interested in knowing the closest point in such a manifold given a new sketched face component, which needs to be refined. Observing that in the embedding spaces semantically {similar} components are close to each other, we assume that the underlying component manifolds are locally linear. We then follow the main idea of the classic locally linear embedding (LLE) algorithm ~\cite{roweis2000nonlinear} to project the feature vector of the sketched face component to its component manifold (Section \ref{sec:feature_space_ref}).

The learned feature embeddings also allow us to guide conditional sketch-to-image synthesis to explicitly exploit the information in the feature space. Unlike traditional sketch-to-image synthesis methods (e.g., \cite{isola2017image, wang2018high}), which learn conditional GANs to translate sketches to images, our approach forces the synthesis pipeline to go through the component feature spaces and then map 1-channel feature vectors to 32-channel feature maps before the use of a conditional GAN (Section \ref{para: module3}). This greatly improves the information flow and benefits component fusion.
Below we first discuss our data preparation procedure (Section \ref{sec:data_preparation}). We then introduce our novel pipeline for sketch-to-image synthesis (Section \ref{sec:framework}), and our approach for manifold projection (Section \ref{sec:feature_space_ref}). Finally present our shadow-guided interface (Section \ref{sec:user_interface}).

\subsection{Data Preparation}\label{sec:data_preparation}
To train our network, it requires a reasonably large-scale dataset of face sketch-image pairs. There exist several relevant datasets like the CUHK face sketch database \cite{wang2008face, zhang2011coupled}. 
However, the sketches in such datasets involve shading effects while we expect a more abstract representation of faces using sparse lines. We thus contribute to a new dataset of pairs of face images and corresponding synthesized sketches. We build this on the face image data of CelebAMask-HQ \cite{CelebAMask-HQ}, which contains high-resolution facial images with semantic masks of facial attributes. For simplicity, we currently focus on front faces, without decorative accessories (e.g., glasses, face masks).

To extract sparse lines from real images, we have tried the following edge detection methods.
As shown in Figure \ref{fig:edge_compare} (b) and (d), the holistically-nested edge detection (HED) method~\cite{xie15hed} and the traditional Canny edge detection algorithm \cite{canny1986a} tend to produce edge maps with discontinuous lines. APDrawingGAN \cite{YiLLR19}, a very recent approach for generating portrait drawings from face photos leads to artistically pleasing results, which, however, are different from our expectation (e.g., see the regional effects in the hair area and missing details around the mouth in Figure \ref{fig:edge_compare} (c)). We also resorted to the Photocopy filter in Photoshop, which preserves facial details but meanwhile brings excessive noise (Figure \ref{fig:edge_compare} (e)).
By applying the sketch simplification method by Simo-Serra et al. \shortcite{SimoSerraSIGGRAPH2016} to the result by the Photocopy filter, we get an edge map with the noise reduced and the lines better resembling hand-drawn sketches (Figure \ref{fig:edge_compare} (f)). We thus adopt this approach (i.e., Photocopy + sketch simplification) to prepare our training dataset, which contains 17K pairs of sketch-image pairs (see an example pair in Figure \ref{fig:edge_compare} (f) and (a)), with 6247 for male subjects and 11456 for female subjects.
Since our dataset is not very large-scale, we reserve the data in the training process as much as possible to provide as many samples as possible to span the component manifolds. Thus we set a training/testing ratio to 20:1 in our experiments. It results in 16,860 samples for training and 842 for testing.

\subsection{Sketch-to-Image Synthesis Architecture}\label{sec:framework}

As illustrated in Figure \ref{fig:pipeline}, our deep learning framework takes as input a sketch image and generates a high-quality facial image of size $512 \times 512$.
It consists of two sub-networks:
The first sub-network is our \moduleOne~module, which is responsible for learning feature embeddings of individual face components using separate auto-encoder networks.
This step turns component sketches into semantically meaningful feature vectors. The second sub-network consists of two modules: \moduleTwo~and \moduleThree. \moduleTwo~turns the component feature vectors to the corresponding feature maps to improve the information flow. The feature maps of individual face components are then combined according to the face structure and finally passed to \moduleThree~for face image synthesis.

\begin{figure}
    \centering
    \setlength{\fboxrule}{0.5pt}
    \setlength{\fboxsep}{-0.01cm}
    \framebox{\includegraphics[width=0.44\linewidth]{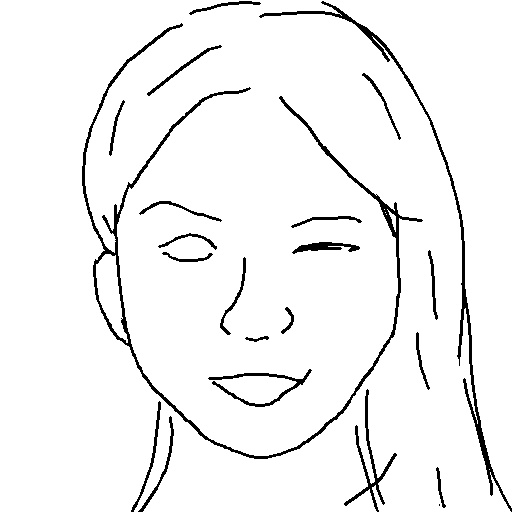}}
    \framebox{\includegraphics[width=0.44\linewidth]{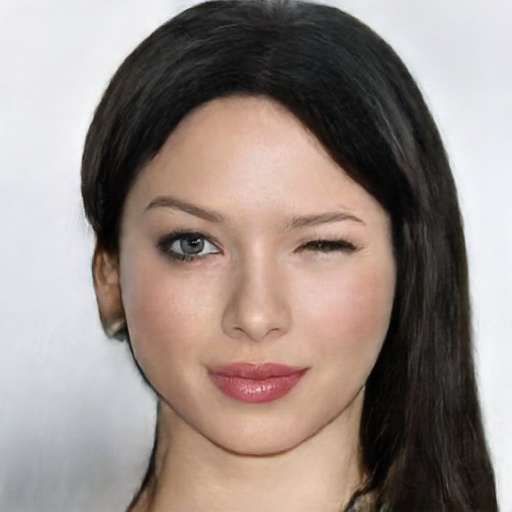}}
    \framebox{\includegraphics[width=0.44\linewidth]{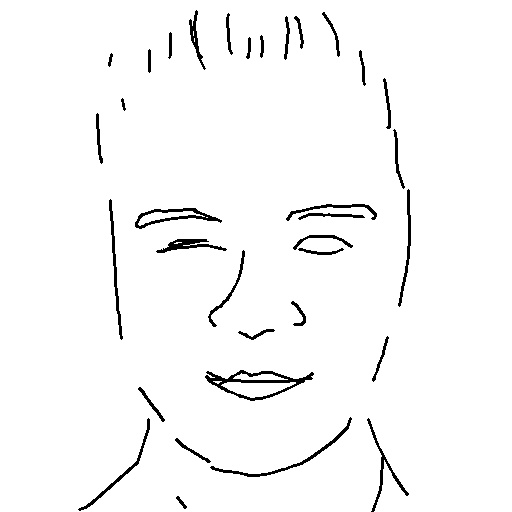}}
    \framebox{\includegraphics[width=0.44\linewidth]{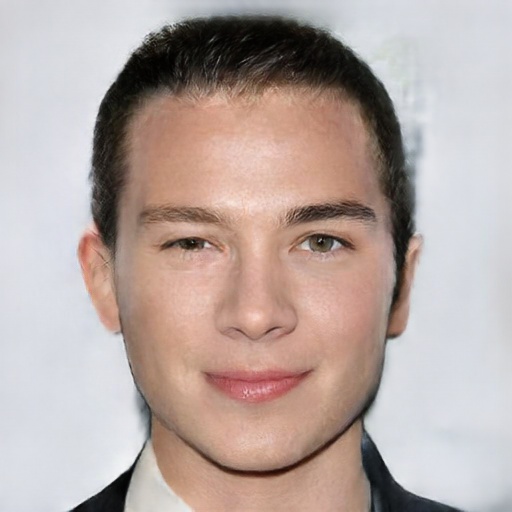}}
    \caption{Two examples of generation flexibility supported by using separate components for the left and right eyes.}
    \label{fig:flexible}
\end{figure}

\paragraph{Component Embedding Module}\label{para: module1}
Since human faces share a clear structure, we decompose a face sketch into five components, denoted as $S^c,\ c \in {\{1,2,3,4,5\}}$ for ``left-eye", ``right-eye", ``nose", ``mouth", and ``remainder", respectively.
To handle the details in-between components, we define the first four components simply by using four overlapping windows centered at individual face components ({derived from the pre-labeled segmentation masks in the dataset}), as illustrated in Figure \ref{fig:pipeline} (Top-Left). A ``remainder'' image corresponding to the ``remainder'' component is the same as the original sketch image but with the eyes, nose and mouth removed.
Here we treat ``left-eye'' and ``right-eye'' separately to best explore the flexibility in the generated faces (see two examples in Figure \ref{fig:flexible}).
To better control of the details of individual components, for each face component type we learn a local feature embedding.
We obtain the feature descriptors of individual components by using
five auto-encoder networks,
denoted as $\{E_c,D_c\}$ with $E_c$ being an encoder and $D_c$ a decoder for component $c$.

Each auto-encoder consists of five encoding layers and five decoding layers. We add a fully connected layer in the middle to ensure the latent descriptor is of 512 dimensions for all the five components. We experimented with different numbers of dimensions for the latent representation (128, 256, 512) -- we found that 512 dimensions are enough for reconstructing and representing the sketch details. Instead, lower-dimensional representations tend to lead to blurry results.
By trial and error, we append a residual block after every convolution/deconvolution operation in each encoding/decoding layer to construct the latent descriptors instead of only using convolution and deconvolution layers. We use Adam solver ~\cite{kingma2014method} in the training process.
Please find the details of the network architectures and the parameter settings in the supplemental materials.

\paragraph{Feature Mapping Module}
Given an input sketch, we can project its individual parts to the component manifolds to increase its plausibility (Section \ref{sec:feature_space_ref}). One possible solution to synthesize a realistic image is to first convert the feature vectors of the projected manifold points back to the component sketches using the learned decoders $\{D_c\}$, then perform component-level sketch-to-image synthesis (e.g., based on ~\cite{wang2018high}), and finally fuse the component images together to get a complete face.
However, this straightforward solution easily leads to inconsistencies in synthesized results in terms of both local details and global styles, {since there is no mechanism to coordinate the individual generation processes}.

\begin{figure}
    \centering
    \subfigure[]{\includegraphics[width=0.44\linewidth]{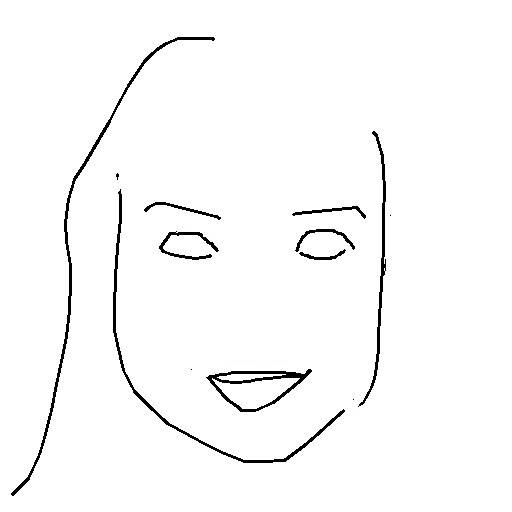}}
    \subfigure[]{\includegraphics[width=0.44\linewidth]{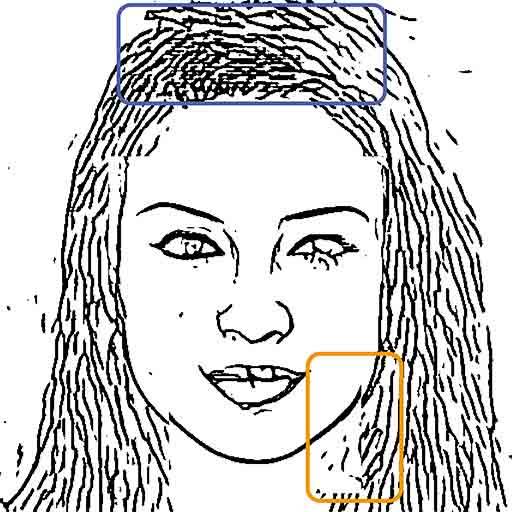}}\\
    \subfigure[]{\includegraphics[width=0.44\linewidth]{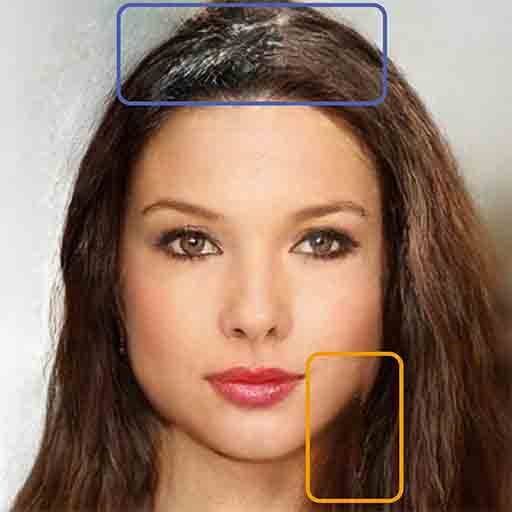}}
    \subfigure[]{\includegraphics[width=0.44\linewidth]{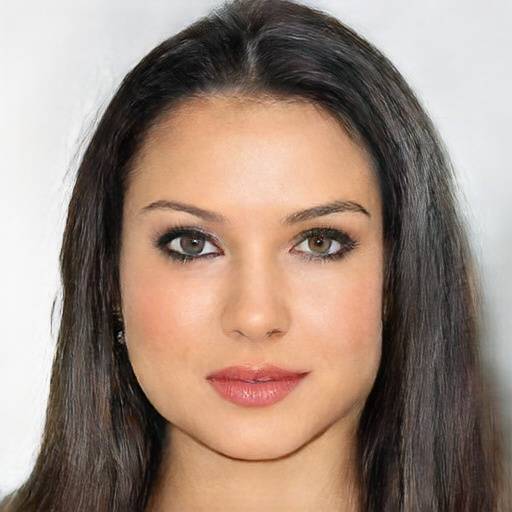}}
    \caption{Given the same input sketch (a), image synthesis conditioned on the feature vectors after manifold projection achieves a more realistic result (d) than that (c) by image synthesis conditioned on an intermediate sketch (b). See the highlighted artifacts in both the intermediate sketch (b) and the corresponding synthesized result (c) by pix2pixHD~\cite{wang2018high}.
    }
    \label{fig:attempt_}
\end{figure}

Another possible solution is to first fuse the decoded component sketches into a complete face sketch (Figure \ref{fig:attempt_} (b)) and then perform sketch-to-image synthesis to get a face image (Figure \ref{fig:attempt_} (c)). It can be seen that this solution also easily causes artifacts (e.g., misalignment between face components, incompatible hair styles) in the synthesized sketch, and such artifacts are inherited to the synthesized image, since existing deep learning solutions for sketch-to-image synthesis tend to use input sketches as rather hard constraints, as discussed previously.

We observe that the above issues mainly happen in the overlapping regions of the cropping windows for individual components. Since sketches only have one channel, the incompatibility of neighboring components in the overlapping regions is thus difficult to automatically resolve by sketch-to-image networks. This motivates us to map the feature vectors of sampled manifold points to {multi-channel} feature maps (i.e., 3D feature tensors).
This significantly improves the information flow, and fusing the feature maps instead of sketch components helps resolve the inconsistency between face components.

Since the descriptors for different components bear different semantic meanings, we design the \moduleTwo~module with five separate decoding models converting feature vectors to spatial feature maps.
Each decoding model consists of a fully connected layer and five decoding layers.
For each feature map, it has 32 channels and is of the same spatial size as the corresponding component in the sketch domain.
The resulting feature maps for ``left-eye'', ``right-eye'', ``nose'', and ``mouth'' are placed back to the ``remainder'' feature maps according to the exact positions of the face components in the input face sketch image to retain the original spatial relations between face components. As illustrated in Figure \ref{fig:pipeline} (Bottom-Center), we use a fixed depth order (i.e., ``left/right eyes" > ``nose" > ``mouth'' > ``remainder") to merge the feature maps.

\paragraph{Image Synthesis Module}\label{para: module3}

Given the combined feature maps, the \moduleThree~module converts them to a realistic face image. We implement this module using a conditional GAN architecture, which takes the feature maps as input to a generator, with the generation guided by a discriminator. Like the global generator in \emph{pix2pixHD}~\cite{wang2018high}, our generator contains an encoding part, a residual block, and a decoding unit. The input feature maps go through these units sequentially. Similar to \cite{wang2018high}, the discriminator is designed to determine the samples in a multi-scale manner: we downsample the input to multiple sizes and use multiple discriminators to process different inputs at different scales.
We use this setting to learn the high-level correlations among parts implicitly.

\paragraph{Two-stage Training.}
As illustrated in Figure \ref{fig:pipeline}, we adopt a two-stage training strategy to train our network using our dataset of sketch-image pairs (Section \ref{sec:data_preparation}). In Stage I, we train only the \moduleOne~module, by using component sketches to train five individual auto-encoders for feature embeddings. The training is done in a self-supervised manner, with the mean square error (MSE) loss between an input sketch image and the reconstructed image.
In Stage II, we fix the parameters of the trained component encoders and train the entire network with the unknown parameters in the \moduleTwo~and \moduleThree~modules together in an end-to-end manner. For the GAN in the \moduleThree, besides the GAN loss, we also incorporate a $L_1$ loss to further guide the generator and thus ensure the pixel-wise quality of generated images. We use the perceptual loss~\cite{johnson2016perceptual} in the discriminator to compare the high-level difference between real and generated images.
Due to the different characteristics of female and male portraits, we train the network using the complete set but constrain the searching space into the male and female spaces for testing.

\begin{figure}
    \centering
    \includegraphics[width=0.9\linewidth]{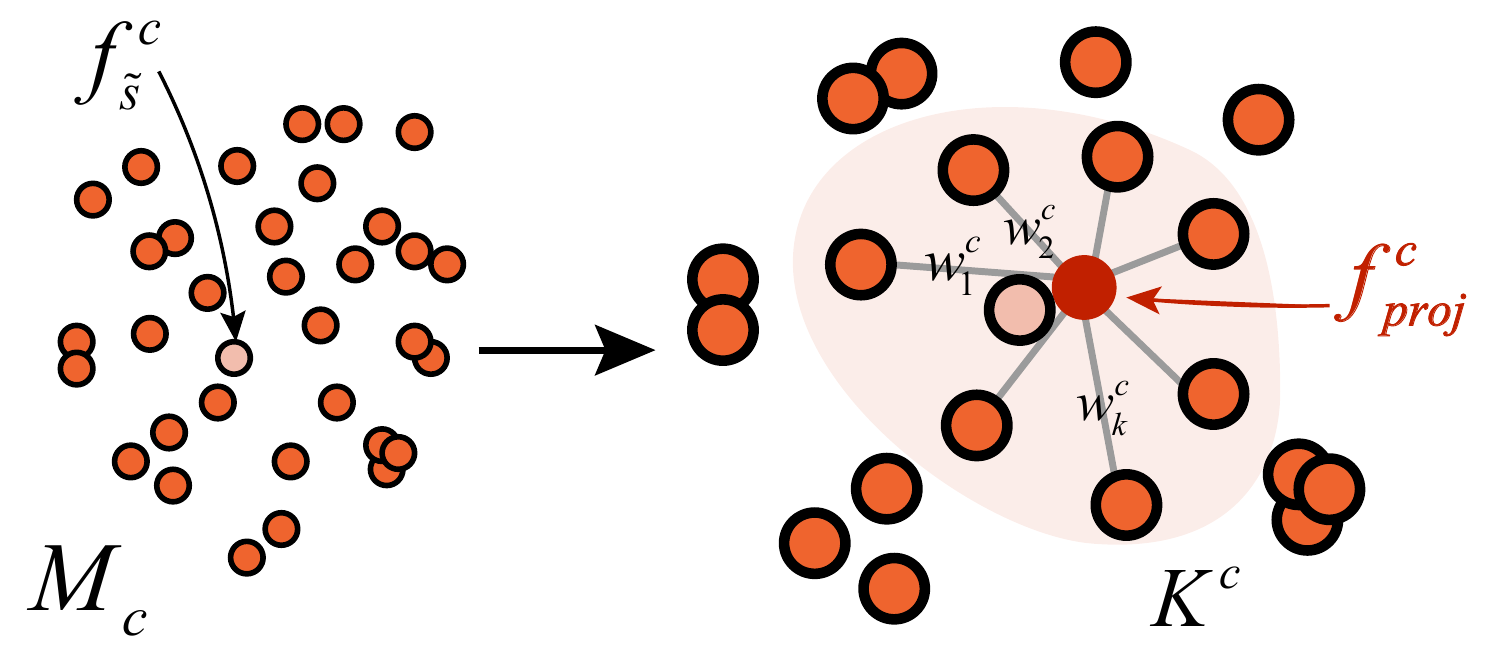}
    \caption{Illustration of manifold projection. Given a new feature vector $f_{\tilde{s}}^c$ , we replace it with the projected feature vector $f_{proj}^c$ using K nearest neighbors of $f_{\tilde{s}}^c$.}
    \label{fig:local_inter}
\end{figure}

\subsection{Manifold Projection}\label{sec:feature_space_ref}
Let $\mathcal{S} = \{s_i\}$ denote a set of sketch images used to train the feature embeddings of face components (Section \ref{para: module1}). For each component $c$, we can get a set of points in the $c$-component feature space by using the trained encoders, denoted as $\mathcal{F}^c = \{f_i^c = E_c(s_i^c)\}$. Although each feature space is 512-dimensional, given that similar component images are placed closely in such feature spaces, we tend to believe that all the points in $\mathcal{F}^c$ are in an underlying low-dimensional manifold, denoted as $\mathcal{M}^c$, and further assume each component manifold is locally linear: each point and its neighbors lie on or close to a locally linear patch of the manifold~\cite{roweis2000nonlinear}.

Given an input sketch $\tilde{s}$, to increase its plausibility as a human face, we project its $c$-th component to $\mathcal{M}^c$. With the locally linear assumption, we follow the main idea of LLE and take a \emph{retrieval-and-interpolation} approach to project the $c$-th component feature vector of $E_c(\tilde{s}^c)$, denoted as $f_{\tilde{s}}^c$ to $\mathcal{M}^c$, as illustrated in Figure \ref{fig:pipeline}.

As illustrated in Figure \ref{fig:local_inter}, given the $c$-th component feature vector $f_{\tilde{s}}^c$, we first find the $K$  nearest samples in $\mathcal{F}^c$ under the Euclidean space. By trial and error, we found that $K$=10 is sufficient in providing face plausibility while maintaining adequate variations.
Let $\mathcal{K}^c = \{s_k^c\}$ (with $\{s_k\} \subset \mathcal{S}$) denote the resulting set of $K$ nearest samples, i.e., the neighbors of $\tilde{s}^c$ on $\mathcal{M}^c$. We then seek a linear combination of these neighbors to reconstruct $\tilde{s}^c$ by minimizing the reconstruction error. This is equivalent to solving for the interpolation weights through the following minimization problem:
\begin{equation}
    \centering
    \begin{split}
      \min ||f_{\tilde{s}}^c-\sum_{k\in \mathcal{K}^c}w^c_k \cdot f_k^c||^2_2,
       \ \ \ \ \ s.t. \sum_{k\in \mathcal{K}}w^c_k=1,
    \end{split}
    \label{equ:weight}
\end{equation}
where $w^c_k$ is the unknown weight for sample $s_k^c$. The weights can be found by solving a constrained least-squares problem for individual components independently.
Given the solved weights $\{w_k^c\}$, the projected point of $\tilde{s}^c$ on $\mathcal{M}^c$ can be computed as
\begin{equation}
f^c_{proj} = \sum_{k\in \mathcal{K}^c} w^c_k \cdot f_k^c.
\end{equation}
$f^c_{proj}$ is the feature vector of the refined version of $\tilde{s}^c$, and can be passed to the \moduleTwo~and \moduleThree~modules for image synthesis.

\begin{figure}
    \centering
    \setlength{\fboxrule}{1pt}
    \setlength{\fboxsep}{-0.03cm}
    \framebox{\includegraphics[width=0.19\linewidth]{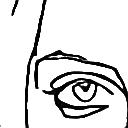}}
    \framebox{\includegraphics[width=0.19\linewidth]{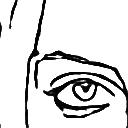}}
    \framebox{\includegraphics[width=0.19\linewidth]{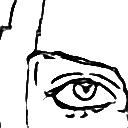}}
    \framebox{\includegraphics[width=0.19\linewidth]{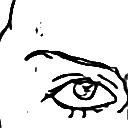}}
    \framebox{\includegraphics[width=0.19\linewidth]{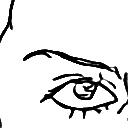}}

    \framebox{\includegraphics[width=0.19\linewidth]{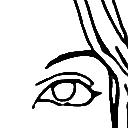}}
    \framebox{\includegraphics[width=0.19\linewidth]{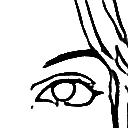}}
    \framebox{\includegraphics[width=0.19\linewidth]{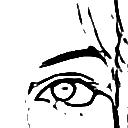}}
    \framebox{\includegraphics[width=0.19\linewidth]{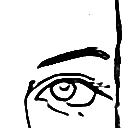}}
    \framebox{\includegraphics[width=0.19\linewidth]{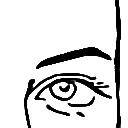}}

    \framebox{\includegraphics[width=0.19\linewidth]{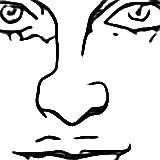}}
    \framebox{\includegraphics[width=0.19\linewidth]{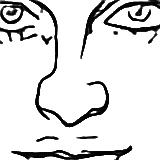}}
    \framebox{\includegraphics[width=0.19\linewidth]{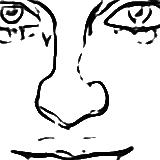}}
    \framebox{\includegraphics[width=0.19\linewidth]{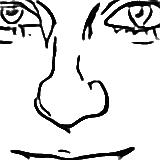}}
    \framebox{\includegraphics[width=0.19\linewidth]{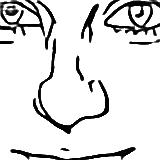}}

    \framebox{\includegraphics[width=0.19\linewidth]{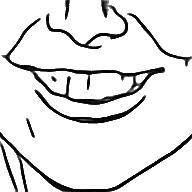}}
    \framebox{\includegraphics[width=0.19\linewidth]{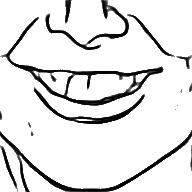}}
    \framebox{\includegraphics[width=0.19\linewidth]{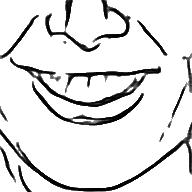}}
    \framebox{\includegraphics[width=0.19\linewidth]{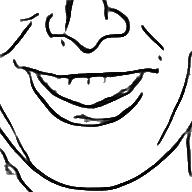}}
    \framebox{\includegraphics[width=0.19\linewidth]{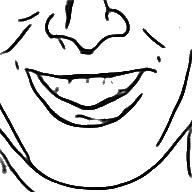}}
    \caption{
    Illustration of linear interpolation between pairs of randomly selected neighboring component sketches (Leftmost and Rightmost) in the corresponding feature spaces. The middle three images are decoded from the uniformly interpolated feature vectors.
    }
    \label{fig:inter}
\end{figure}

To verify the local continuity of the underlying manifolds, we first randomly select a sample from $\mathcal{S}$, and for its $c$-th component randomly select one of its nearest neighbors in the correspondence feature space. We then perform linear interpolation between such a pair of component sketches in the $c$-th feature space, and reconstruct the interpolated component sketches using the learned $c$-th decoder $D_c$.
The reconstructed results are shown in Figure \ref{fig:inter}.
It can be seen that as we change the interpolation weight continuously, it results in smooth changes between the consecutive reconstructed component sketches from a pair of selected sketches. This shows the feasibility of our descriptor interpolation.

\begin{figure}[h]
    \centering
    \setlength{\fboxrule}{0.1pt}
    \setlength{\fboxsep}{-0.03cm}
    \framebox{\includegraphics[width=1\linewidth]{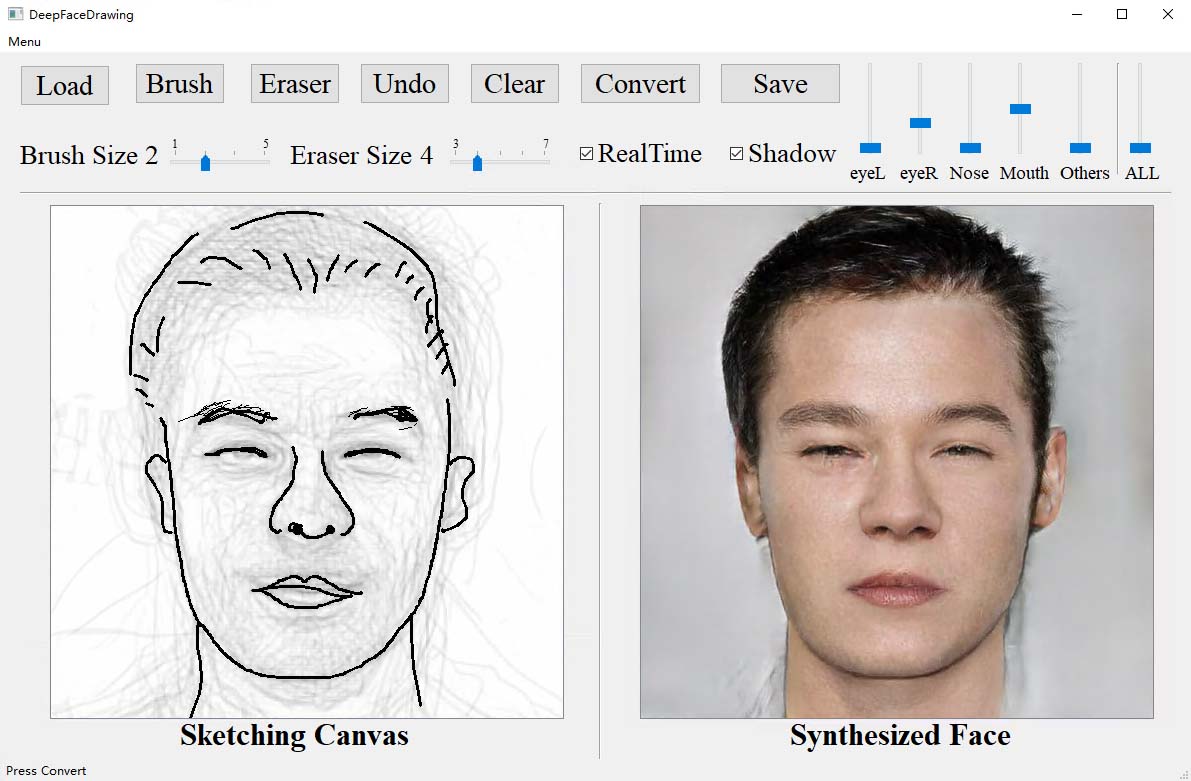}}
    \caption{A screenshot of our shadow-guided sketching interface (Left) for facial image synthesis (Right). The sliders at the up-right corner can be used to control the degree of interpolation between an input sketch and a refined version after manifold project for individual components.}
    \label{fig:interface}
\end{figure}

\subsection{Shadow-guided Sketching Interface}\label{sec:user_interface}
To assist users, especially those with little training in drawing, inspired by \emph{ShadowDraw} \cite{lee2011shadowdraw}, we provide a shadow-guided sketching interface.
Given a current sketch $ \tilde{s}$, we first find $K$ ($K = 10$ in our implementation) most similar sketch component images from $\mathcal{S}$ according to ${\tilde{s}}^c$ by using the Euclidean distance in the feature space. The found component images are then blended as shadow and placed at the corresponding components' positions for sketching guidance (Figure \ref{fig:interface} (Left)).
Initially when the canvas is empty, the shadow is more blurry. The shadow is updated instantly for every new input stroke. The synthesized image is displayed in the window on the right. Users may choose to update the synthesized image instantly or trigger an ``Convert'' command. We show two sequences of sketching and synthesis results in Figure \ref{fig:userStudy1}.

Users with good drawing skills tend to trust their own drawings more than those with little training in drawing. We thus provide a slider for each component type to control the blending weights between a sketched component and its refined version after manifold projection. Let $wb^c$ denote the blending weight for component $c$. The feature vector after blending can be calculated as:
\begin{equation}
    f^c_{blend} = wb^c \times f_{\tilde{s}}^c + ( 1-wb^c )\times f^c_{proj}.
\end{equation}
Feeding $f^c_{blend}$ to the subsequent trained modules, we get a new synthesized image.
\begin{figure}
    \centering
    \includegraphics[width=0.32\linewidth]{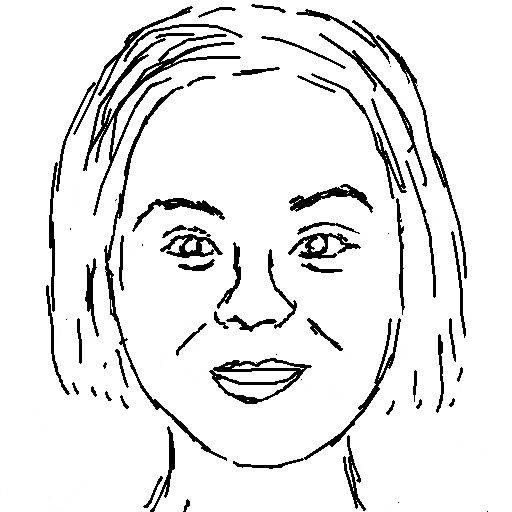}  \includegraphics[width=0.32\linewidth]{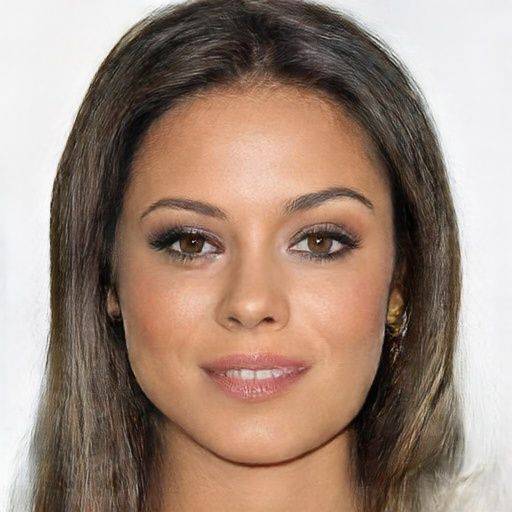} \includegraphics[width=0.32\linewidth]{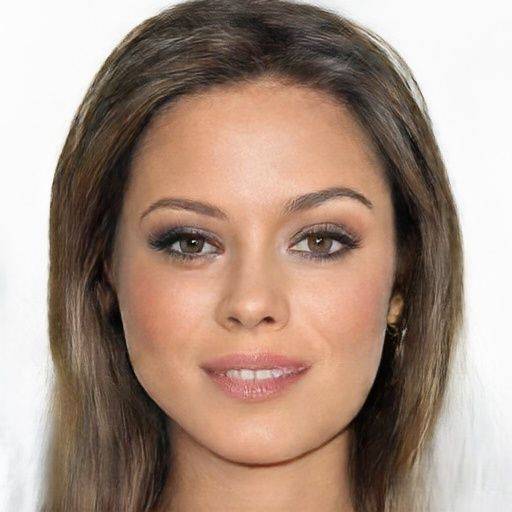} \\

    Input Sketch \qquad  \qquad    $wb^5=0.00$ \qquad \qquad $wb^5=0.25$ \\

    \includegraphics[width=0.32\linewidth]{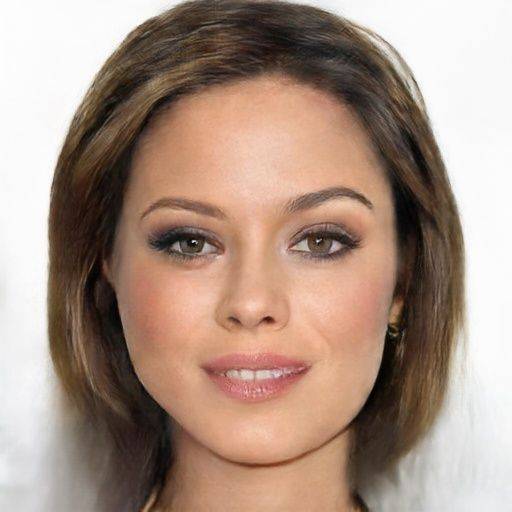}  \includegraphics[width=0.32\linewidth]{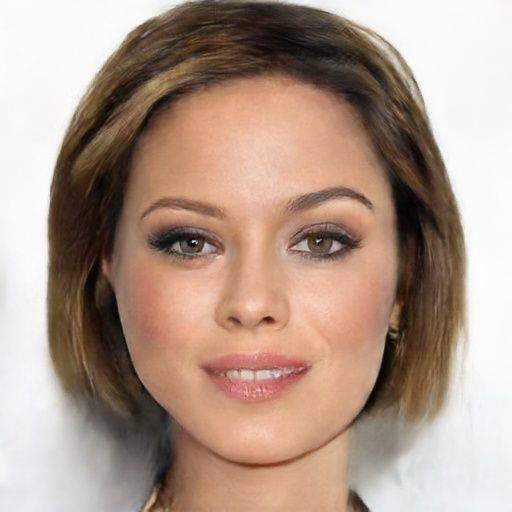} \includegraphics[width=0.32\linewidth]{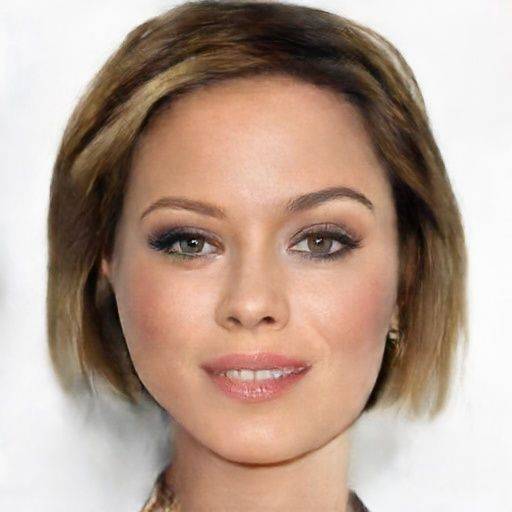}

    $wb^5=0.50$ \qquad \qquad  $wb^5=0.75$ \qquad \qquad  $wb^5=1.00$

    \caption{Interpolating an input sketch and its refined version (for the ``remainder'' component in this example) after manifold projection under different blending weight values. $wb^c = 1$ means a full use of an input sketch for image synthesis, while by setting $wb^c = 0$ we fully trust the data for interpolation.
    }
    \label{fig:my_label}
\end{figure}

\begin{figure}
    \centering
    \setlength{\fboxrule}{1pt}
    \setlength{\fboxsep}{-0.03cm}
    {{\includegraphics[width=0.45\linewidth]{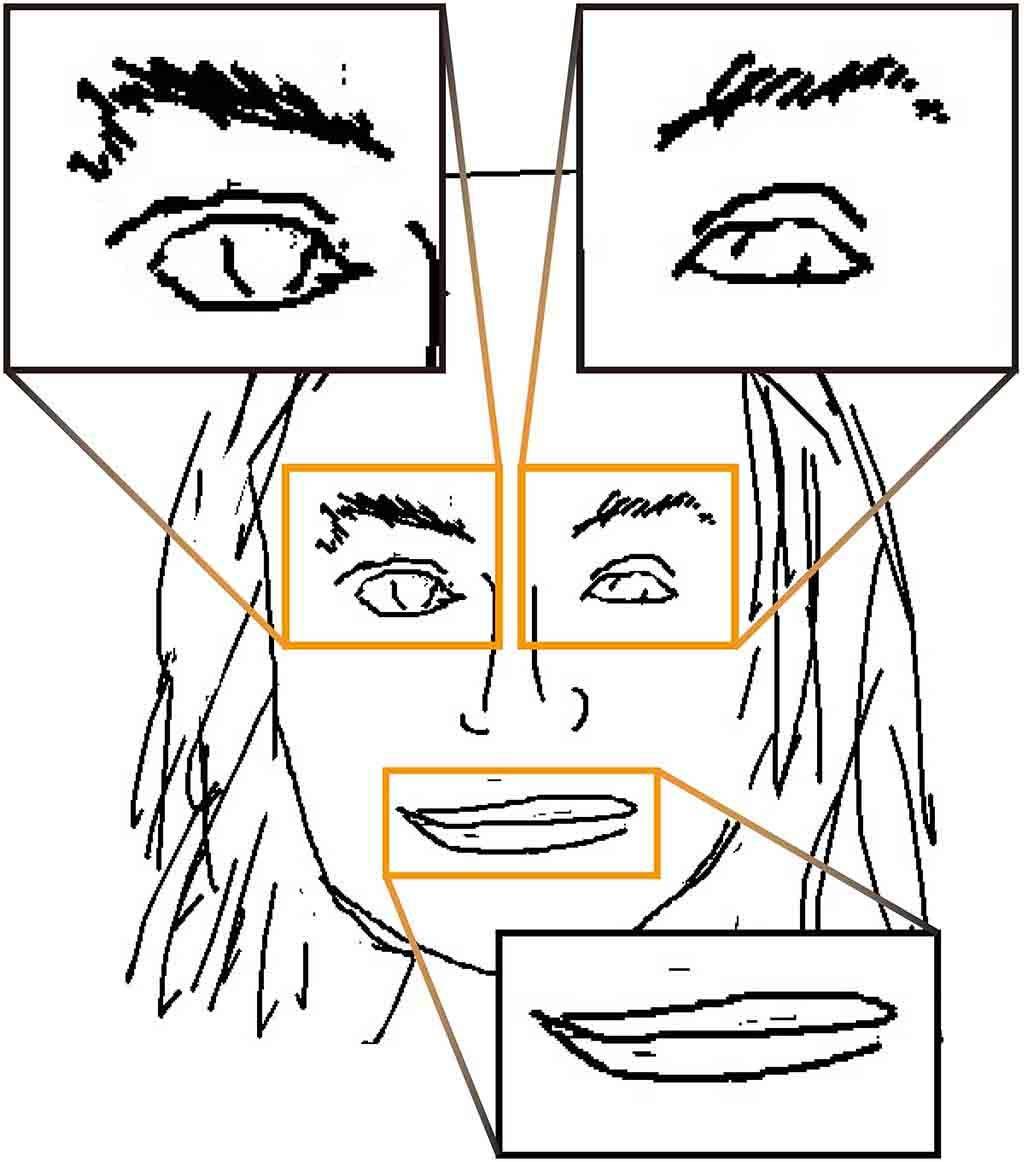}}}
    {{\includegraphics[width=0.45\linewidth]{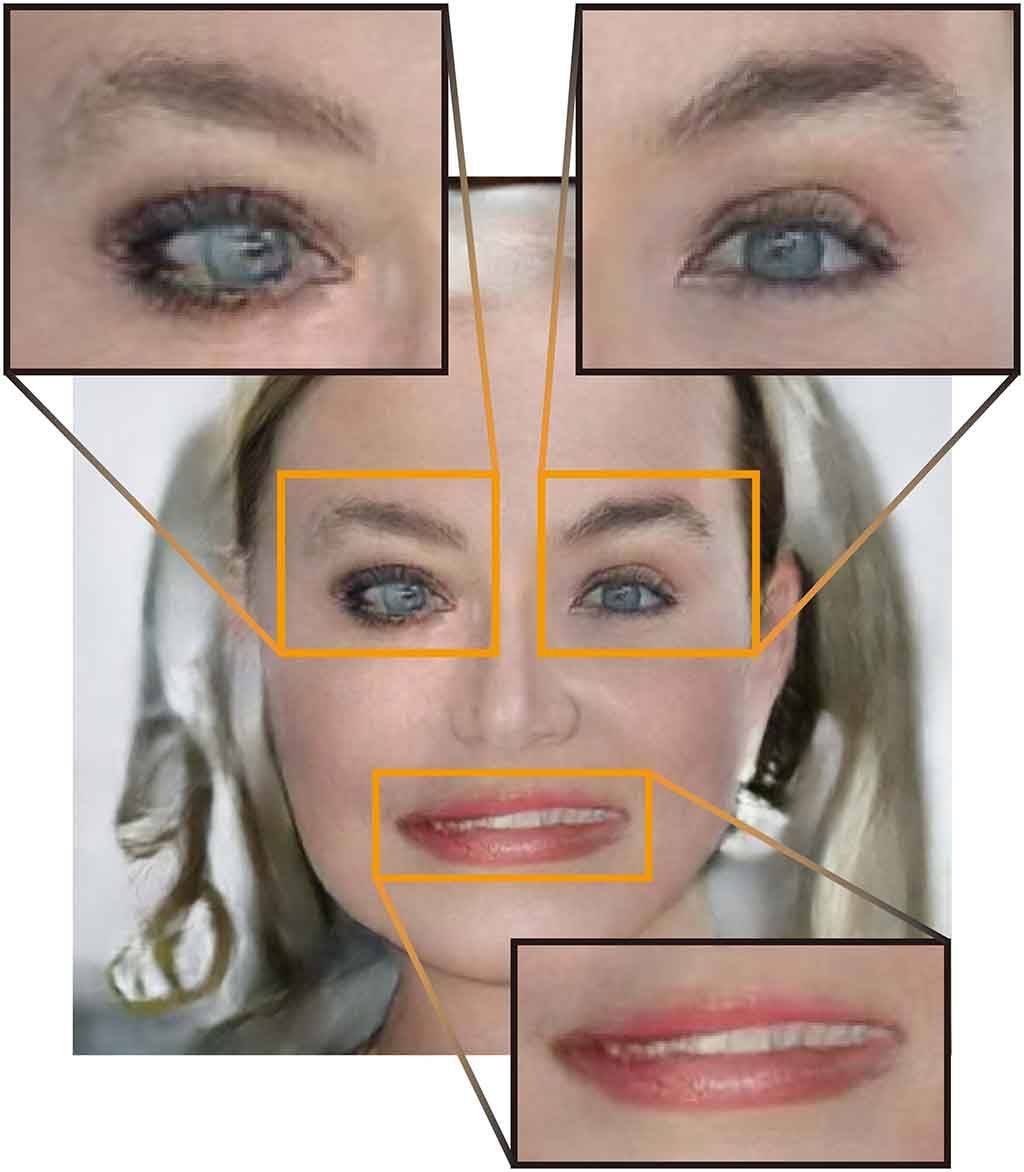}}}\\
    \qquad \qquad Input Sketch \qquad \qquad Without Refinement $wb=1.0$ \\
    {{\includegraphics[width=0.45\linewidth]{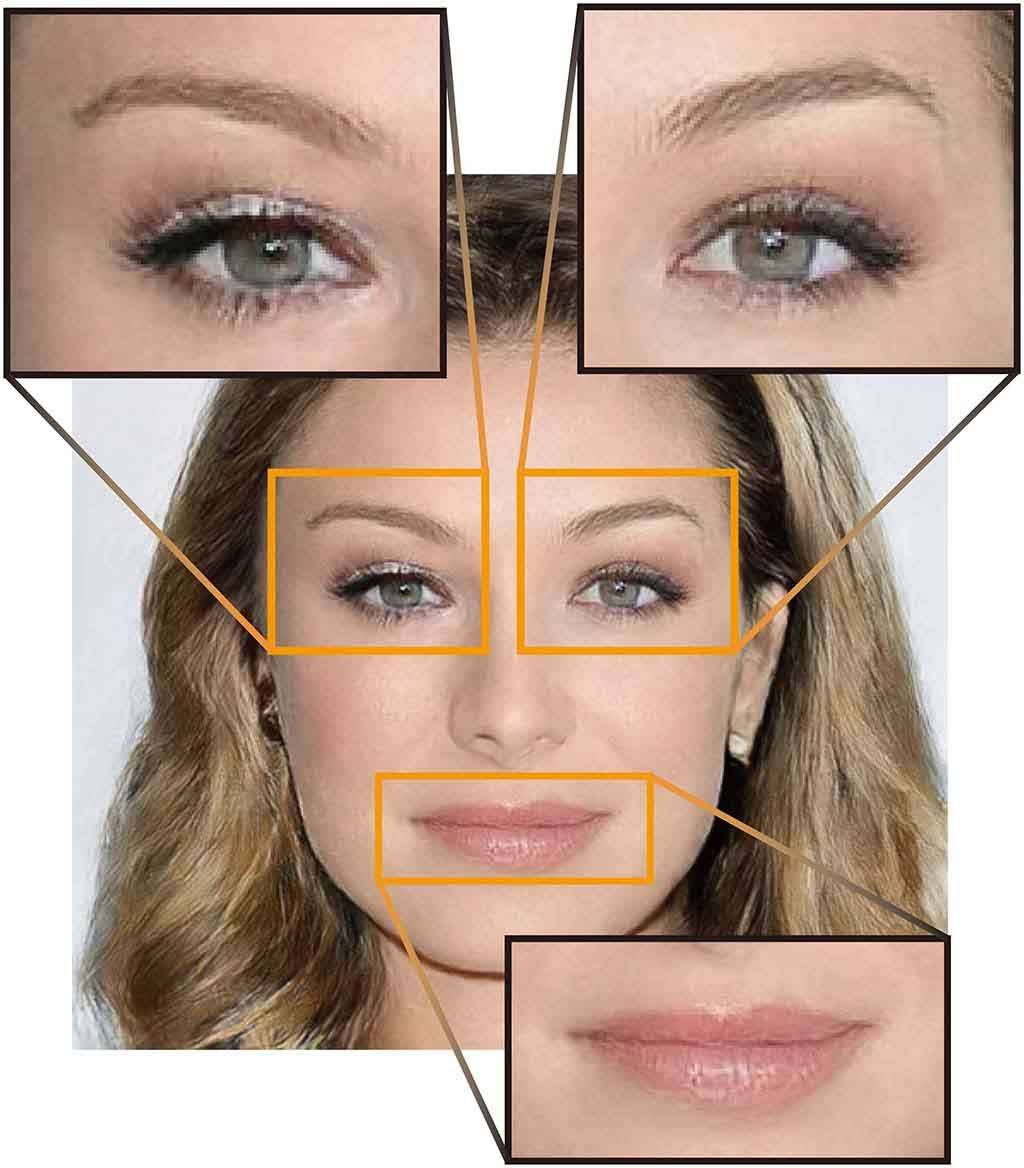}}}
    {{\includegraphics[width=0.45\linewidth]{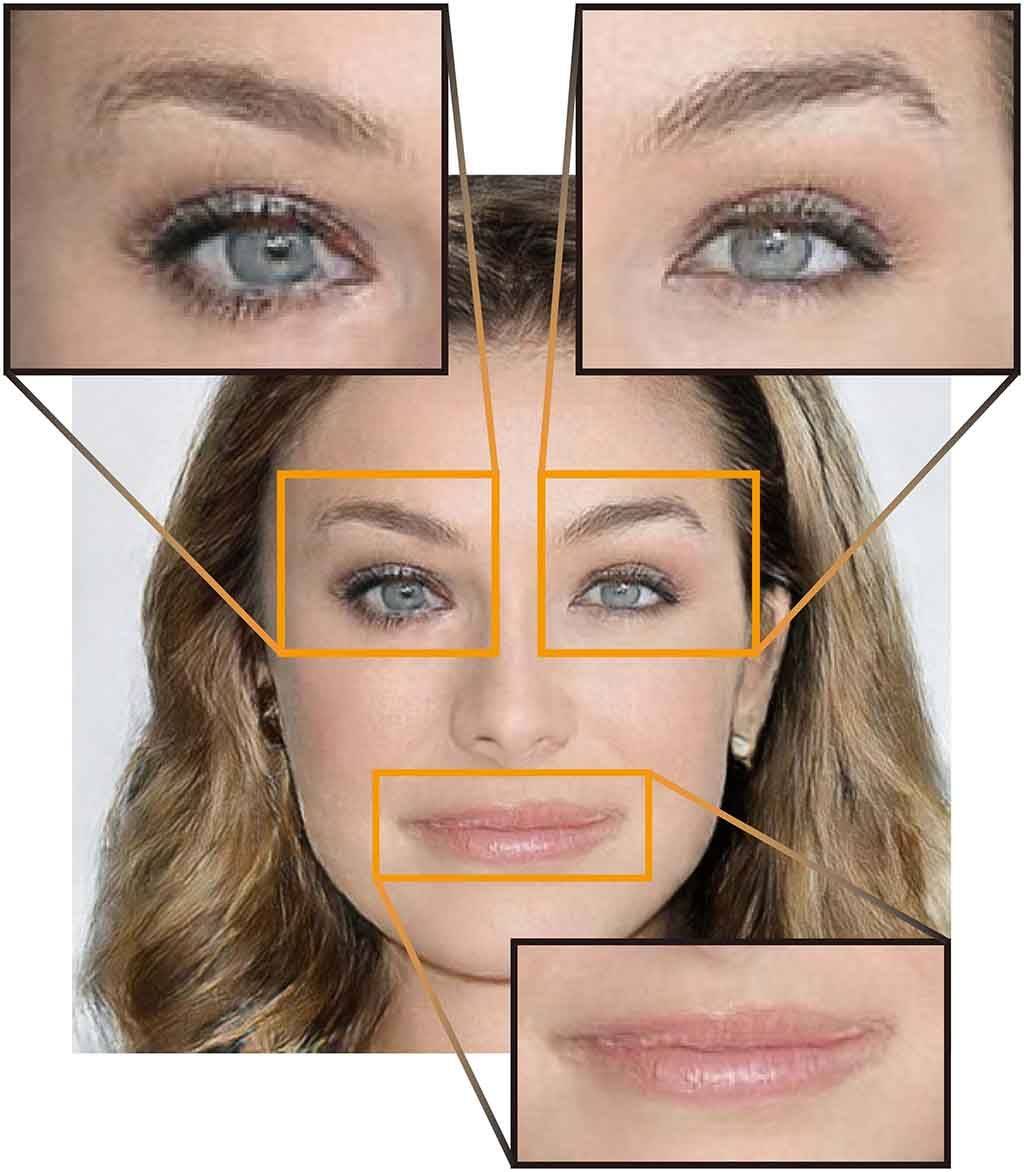}}}
    \\
    Full Refinement $wb=0.0$
    \qquad
    $wb^{1,2,4}=\{0.7,0.4,0.3\}$  \\
    \caption{Blending an input sketch and its refined version after manifold projection for the ``left-eye'', ``right-eye'', and ``mouth'' components. Upper Right: result without any sketch refinement; Lower Left: result with full-degree sketch refinement; Lower Right: result with partial-degree sketch refinement.
    }
    \label{fig:userAdjust}
\end{figure}

Figure \ref{fig:my_label} shows an example of synthesized results under different values of $wb^c$.
This blending feature is particularly useful for creating faces that are very different from any existing samples or their blending. For example, for the female data in our training set, most of the subjects have long hairstyles. Always pushing our input sketch to such samples would not allow us to create short-hairstyle effects. This is solved by trusting the input sketch for its ``remainder'' component by adjusting its corresponding blending weight.
Figure \ref{fig:userAdjust} shows another example with different blending weights for different components. It can be easily seen that the result with automatic refinement (lower left) is visually more realistic than that without any refinement (upper right). Fine-tuning of the blending weights leads to a result better reflecting the input sketch more faithfully.

\section{Experiments}\label{sec:experiments}

We have done extensive evaluations to show the effectiveness of our sketch-to-image face synthesis system and its usability via a pilot study. Below we present some of the obtained results. Please refer to the supplemental materials for more results and an accompanying video for sketch-based image synthesis in action.

\begin{figure}
    \centering
    \setlength{\fboxrule}{0.5pt}
    \setlength{\fboxsep}{-0.01cm}
    \framebox{\includegraphics[width=0.32\linewidth]{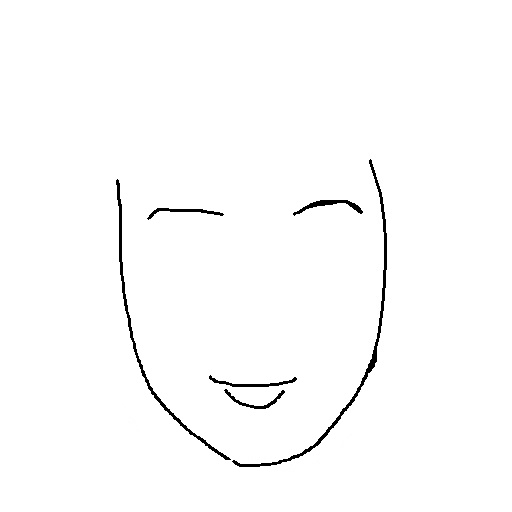}}
    \framebox{\includegraphics[width=0.32\linewidth]{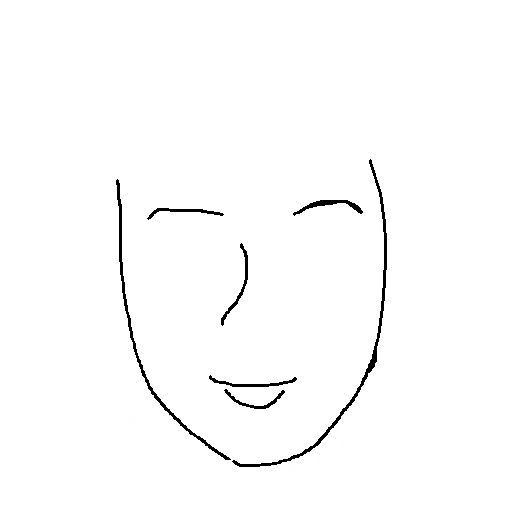}}
    \framebox{\includegraphics[width=0.32\linewidth]{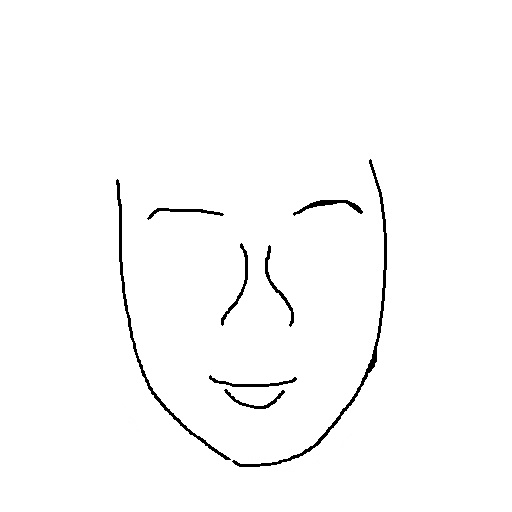}}

    \framebox{\includegraphics[width=0.32\linewidth]{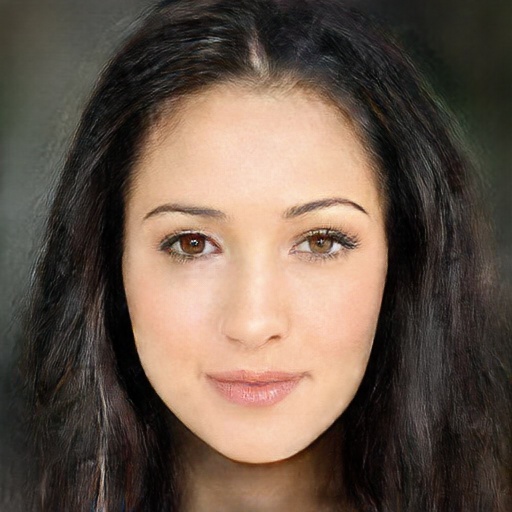}}
    \framebox{\includegraphics[width=0.32\linewidth]{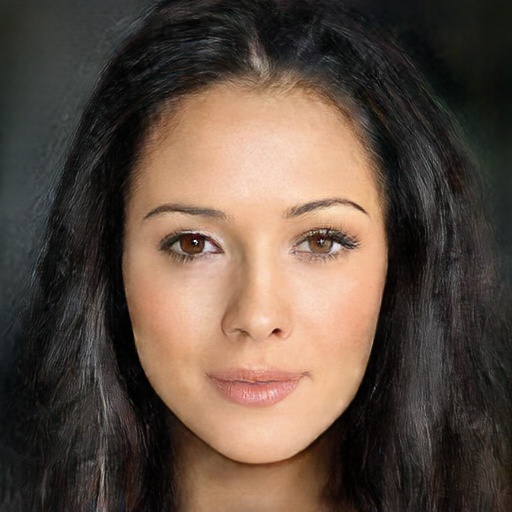}}
    \framebox{\includegraphics[width=0.32\linewidth]{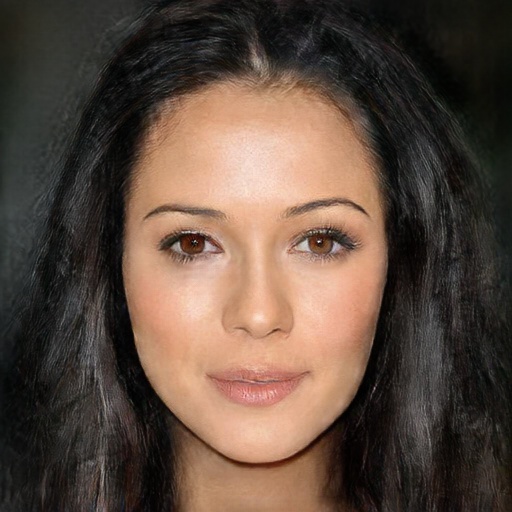}}

    \framebox{\includegraphics[width=0.32\linewidth]{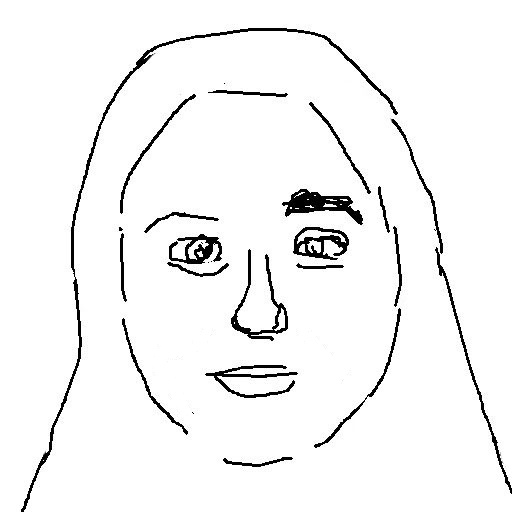}}
    \framebox{\includegraphics[width=0.32\linewidth]{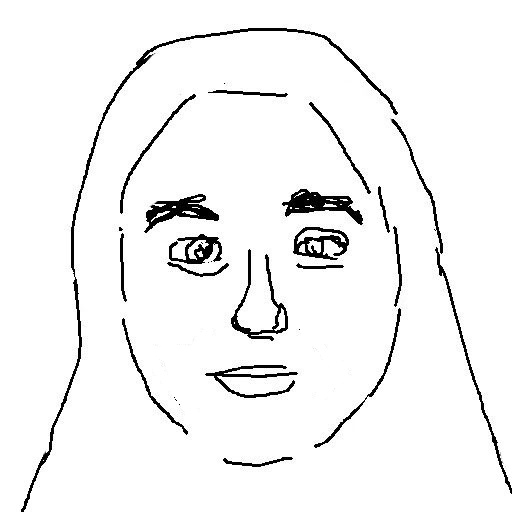}}
    \framebox{\includegraphics[width=0.32\linewidth]{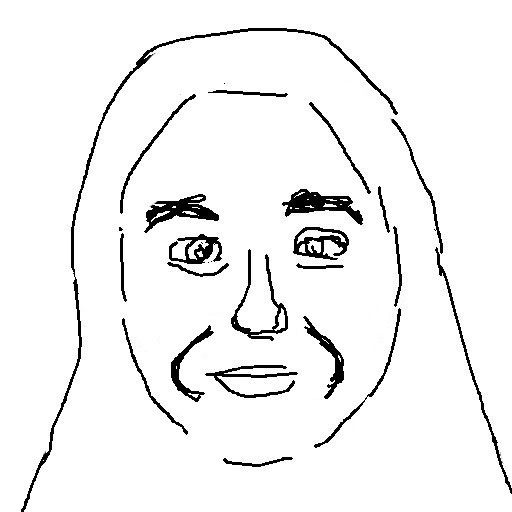}}

    \framebox{\includegraphics[width=0.32\linewidth]{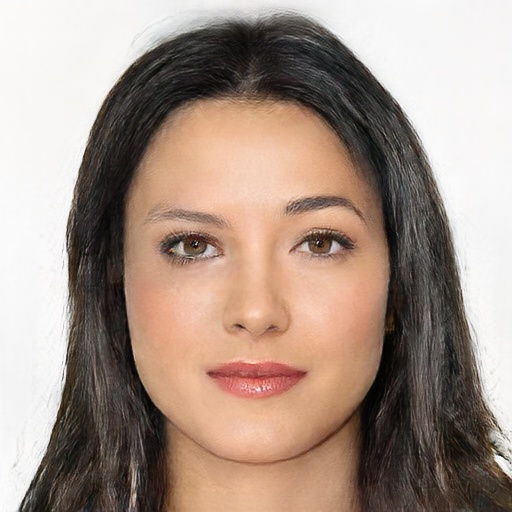}}
    \framebox{\includegraphics[width=0.32\linewidth]{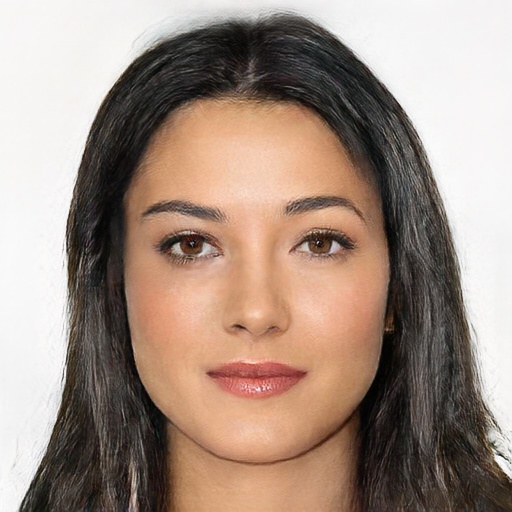}}
    \framebox{\includegraphics[width=0.32\linewidth]{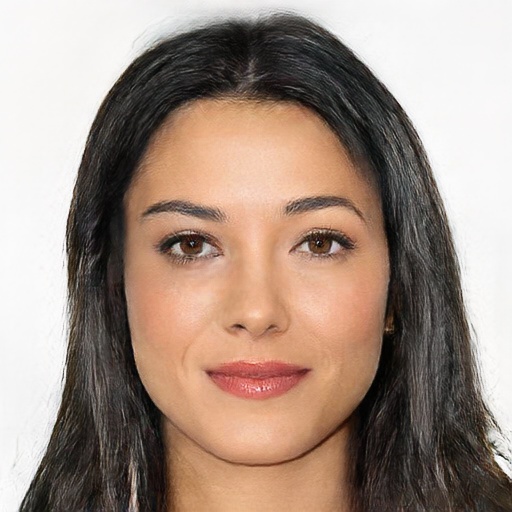}}
    \caption{Representative results through progressive sketching for adding details (1st example) and stressing local details (2nd example).
    }
    \label{fig:local_edit}
\end{figure}

\begin{figure*}
    \centering
    \setlength{\fboxrule}{0.5pt}
    \setlength{\fboxsep}{-0.01cm}

    \begin{minipage}{\linewidth}
    \framebox{\includegraphics[width=0.1965\linewidth]{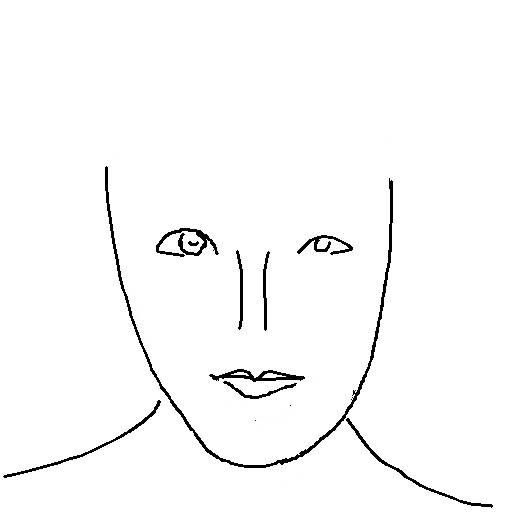}}
    \framebox{\includegraphics[width=0.1965\linewidth]{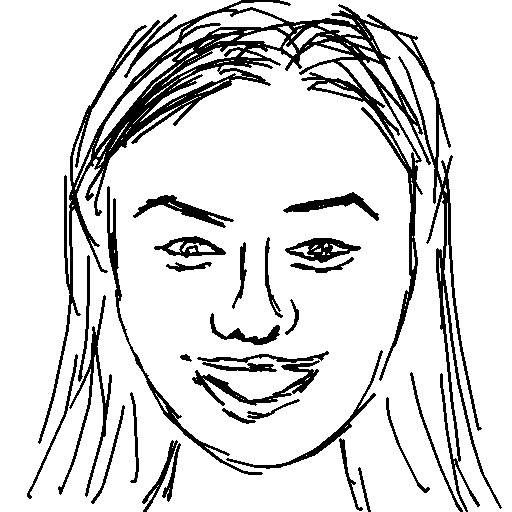}}
    \framebox{\includegraphics[width=0.1965\linewidth]{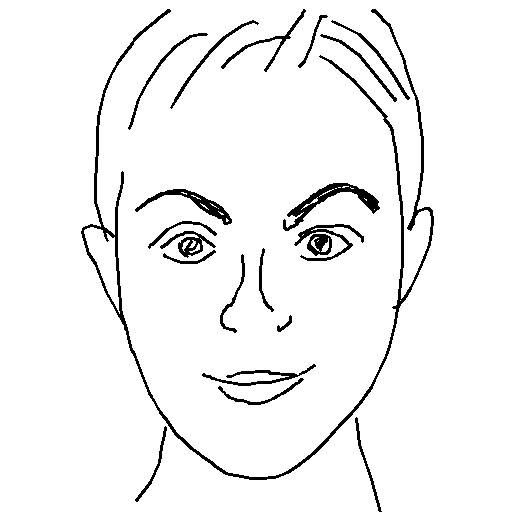}}
    \framebox{\includegraphics[width=0.1965\linewidth]{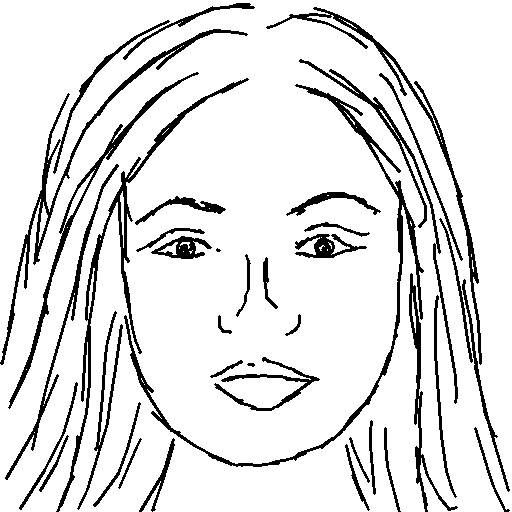}}
    \framebox{\includegraphics[width=0.1965\linewidth]{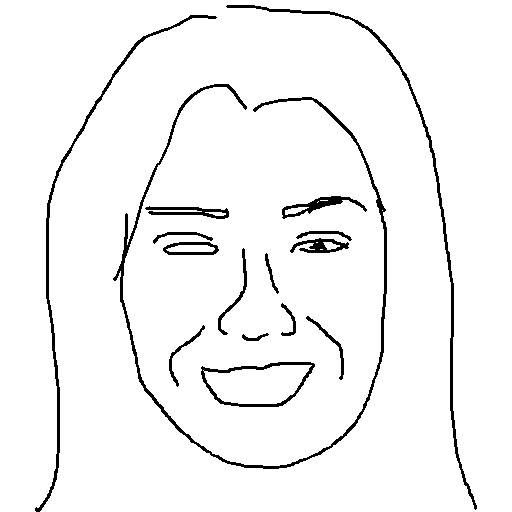}}
    \end{minipage}
    \begin{minipage}{\linewidth}
    \framebox{\includegraphics[width=0.1965\linewidth]{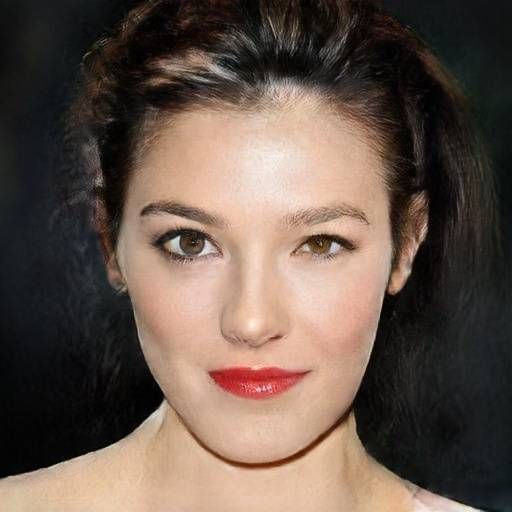}}
    \framebox{\includegraphics[width=0.1965\linewidth]{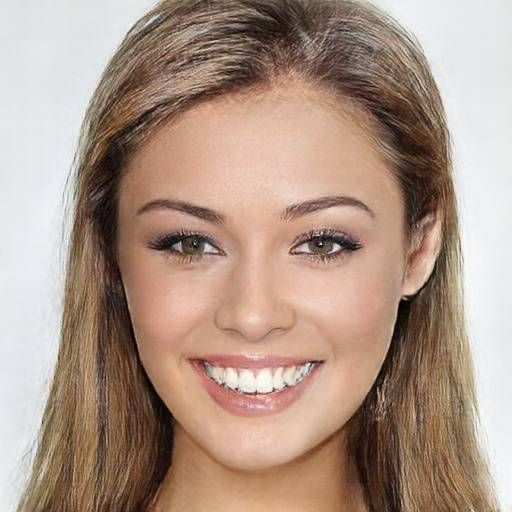}}
    \framebox{\includegraphics[width=0.1965\linewidth]{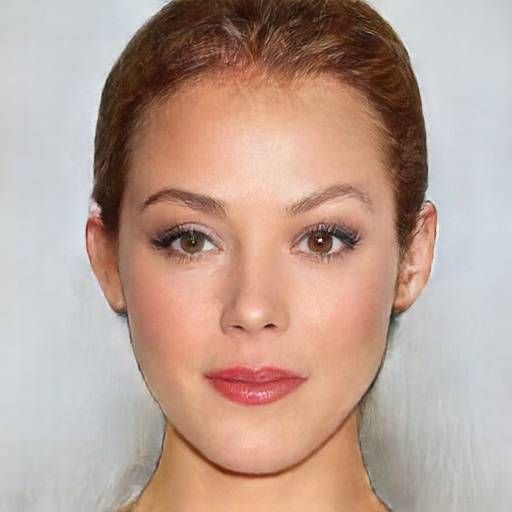}}
    \framebox{\includegraphics[width=0.1965\linewidth]{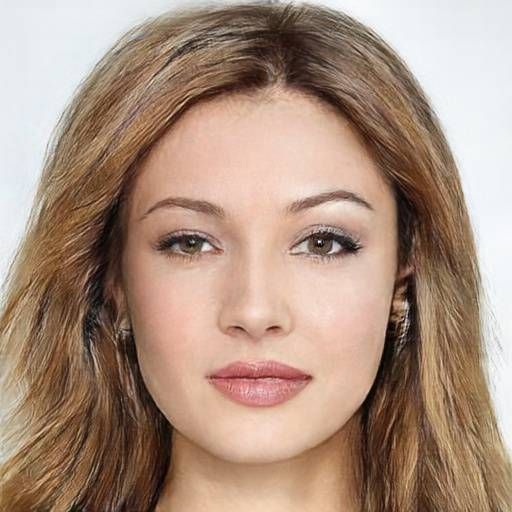}}
    \framebox{\includegraphics[width=0.1965\linewidth]{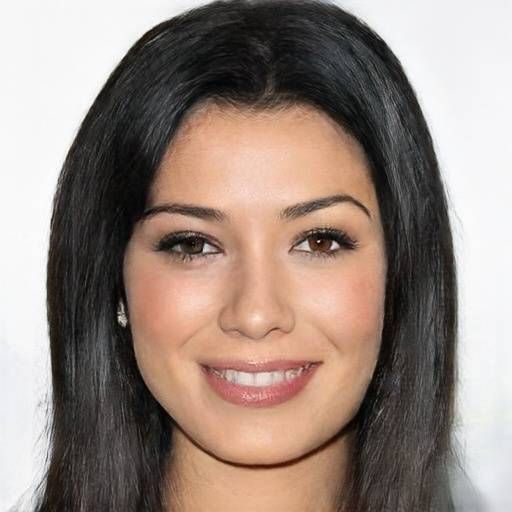}}
    \end{minipage}

    \begin{minipage}{\linewidth}
    \framebox{\includegraphics[width=0.1965\linewidth]{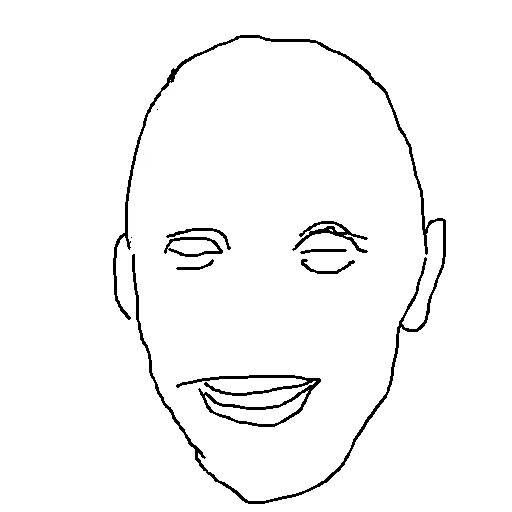}}
    \framebox{\includegraphics[width=0.1965\linewidth]{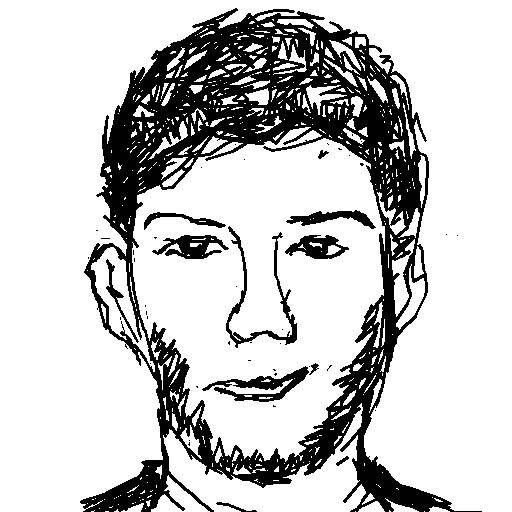}}
    \framebox{\includegraphics[width=0.1965\linewidth]{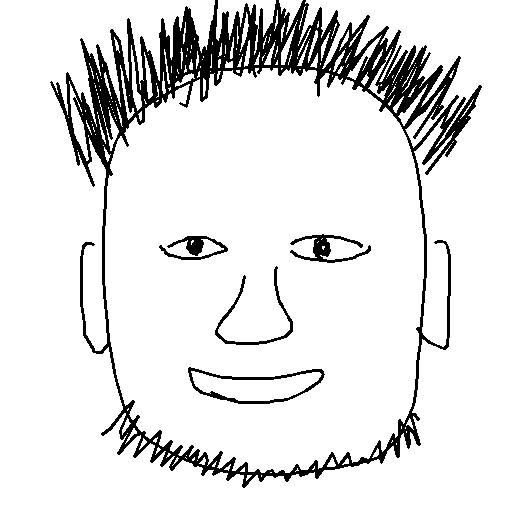}}
    \framebox{\includegraphics[width=0.1965\linewidth]{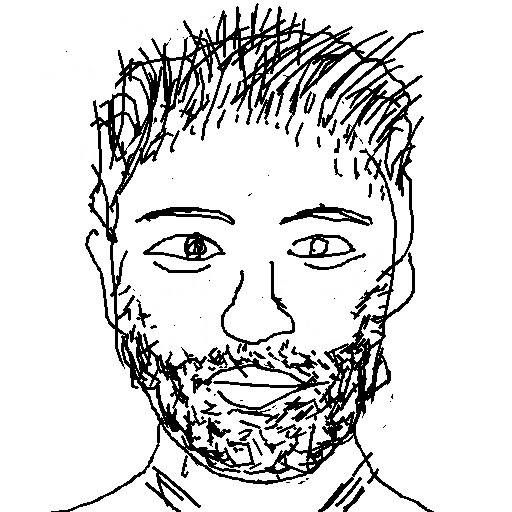}}
    \framebox{\includegraphics[width=0.1965\linewidth]{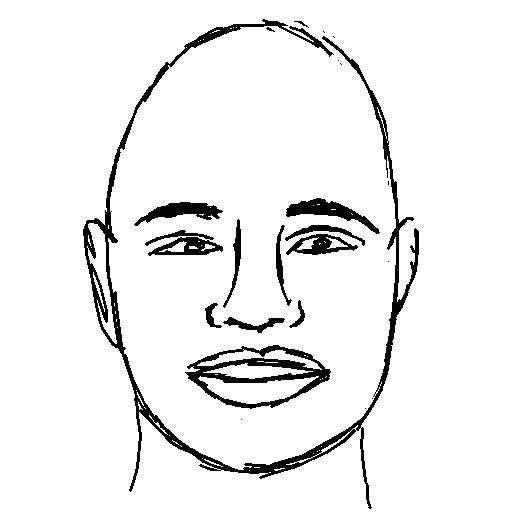}}
    \end{minipage}
    \begin{minipage}{\linewidth}
    \framebox{\includegraphics[width=0.1965\linewidth]{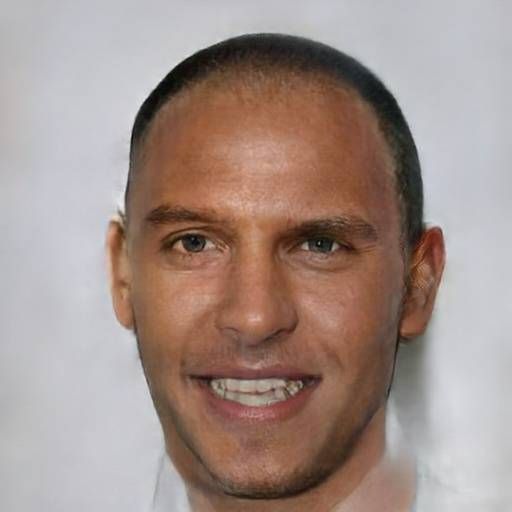}}
    \framebox{\includegraphics[width=0.1965\linewidth]{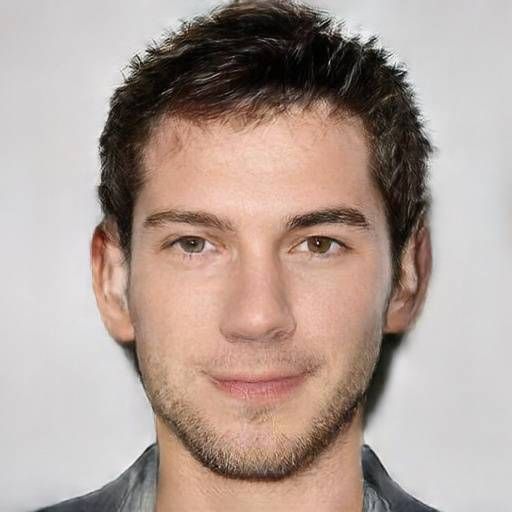}}
    \framebox{\includegraphics[width=0.1965\linewidth]{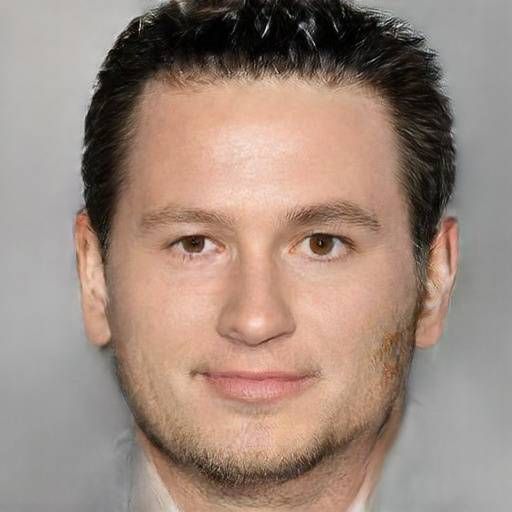}}
    \framebox{\includegraphics[width=0.1965\linewidth]{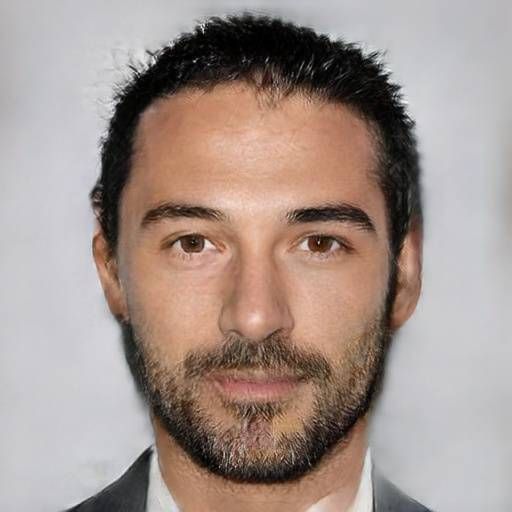}}
    \framebox{\includegraphics[width=0.1965\linewidth]{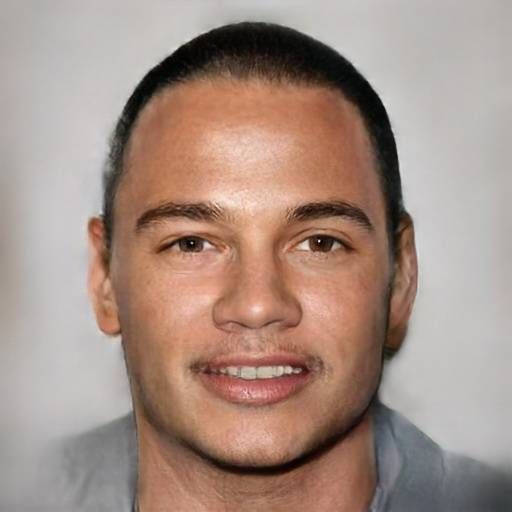}}
    \end{minipage}

    \begin{minipage}{\linewidth}
    \framebox{\includegraphics[width=0.1965\linewidth]{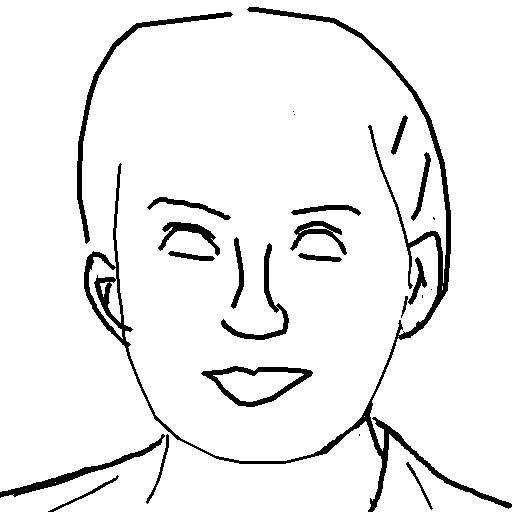}}
    \framebox{\includegraphics[width=0.1965\linewidth]{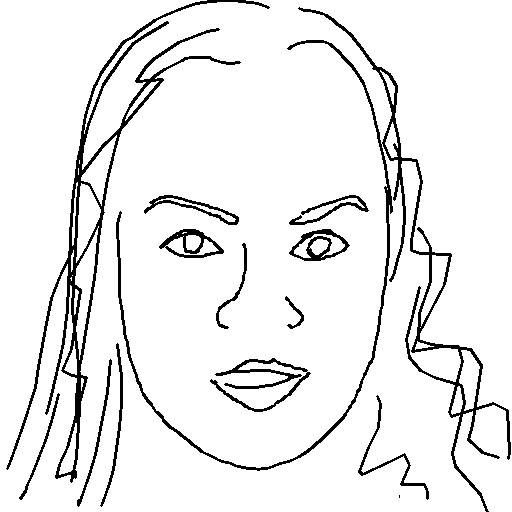}}
    \framebox{\includegraphics[width=0.1965\linewidth]{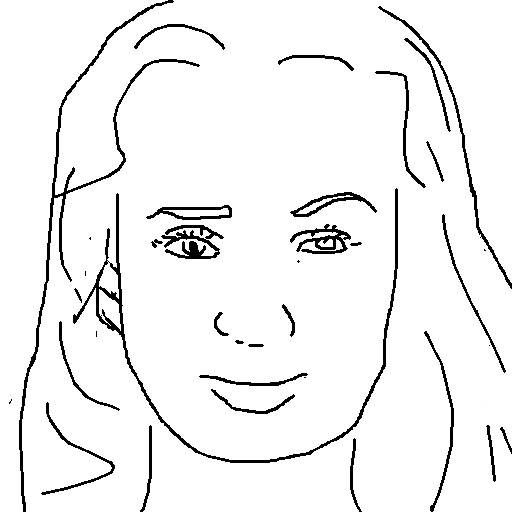}}
    \framebox{\includegraphics[width=0.1965\linewidth]{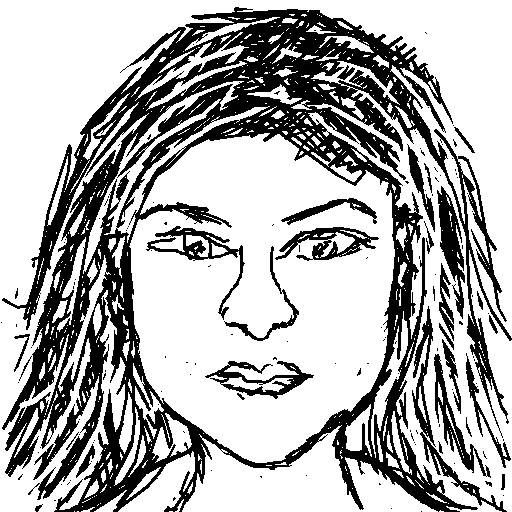}}
    \framebox{\includegraphics[width=0.1965\linewidth]{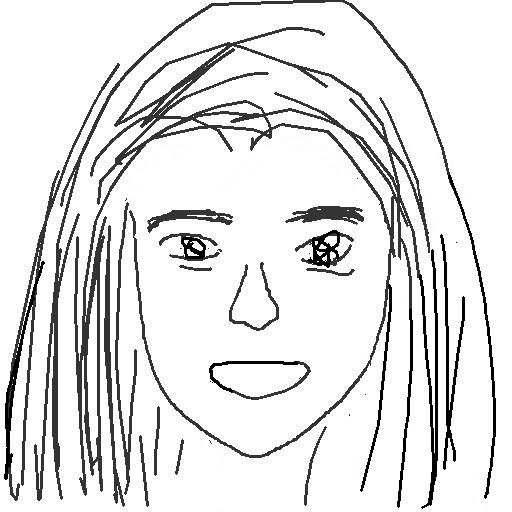}}
    \end{minipage}
    \begin{minipage}{\linewidth}
    \framebox{\includegraphics[width=0.1965\linewidth]{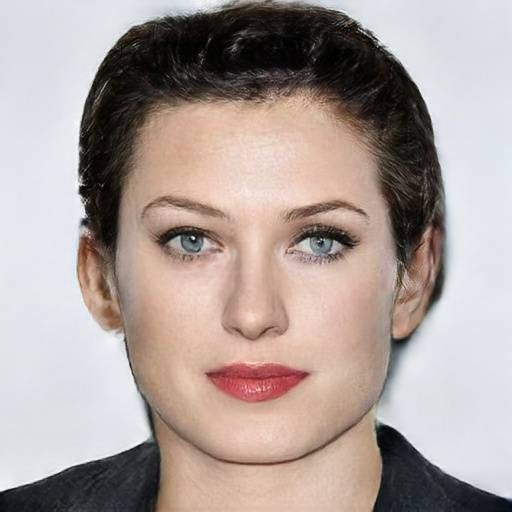}}
    \framebox{\includegraphics[width=0.1965\linewidth]{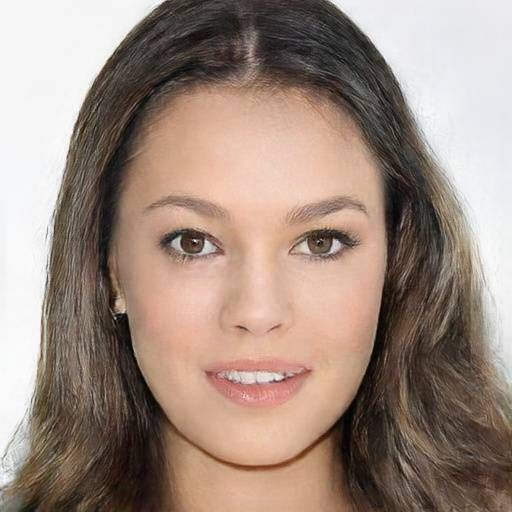}}
    \framebox{\includegraphics[width=0.1965\linewidth]{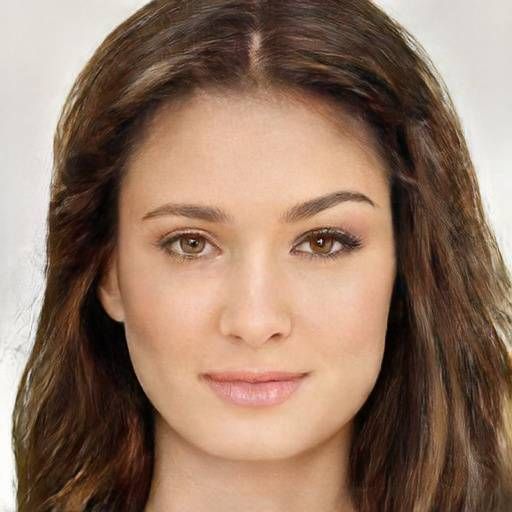}}
    \framebox{\includegraphics[width=0.1965\linewidth]{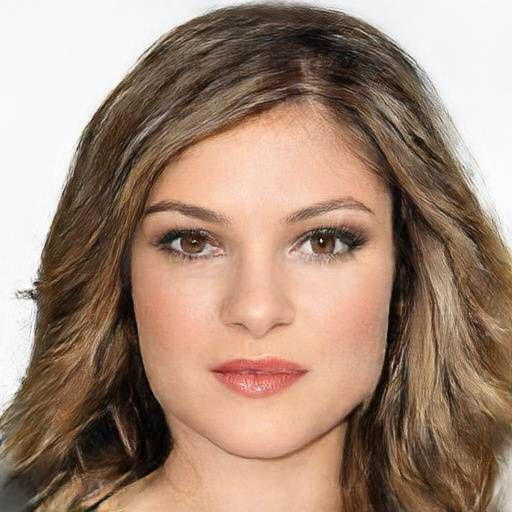}}
    \framebox{\includegraphics[width=0.1965\linewidth]{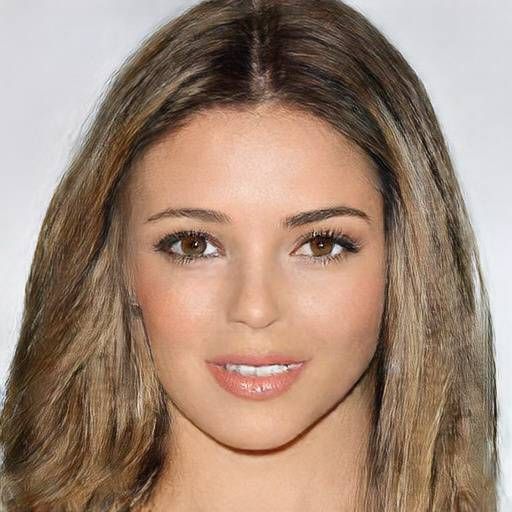}}
    \end{minipage}
    \caption{Gallery of input sketches and synthesized results in the usability study.
    }
    \label{fig:results_man}
\end{figure*}

\begin{figure*}
    \centering
    {\includegraphics[width=0.12\linewidth]{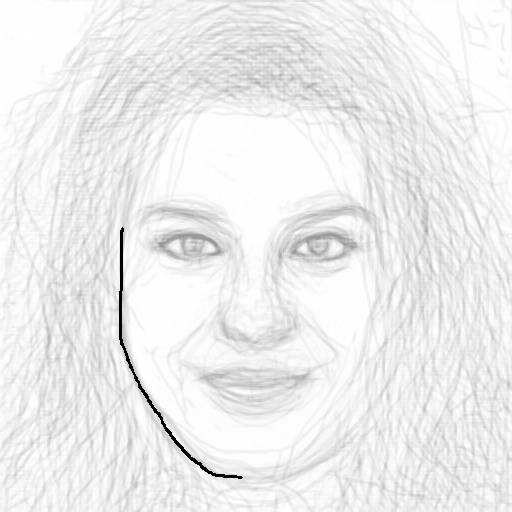}}
    {\includegraphics[width=0.12\linewidth]{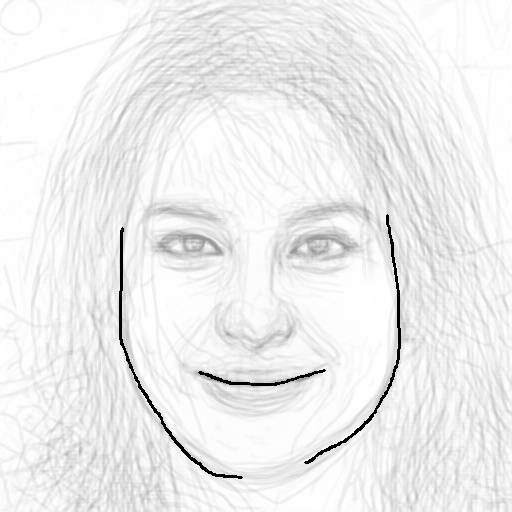}}
    {\includegraphics[width=0.12\linewidth]{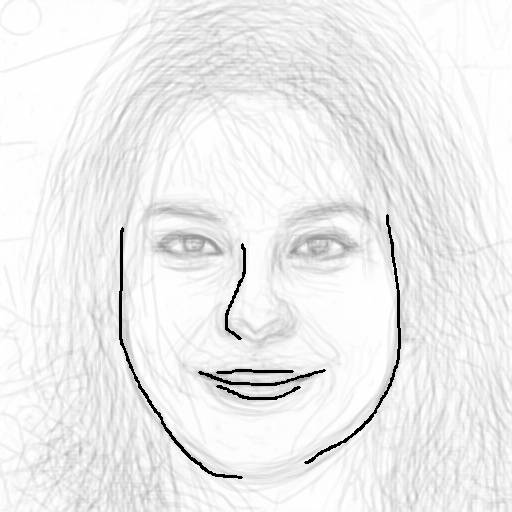}}
    {\includegraphics[width=0.12\linewidth]{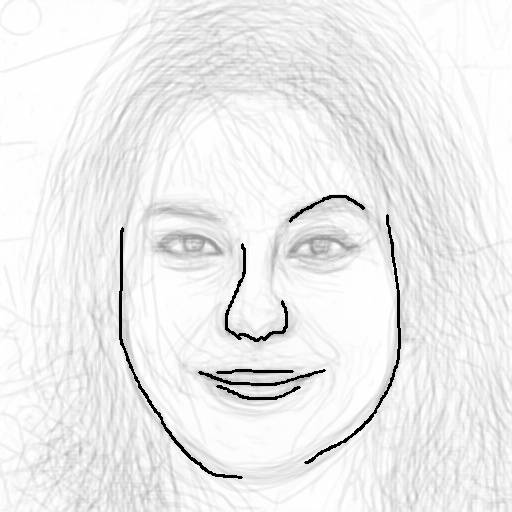}}
    {\includegraphics[width=0.12\linewidth]{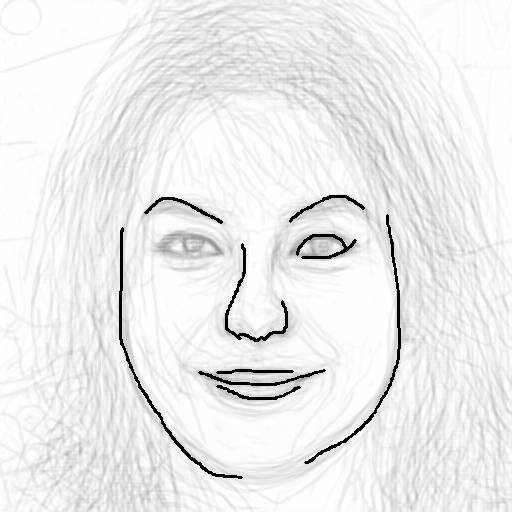}}
    {\includegraphics[width=0.12\linewidth]{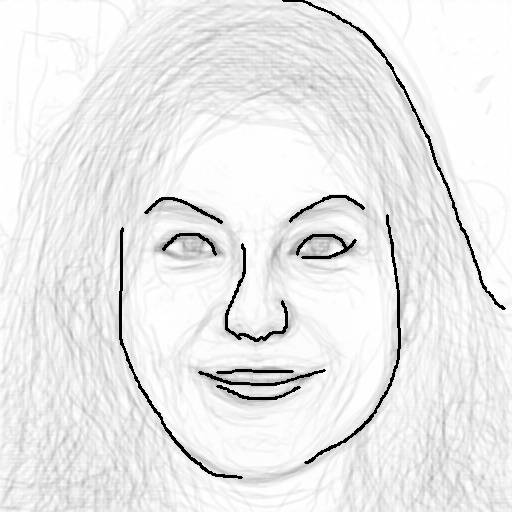}}
    {\includegraphics[width=0.12\linewidth]{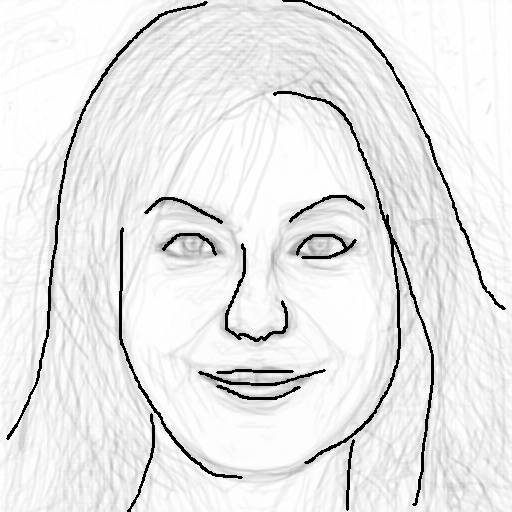}}
    {\includegraphics[width=0.12\linewidth]{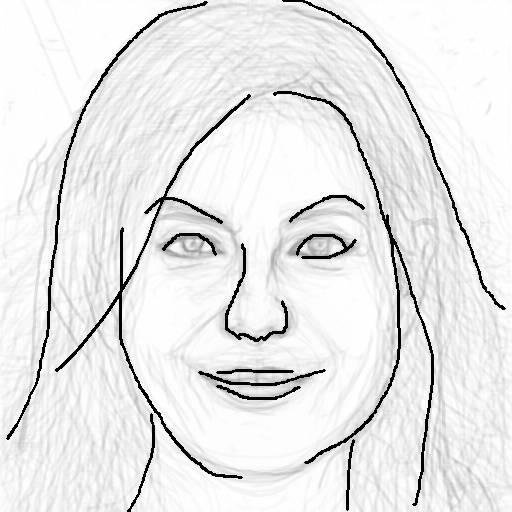}}

    {\includegraphics[width=0.12\linewidth]{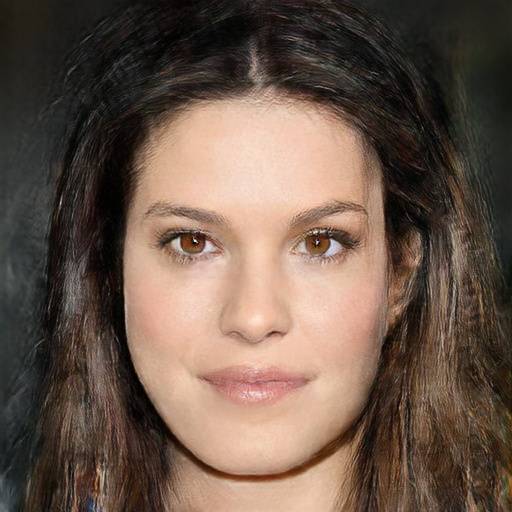}}
    {\includegraphics[width=0.12\linewidth]{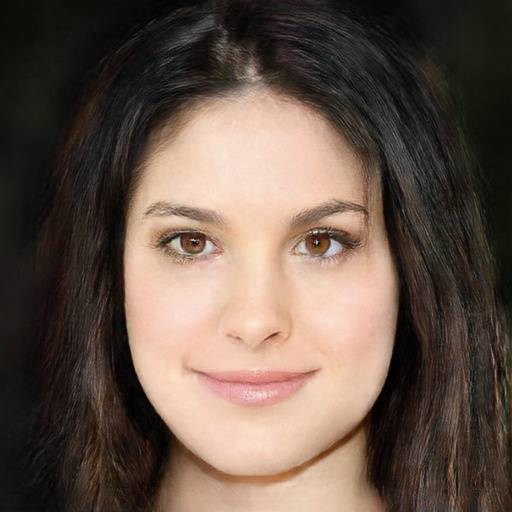}}
    {\includegraphics[width=0.12\linewidth]{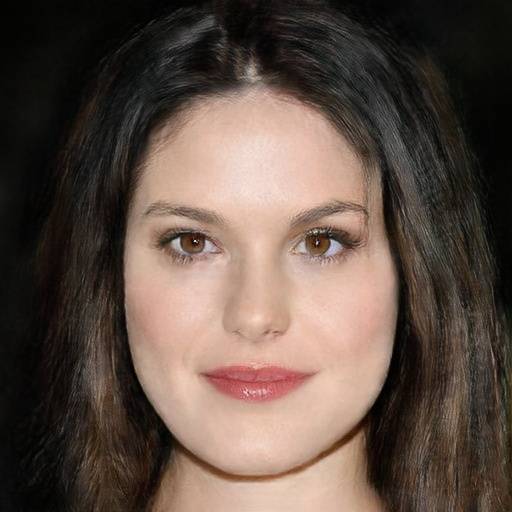}}
    {\includegraphics[width=0.12\linewidth]{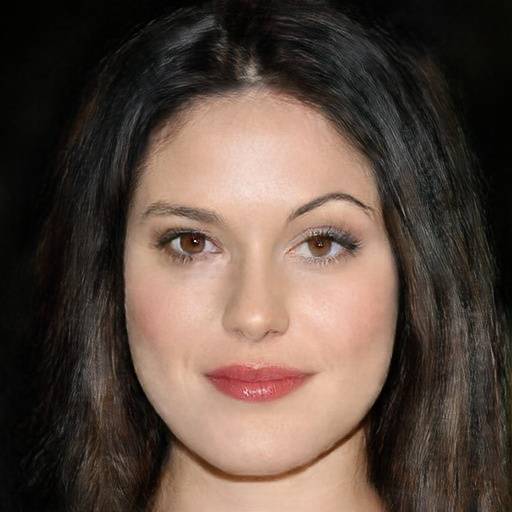}}
    {\includegraphics[width=0.12\linewidth]{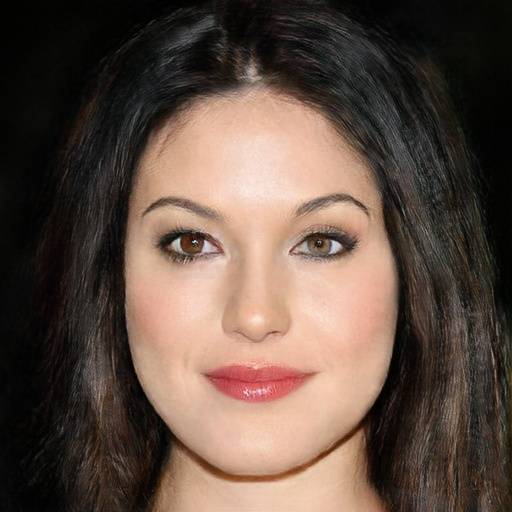}}
    {\includegraphics[width=0.12\linewidth]{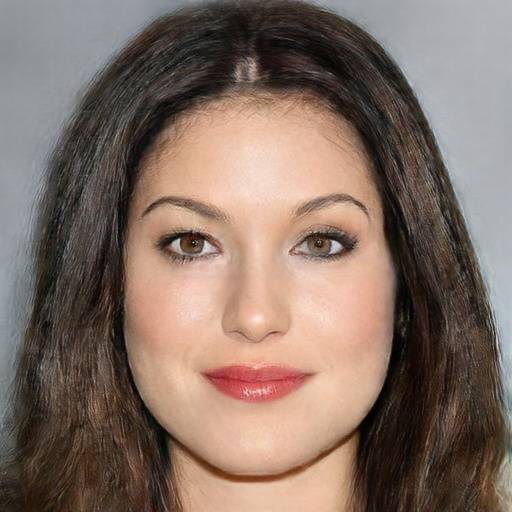}}
    {\includegraphics[width=0.12\linewidth]{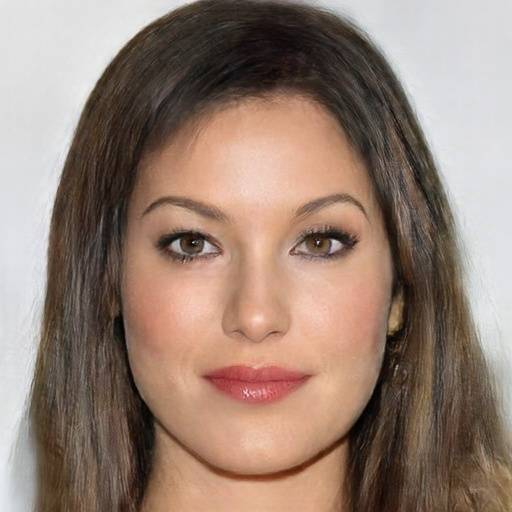}}
    {\includegraphics[width=0.12\linewidth]{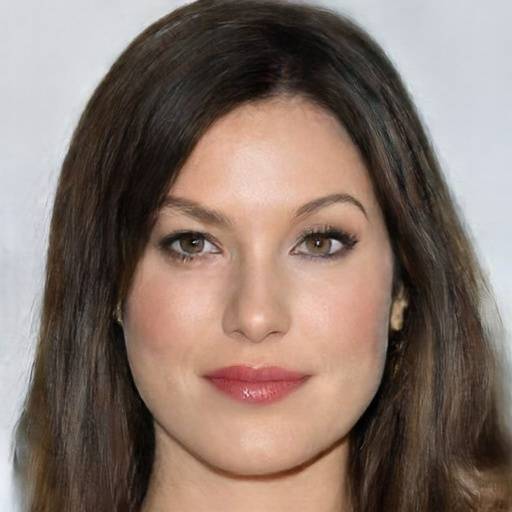}}

    {\includegraphics[width=0.12\linewidth]{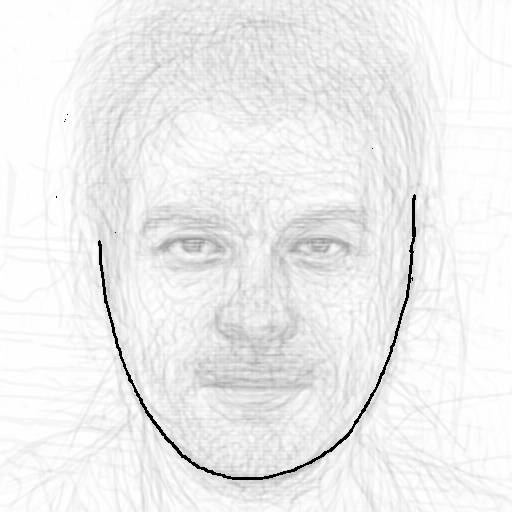}}
    {\includegraphics[width=0.12\linewidth]{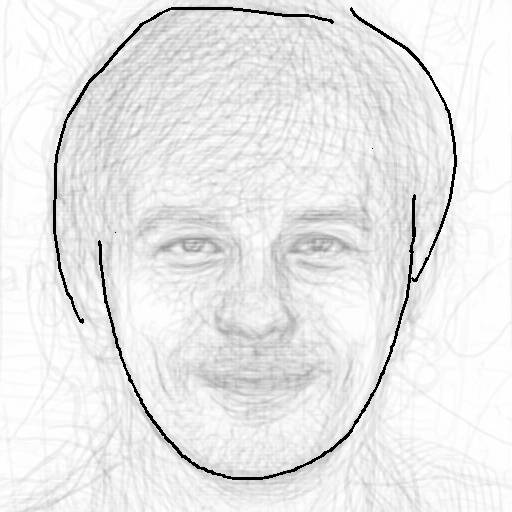}}
    {\includegraphics[width=0.12\linewidth]{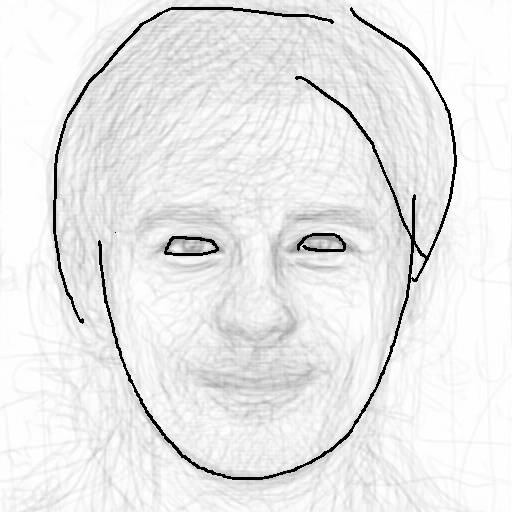}}
    {\includegraphics[width=0.12\linewidth]{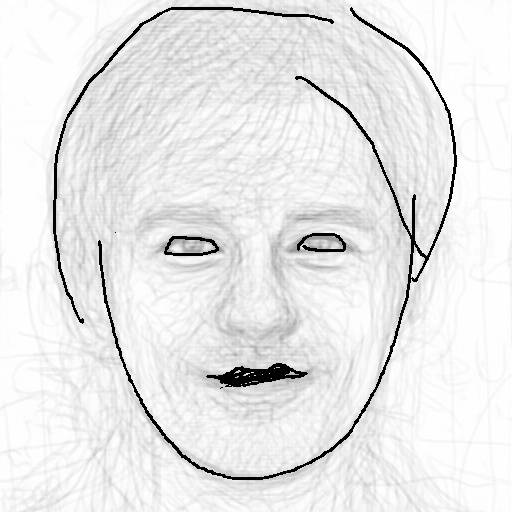}}
    {\includegraphics[width=0.12\linewidth]{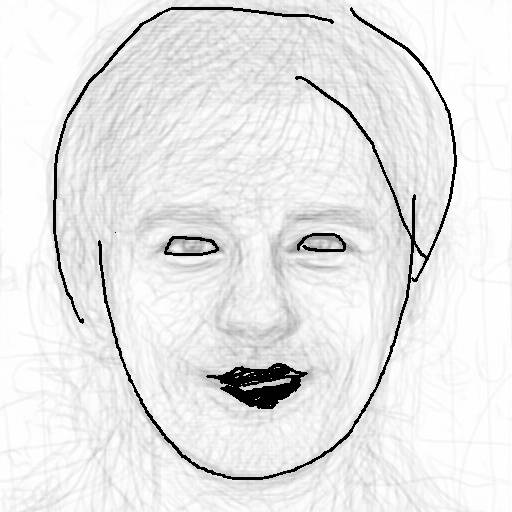}}
    {\includegraphics[width=0.12\linewidth]{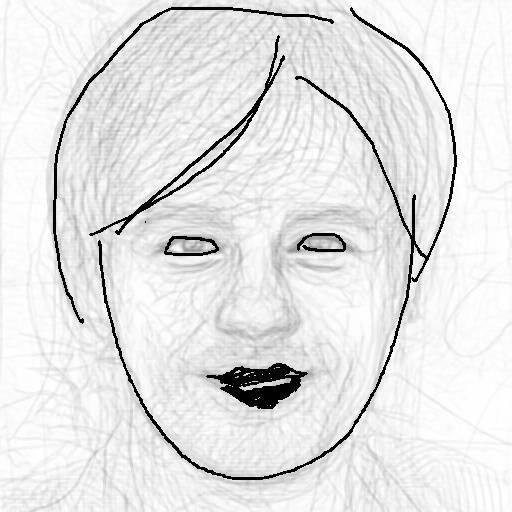}}
    {\includegraphics[width=0.12\linewidth]{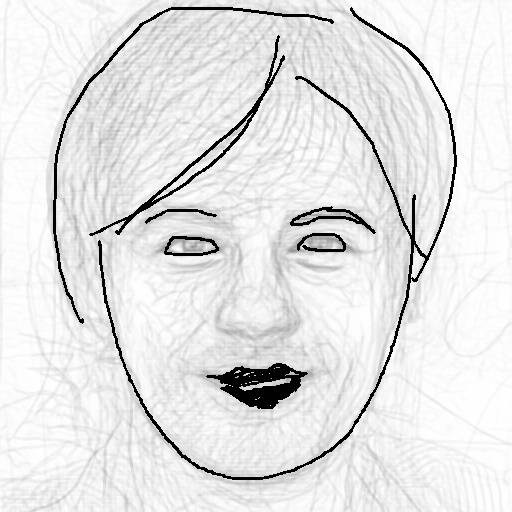}}
    {\includegraphics[width=0.12\linewidth]{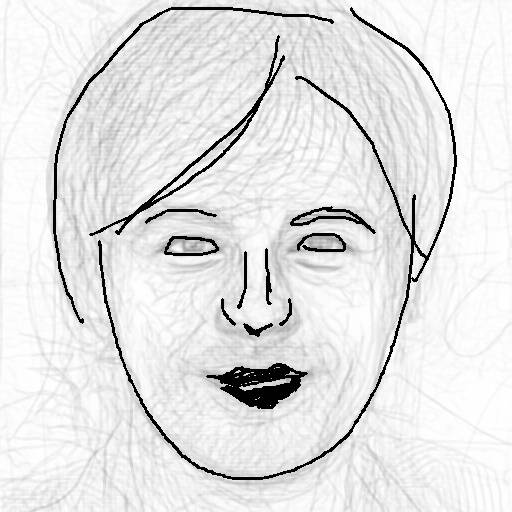}}

    {\includegraphics[width=0.12\linewidth]{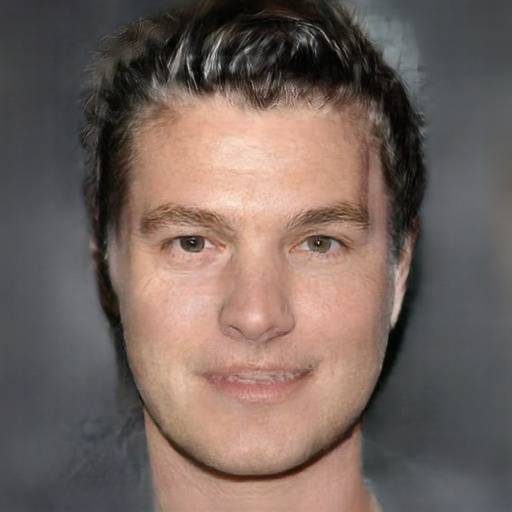}}
    {\includegraphics[width=0.12\linewidth]{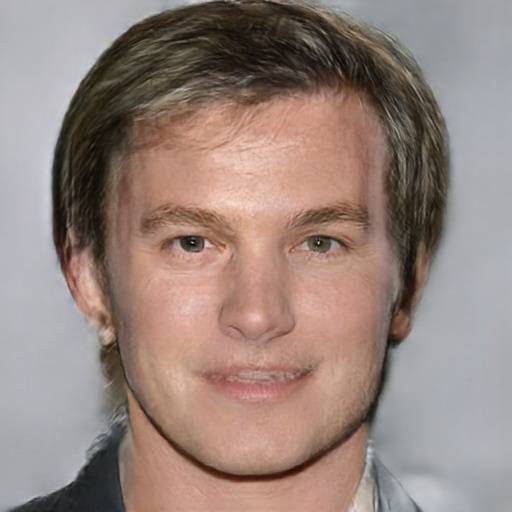}}
    {\includegraphics[width=0.12\linewidth]{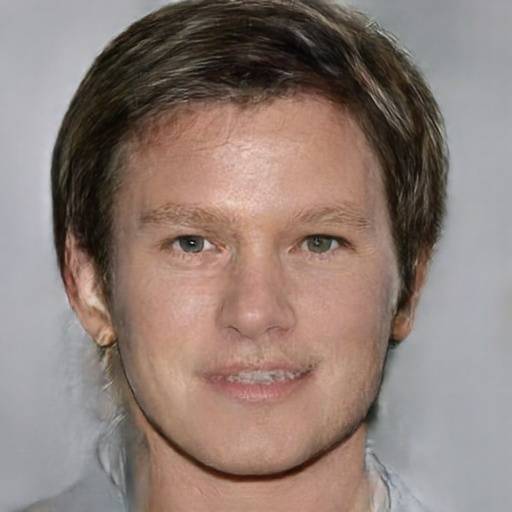}}
    {\includegraphics[width=0.12\linewidth]{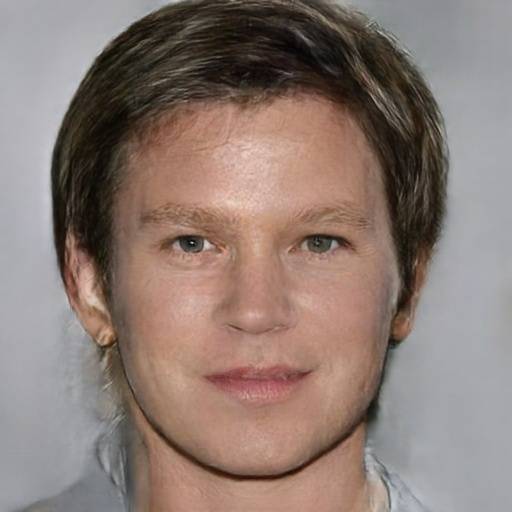}}
    {\includegraphics[width=0.12\linewidth]{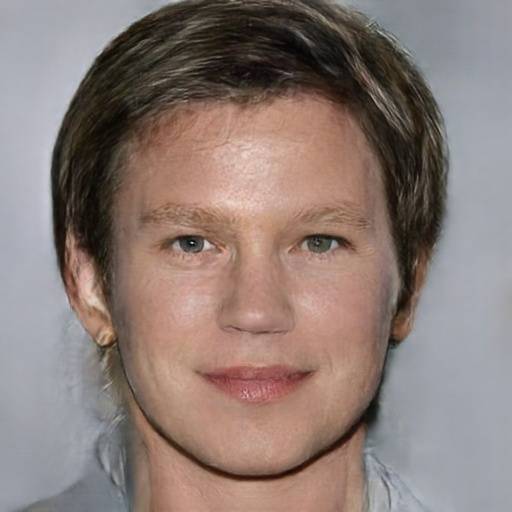}}
    {\includegraphics[width=0.12\linewidth]{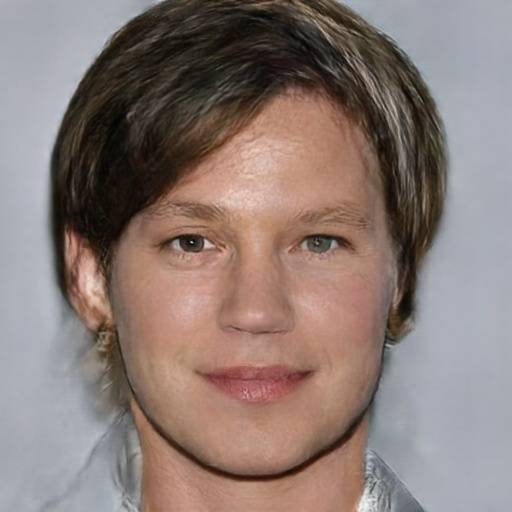}}
    {\includegraphics[width=0.12\linewidth]{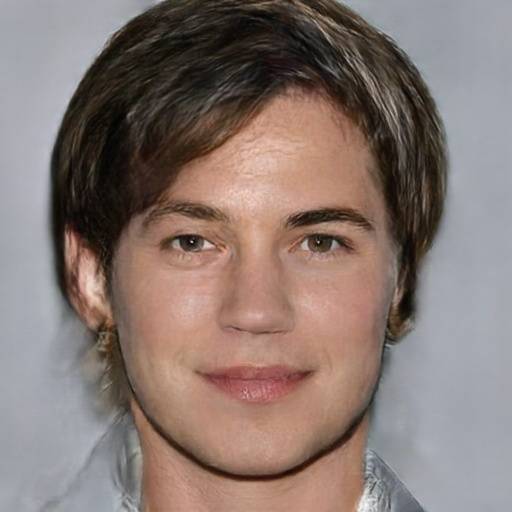}}
    {\includegraphics[width=0.12\linewidth]{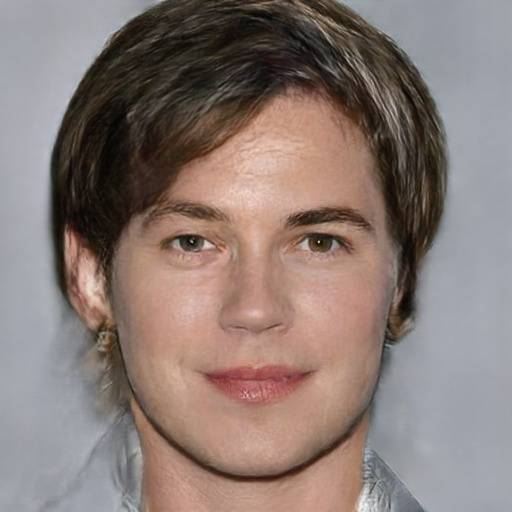}}
    \caption{Two sequences of progressive sketching (under shadow guidance) and synthesis results.
    }
    \label{fig:userStudy1}
\end{figure*}

Figure \ref{fig:local_edit} shows two representative results where users progressively introduce new strokes to add or stress local details. As shown in the demo video, running on a PC {with an Intel i7-7700 CPU, 16GB RAM and a single Nvidia GTX 1080Ti GPU}, our method achieves real-time feedback. Thanks to our local-to-global approach, generally more strokes lead to new or refined details (e.g., the nose in the first example, and the eyebrows and wrinkles in the second example), with other areas largely unchanged. Still due to the combination step, local editing might still introduce subtle but global changes. For example, for the first example, the local change of lighting in the nose area leads to the change of highlight in the whole face (especially in the forehead region).
Figure \ref{fig:userStudy1} shows two more complete sequences of progressive sketching and synthesis, with our shadow-guided interface.

\subsection{Usability Study}\label{sec:usability}

\begin{figure}
    \centering
    \includegraphics[width=0.6\linewidth]{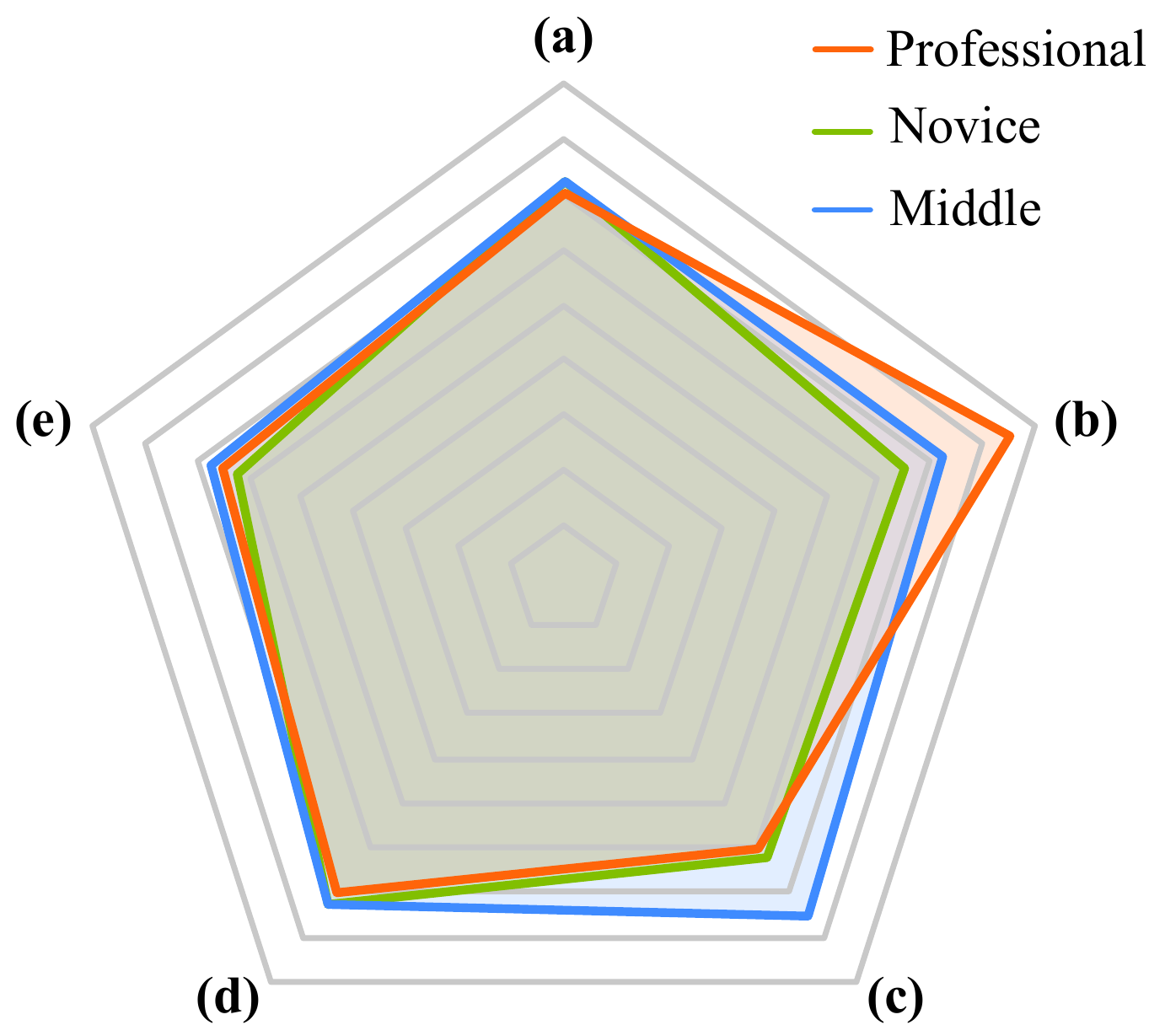}
    \caption{The summary of quantitative feedback in the usability study. (a) Ease of use. (b) Controllability. (c) Variance of results. (d) Quality of results. (e) Expectation fitness.
    }
    \label{fig:userStudy_ui}
\end{figure}

We conducted a usability study to evaluate the usefulness and effectiveness of our system. 10 subjects (9 male and 1 female, aged from 20 to 26) were invited to participate in this study. We first asked them to self-assess their drawing skills through a nine-point Likert scale (1: novice to 9: professional), and divided them into three groups: 4 novice users (drawing skill score: 1 -- 3), 4 middle users (4 -- 6), and 2 professional users (7 -- 9). Before the drawing session, each participant was given a short tutorial about our system (about 10 minutes). {The participants used an iPad with iPencil to remotely control the server PC for drawing.} Then each of them was asked to create at least 3 faces using our system. The study ended with a questionnaire to get user feedbacks on \emph{ease-of-use}, \emph{controllability}, \emph{variance of results}, \emph{quality of results}, and \emph{expectation fitness}. The additional comments on our system were also welcome.

Figure \ref{fig:results_man} gives a gallery of sketches and synthesized faces by the participants. It can be seen that our system consistently produce realistic results given input sketches with different styles and levels of abstraction. For several examples, the participants attempted to depict beard styles via hatching and our system captured the users' intention very well.

Figure \ref{fig:userStudy_ui} shows a radar plot, summarizing quantitative feedbacks on our system for participant groups with different levels of drawing skills. The feedbacks for all the groups of participants were positive in all the measured aspects. Particularly, the participants with good drawing skills felt a high level of controllability, while they gave slightly lower scores for the degree of result variance.
Using our system, the average time needed for drawing a face sketch among the participants with different drawing abilities are: $17'14''$ (professional), $3'17''$ (middle) and $2'26''$ (novice).
It took much longer for professionals, since they spent more time sketching and refining details.
For the refinement sliders, the most frequently used slider was for the ``remainder'' component (56.69\%), which means for more than half of the results, the ``remainder'' slider was manipulated. In contrast, for the other components we have 21.66\% for ``left-eye'', 12.74\% for ``right-eye'', 12.10\% for ``nose'' and 19.75\% for ``mouth''. For all the adjustments made in the components, participants trust the ``remainder'' component most, with the averaged confidence $0.78$; The least trusted component is ``mouth'' ($0.56$); other component confidences are $0.70$ (``left-eye''), $0.61$ (``right-eye'') and $0.58$ (``nose'').
The averaged confidences implied the importance of sketch refinement in creating the synthesized faces in this study.

All of the participants felt that our system was powerful to create realistic faces using such sparse sketches. They liked the intuitive shadow-guided interface, which was quite helpful for them to construct face sketches with proper structures and layouts. On the other hand, some users, particularly those with good drawing skills, felt that the shadow guidance was sometimes distracting when editing details. This finding is consistent with the conclusions in the original \emph{ShadowDraw} paper ~\cite{lee2011shadowdraw}.
One of the professional users mentioned that automatic synthesis of face images given sparse inputs saved a lot of efforts and time compared to traditional painting software.
One professional user mentioned that it would be better if our system could provide color control.
\subsection{Comparison with Alternative Refinement Strategies}\label{sec:alternative}
\begin{figure}
    \centering
    \setlength{\fboxrule}{0.5pt}
    \setlength{\fboxsep}{-0.01cm}

    \begin{minipage}{0.17\linewidth}
		\framebox{\includegraphics[width=1\linewidth]{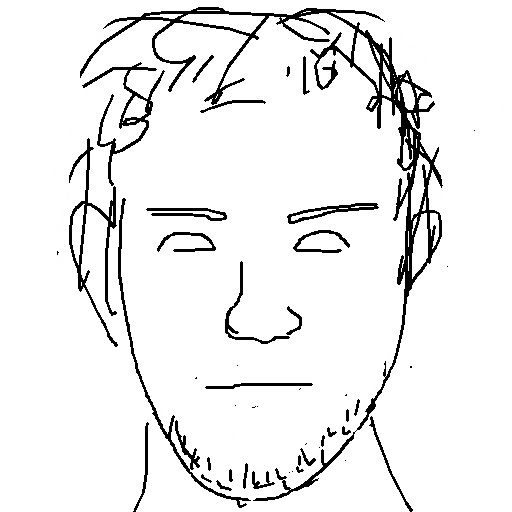}}
		\centerline{Input Sketch}
	\end{minipage}
    \begin{minipage}{0.03\linewidth}
    (a)\\
    \vspace{7mm}
    (b)\\
    \vspace{7mm}
    (c)
    \end{minipage}
    \begin{minipage}{0.145\linewidth}
		\framebox{\includegraphics[width=0.99\linewidth]{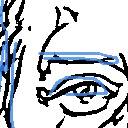}}
		\framebox{\includegraphics[width=0.99\linewidth]{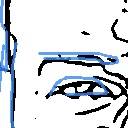}}
		\framebox{\includegraphics[width=0.99\linewidth]{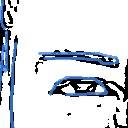}}
		\centerline{Part 1}
	\end{minipage}
    \begin{minipage}{0.145\linewidth}
		\framebox{\includegraphics[width=0.99\linewidth]{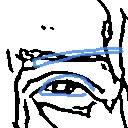}}
		\framebox{\includegraphics[width=0.99\linewidth]{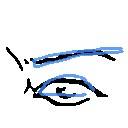}}
		\framebox{\includegraphics[width=0.99\linewidth]{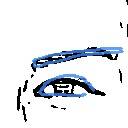}}
		\centerline{Part 2}
	\end{minipage}
    \begin{minipage}{0.145\linewidth}
		\framebox{\includegraphics[width=0.99\linewidth]{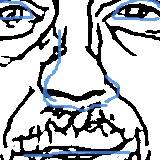}}
		\framebox{\includegraphics[width=0.99\linewidth]{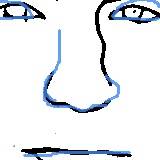}}
		\framebox{\includegraphics[width=0.99\linewidth]{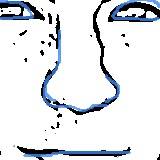}}
		\centerline{Part 3}
	\end{minipage}
    \begin{minipage}{0.145\linewidth}
		\framebox{\includegraphics[width=0.99\linewidth]{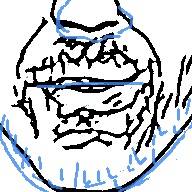}}
		\framebox{\includegraphics[width=0.99\linewidth]{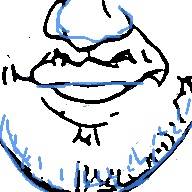}}
		\framebox{\includegraphics[width=0.99\linewidth]{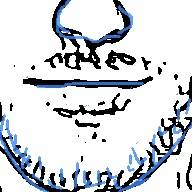}}
		\centerline{Part 4}
	\end{minipage}
    \begin{minipage}{0.145\linewidth}
        \vspace{0.75mm}
		\framebox{\includegraphics[width=0.99\linewidth]{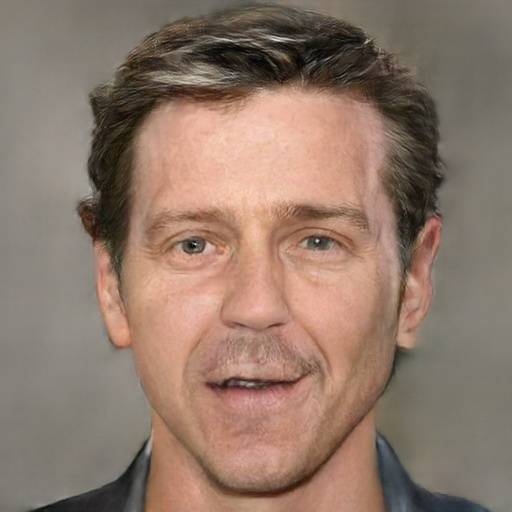}}
		\framebox{\includegraphics[width=0.99\linewidth]{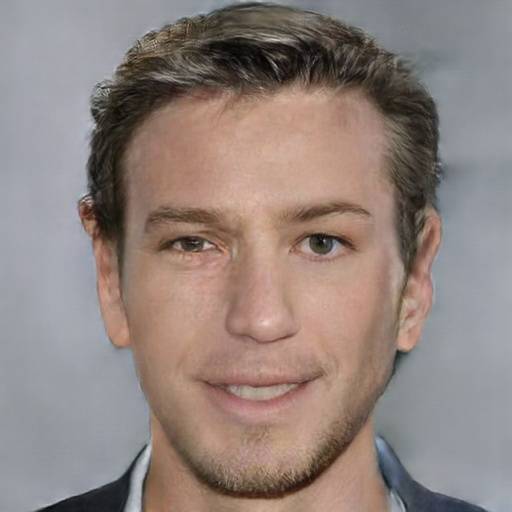}}
		\framebox{\includegraphics[width=0.99\linewidth]{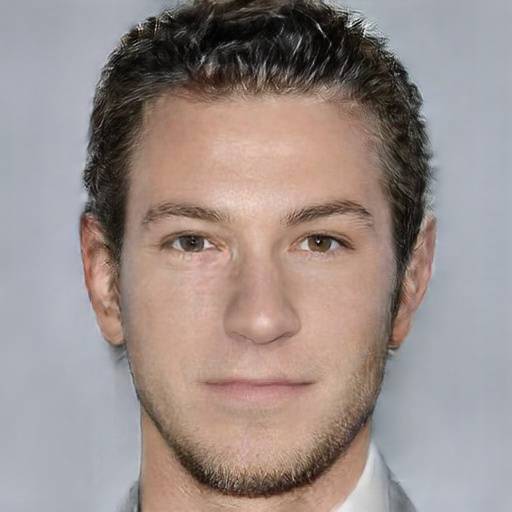}}
		\centerline{Image}
	\end{minipage}

    \begin{minipage}{0.17\linewidth}
		\framebox{\includegraphics[width=1\linewidth]{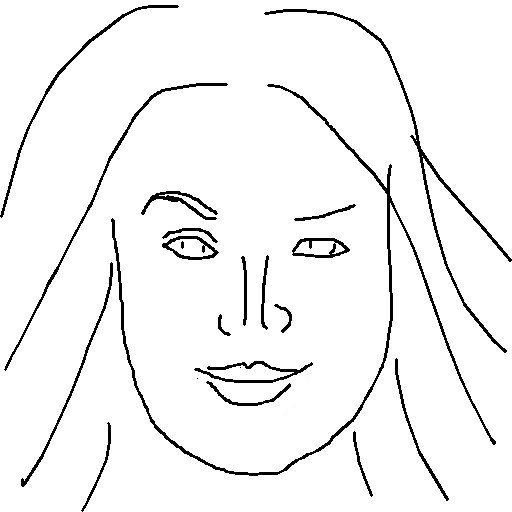}}
		\centerline{Input Sketch}
	\end{minipage}
    \begin{minipage}{0.03\linewidth}
    (a)\\
    \vspace{7mm}
    (b)\\
    \vspace{7mm}
    (c)
    \end{minipage}
    \begin{minipage}{0.145\linewidth}
		\framebox{\includegraphics[width=0.99\linewidth]{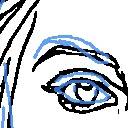}}
		\framebox{\includegraphics[width=0.99\linewidth]{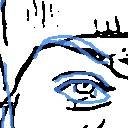}}
		\framebox{\includegraphics[width=0.99\linewidth]{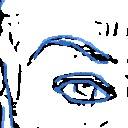}}
		\centerline{Part 1}
	\end{minipage}
    \begin{minipage}{0.145\linewidth}
		\framebox{\includegraphics[width=0.99\linewidth]{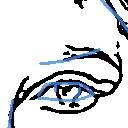}}
		\framebox{\includegraphics[width=0.99\linewidth]{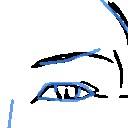}}
		\framebox{\includegraphics[width=0.99\linewidth]{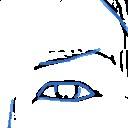}}
		\centerline{Part 2}
	\end{minipage}
    \begin{minipage}{0.145\linewidth}
		\framebox{\includegraphics[width=0.99\linewidth]{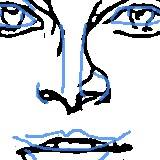}}
		\framebox{\includegraphics[width=0.99\linewidth]{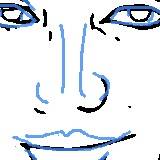}}
		\framebox{\includegraphics[width=0.99\linewidth]{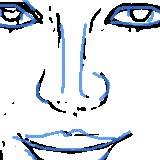}}
		\centerline{Part 3}
	\end{minipage}
    \begin{minipage}{0.145\linewidth}
		\framebox{\includegraphics[width=0.99\linewidth]{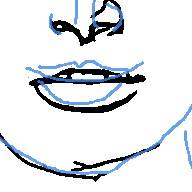}}
		\framebox{\includegraphics[width=0.99\linewidth]{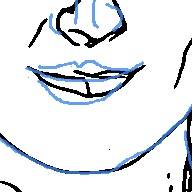}}
		\framebox{\includegraphics[width=0.99\linewidth]{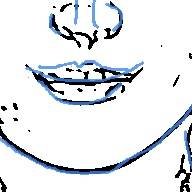}}
		\centerline{Part 4}
	\end{minipage}
    \begin{minipage}{0.145\linewidth}
        \vspace{0.75mm}
		\framebox{\includegraphics[width=0.99\linewidth]{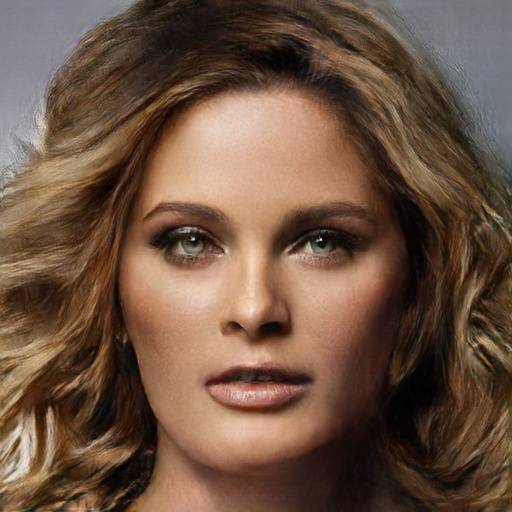}}
		\framebox{\includegraphics[width=0.99\linewidth]{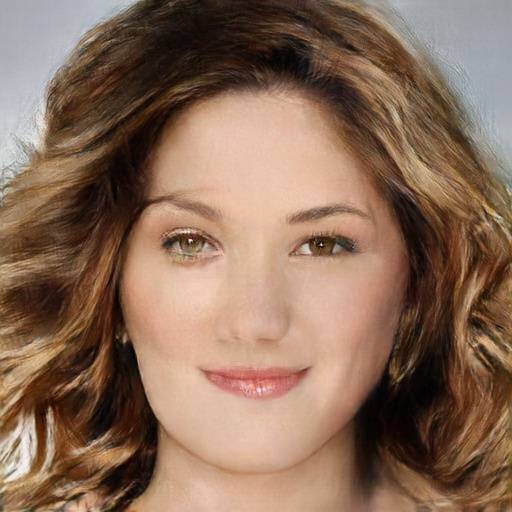}}
		\framebox{\includegraphics[width=0.99\linewidth]{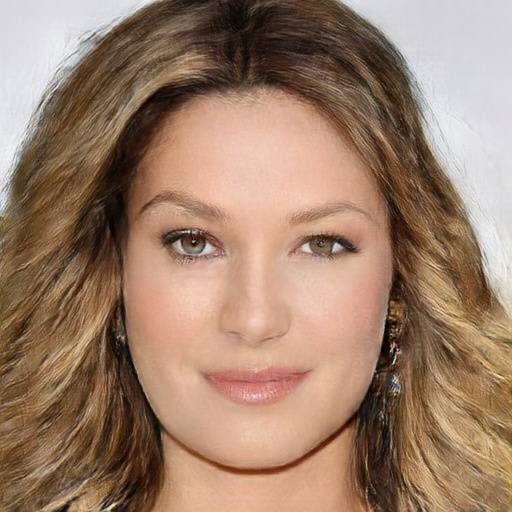}}
		\centerline{Image}
	\end{minipage}
    \caption{
    Comparisons of using global retrieval (a), component-level retrieval (b), and our method (essentially component-level retrieval followed by interpolation) (c) for sketch refinement. The right column shows the corresponding synthesized results. For easy comparison we overlay input sketches (in light blue) on top of the retrieved or interpolated sketches by different methods.
    }
    \label{fig:retrieval_result}
\end{figure}

To refine an input sketch, we essentially take a component-level retrieval-and-interpolation approach. We compare this method with two alternative sketch refinement methods by globally or locally retrieving the most similar sample in the training data. For fair comparison, we use the same \moduleTwo~and \moduleThree~modules for image synthesis.
For the local retrieval method, it is the same as our method except that for manifold projection we simply retrieve the closest (i.e., top-1 instead of top-$K$) component sample in each component-level feature space without any interpolation. For the global retrieval method, we replace the \moduleOne~module with a new module for the feature embeddings of entire face sketches. Specifically, we first learn the feature embeddings of the entire face sketch images, and given a new sketch we find the most similar (i.e., top-1) sample in the whole-face feature space. For each component in the globally retrieved sample image, we then encode it using the corresponding trained component-level encoder (i.e., $E_c$), and pass all the component-level feature vectors to our \moduleTwo~and \moduleThree~for image synthesis. Note that we do not globally retrieve real face images, since our goal here is for a fair comparison of the sketch refinement methods.

Figure \ref{fig:retrieval_result} shows comparison results. From the overlay of input sketches and the retrieved or interpolated sketches, it can be easily seen that the component-level retrieval method returns samples closer to the input component sketches than the global-retrieval method, mainly due to the limited sample data. Thanks to the interpolation step, the interpolated sketches almost perfectly fit the input sketches. Note that we show the decoded sketches after interpolation here only for the comparison purpose, and our conditional image synthesis sub-network takes the interpolated feature vectors as input (Section \ref{sec:framework}).

\subsection{Perceptive Evaluation Study}
As shown in Figure \ref{fig:retrieval_result} (Right), the three refinement methods all lead to realistic face images. To evaluate the visual quality and the faithfulness (i.e., the degree of fitness to input sketches) of synthesize results, we conducted a user study.

\begin{figure}
    \centering
    {\includegraphics[width=0.24\linewidth]{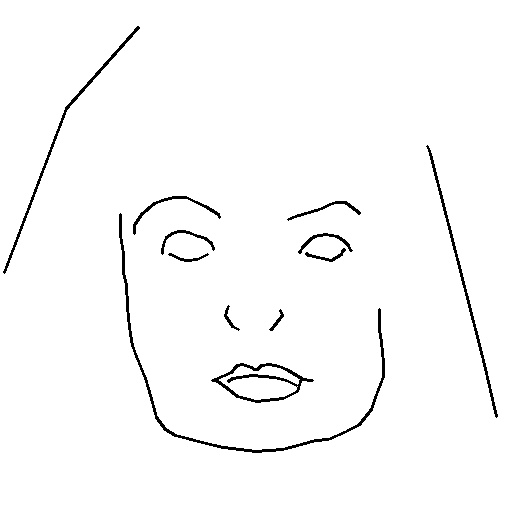}}
    {\includegraphics[width=0.24\linewidth]{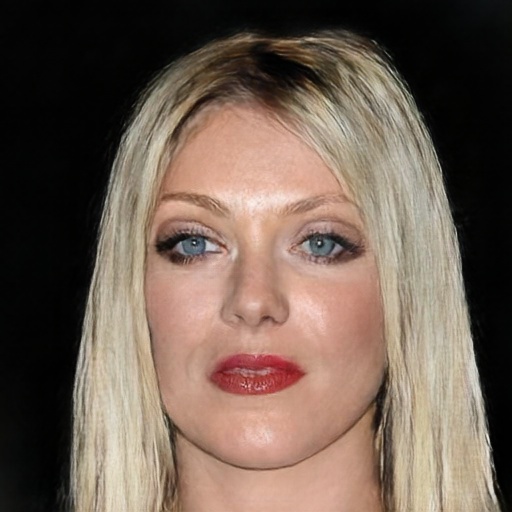}}
    {\includegraphics[width=0.24\linewidth]{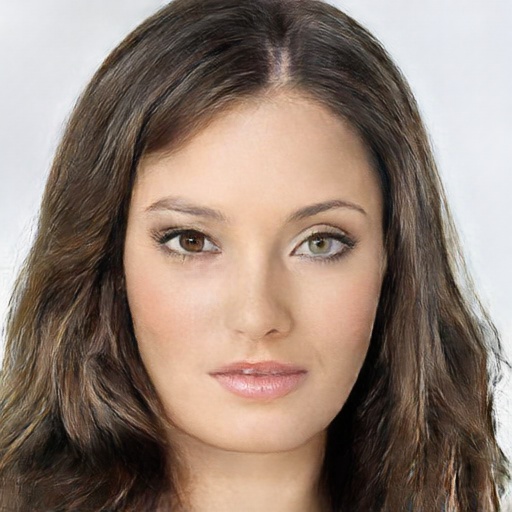}}
    {\includegraphics[width=0.24\linewidth]{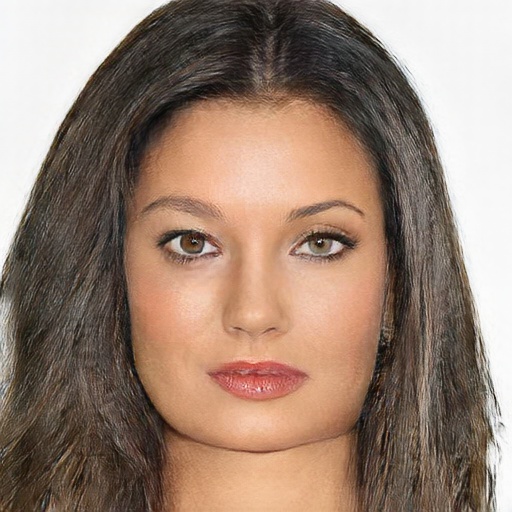}}

    {\includegraphics[width=0.24\linewidth]{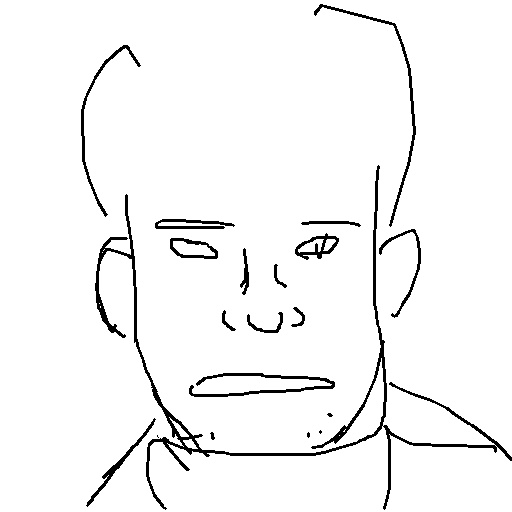}}
    {\includegraphics[width=0.24\linewidth]{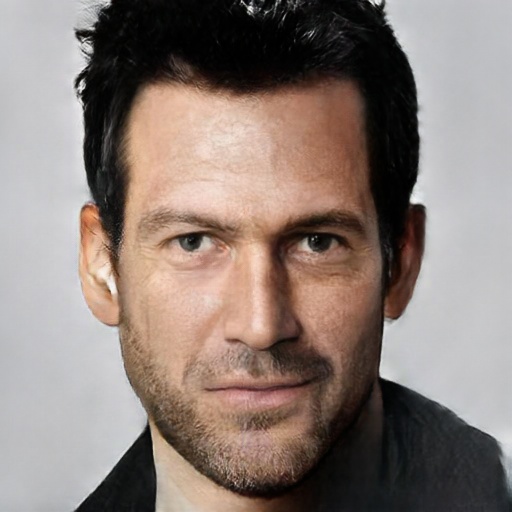}}
    {\includegraphics[width=0.24\linewidth]{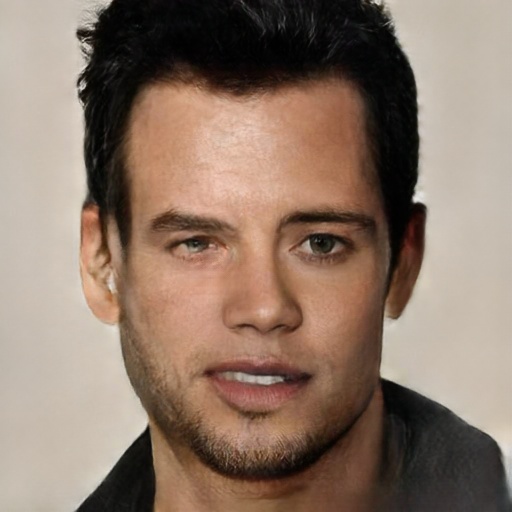}}
    {\includegraphics[width=0.24\linewidth]{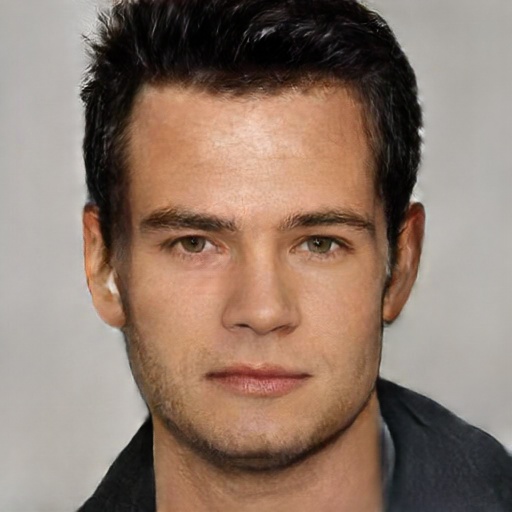}}
    \caption{Two representative sets of input sketches and synthesized results used in the perceptive evaluation study.
    From left to right: input sketch, the results by sketch refinement through global retrieval, local retrieval, and local retrieval with interpolation (our method).}
    \label{fig:userStudy-image}
\end{figure}

We prepared a set of input sketches, containing in total 22 sketches, including 9 from the usability study (Section \ref{sec:usability}) and 13 from {the authors}. This sketch set (see the supplementary materials) covered inputs with different levels of abstraction and different degrees of roughness. We applied the three refinement methods to each input sketch to generate the corresponding synthesized results (see two representative sets in Figure \ref{fig:userStudy-image}).

The evaluation was done through an online questionnaire. 60 participants (39 male, 21 female, aged from 18 to 32) participated in this study. Most of them had no professional training in drawing.
We showed each participant four images including input sketch and the three synthesized images, placed side by side in a random order.
Each participant was asked to evaluate the quality and faithfulness both in a seven-point Likert scale (1 = strongly negative to 7 = strongly positive). In total, to evaluate either the faithfulness or the quality, we got 60 (participants) $\times$ 22 (sketches) = 1,320 subjective evaluations for each method.

\begin{figure}
    \centering
    {\includegraphics[width=0.45\linewidth]{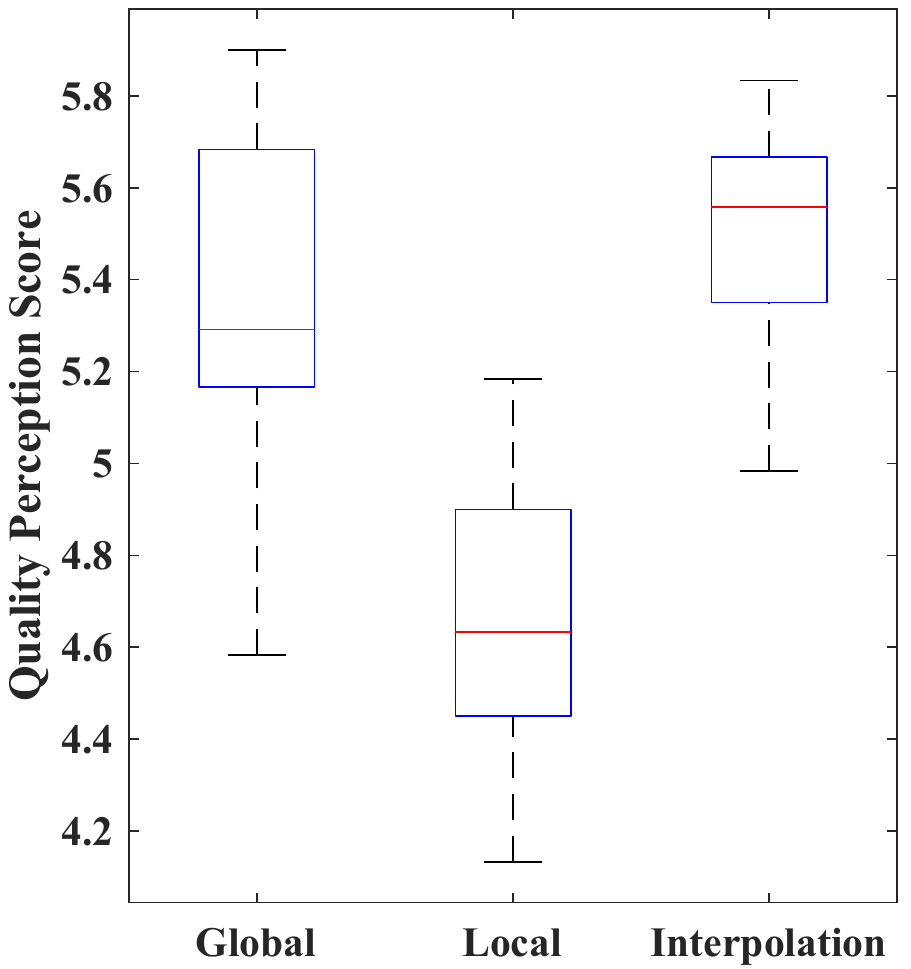}}
    {\includegraphics[width=0.45\linewidth]{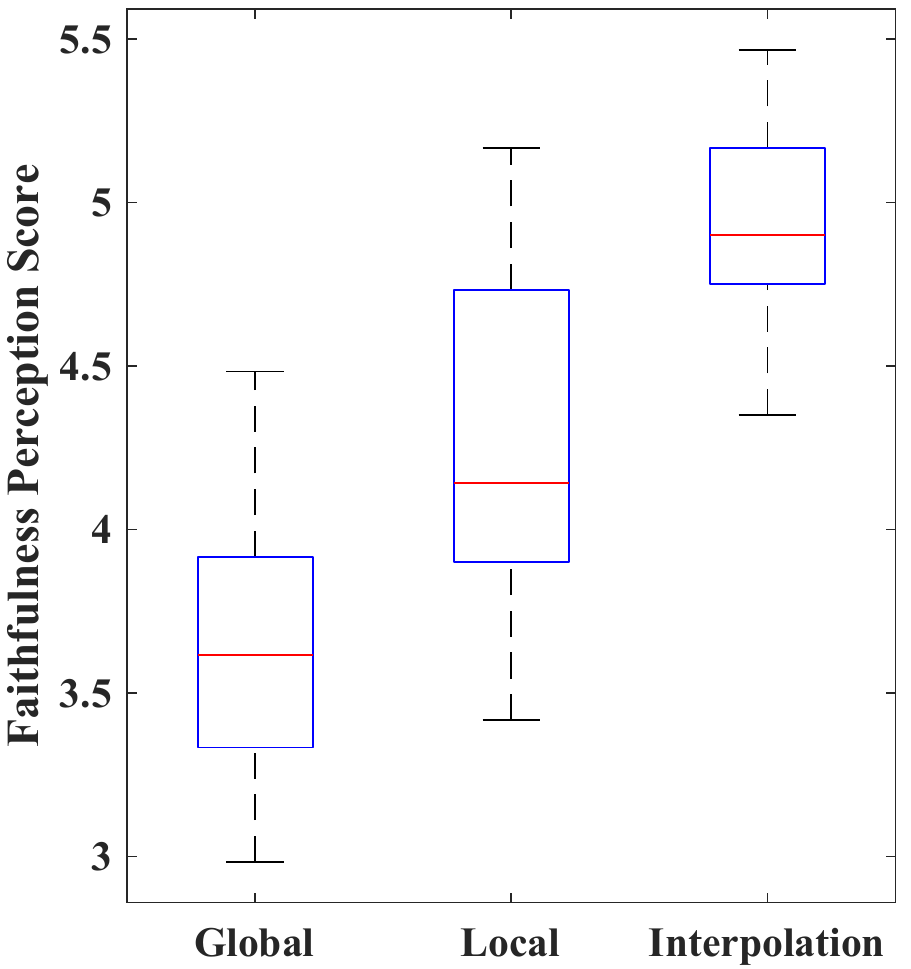}}
    \caption{Box plots of the average quality and faithfulness perception scores over the participants for each method.}
    \label{fig:userStudy}
\end{figure}

Figure \ref{fig:userStudy} plots the statistics of the evaluation results. We performed one-way ANOVA tests on the quality and faithfulness scores, and found significant effects for both quality ($F_{(2, 63)}=47.26, p < 0.001$) and faithfulness ($F_{(2, 63)}=51.72, p < 0.001$).
Paired t-tests further confirmed that our method (mean: 4.85) led to significantly more faithful results than both the global (mean: 3.65; $[t = -29.77$,  $p < 0.001]$ and local (mean: 4.23; $[t = -16.34$, $p<0.001]$) retrieval methods. This is consistent with our expectation, since our method provides the largest flexibility to fit to input sketches.

In terms of visual quality our method (mean: 5.50) significantly outperformed the global retrieval method (mean: 5.37; $[t = -3.94$, $p < 0.001]$) and the local retrieval method (mean: 4.68; $[t = -24.60$, $p < 0.001]$). It is surprisingly to see that the quality of results by our method was even higher than the global retrieval method, since we had expected that the visual quality of the results by the global method and ours would be similar. This is possibly because some information is lost after first decomposing an entire sketch into components and then recombining the corresponding feature maps.

\begin{figure*}
    \centering
    \setlength{\fboxrule}{0.5pt}
    \setlength{\fboxsep}{-0.01cm}
    \begin{tabular}{cc}
         \hspace{-3mm}
        \rotatebox{90}{\quad \large{Input Sketch}}
         &
         \hspace{-3mm}
        \framebox{\includegraphics[width=0.16\linewidth]{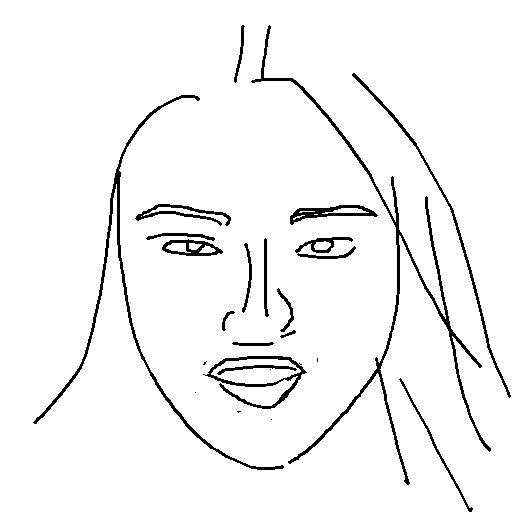}}
        \framebox{\includegraphics[width=0.16\linewidth]{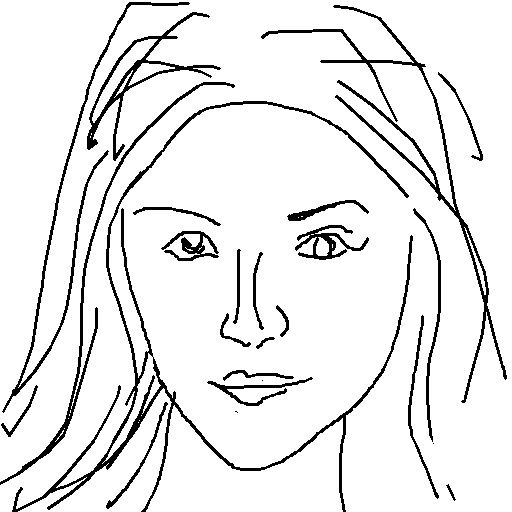}}
        \framebox{\includegraphics[width=0.16\linewidth]{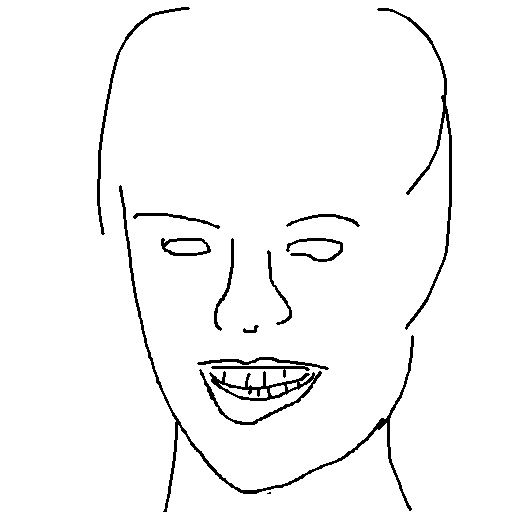}}
        \framebox{\includegraphics[width=0.16\linewidth]{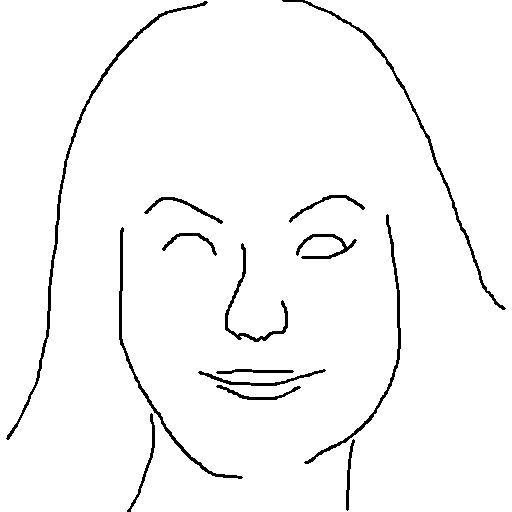}}
        \framebox{\includegraphics[width=0.16\linewidth]{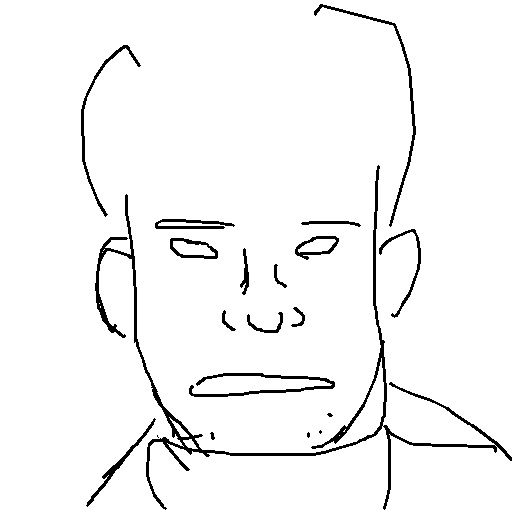}}
        \framebox{\includegraphics[width=0.16\linewidth]{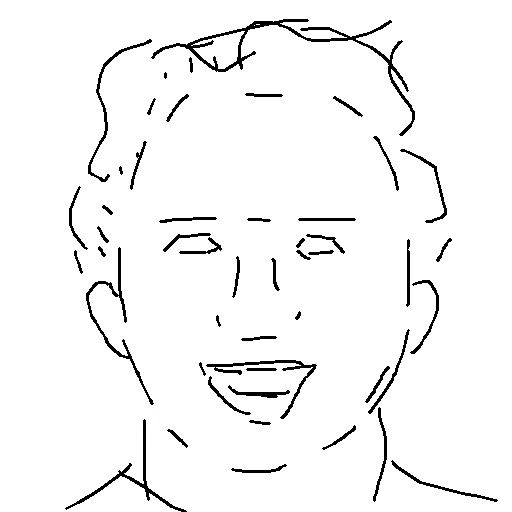}} \\

         \hspace{-3mm}
         \rotatebox{90}{\qquad \quad \large{Pix2pix}}
         &
         \hspace{-3mm}
        \framebox{\includegraphics[width=0.16\linewidth]{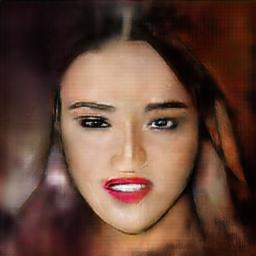}}
        \framebox{\includegraphics[width=0.16\linewidth]{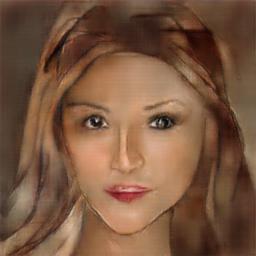}}
        \framebox{\includegraphics[width=0.16\linewidth]{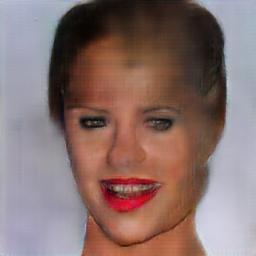}}
        \framebox{\includegraphics[width=0.16\linewidth]{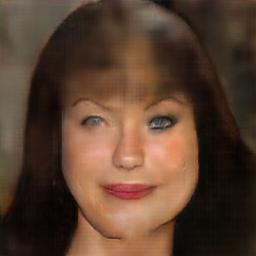}}
        \framebox{\includegraphics[width=0.16\linewidth]{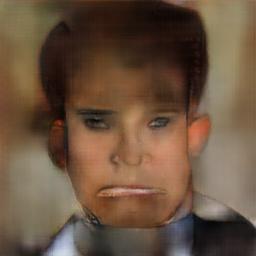}}
        \framebox{\includegraphics[width=0.16\linewidth]{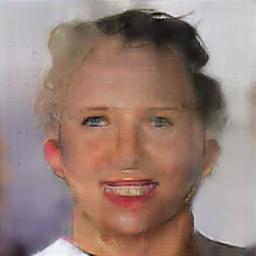}}\\

         \hspace{-3mm}
        \rotatebox{90}{\quad \large{Lines2FacePhoto}}
         &
         \hspace{-3mm}
        \framebox{\includegraphics[width=0.16\linewidth]{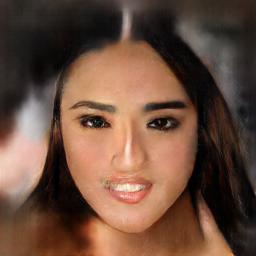}}
        \framebox{\includegraphics[width=0.16\linewidth]{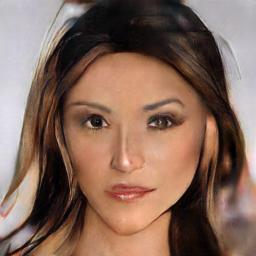}}
        \framebox{\includegraphics[width=0.16\linewidth]{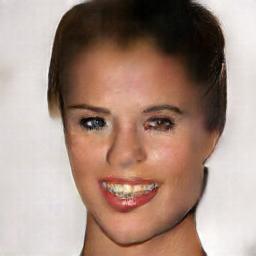}}
        \framebox{\includegraphics[width=0.16\linewidth]{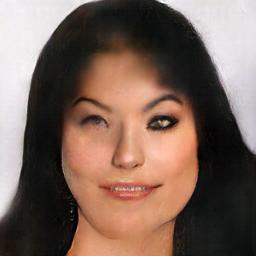}}
        \framebox{\includegraphics[width=0.16\linewidth]{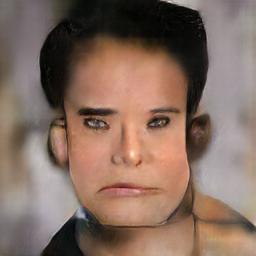}}
        \framebox{\includegraphics[width=0.16\linewidth]{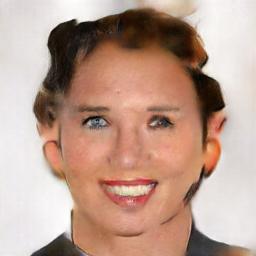}}\\

         \hspace{-3mm}
        \rotatebox{90}{\qquad \large{Pix2pixHD}}
         &
         \hspace{-3mm}
        \framebox{\includegraphics[width=0.16\linewidth]{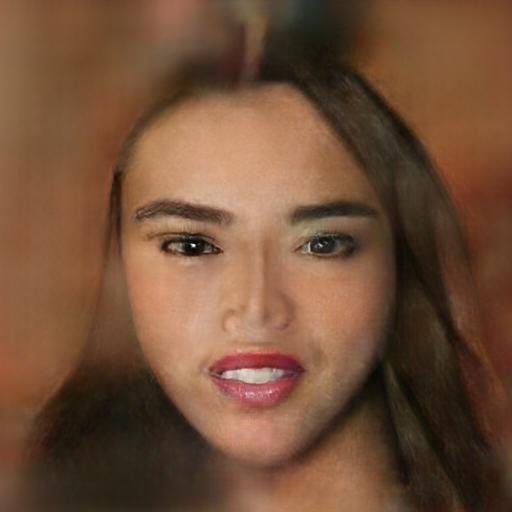}}
        \framebox{\includegraphics[width=0.16\linewidth]{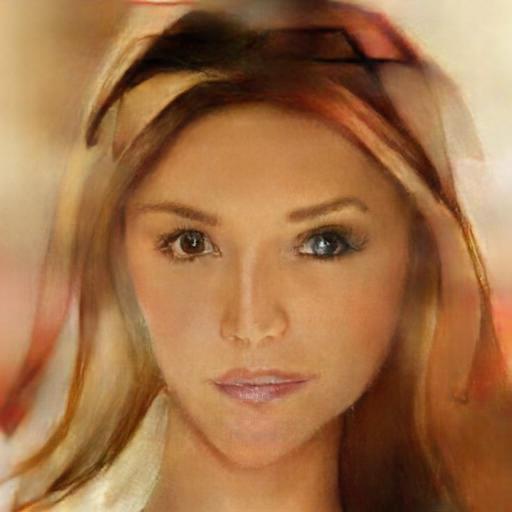}}
        \framebox{\includegraphics[width=0.16\linewidth]{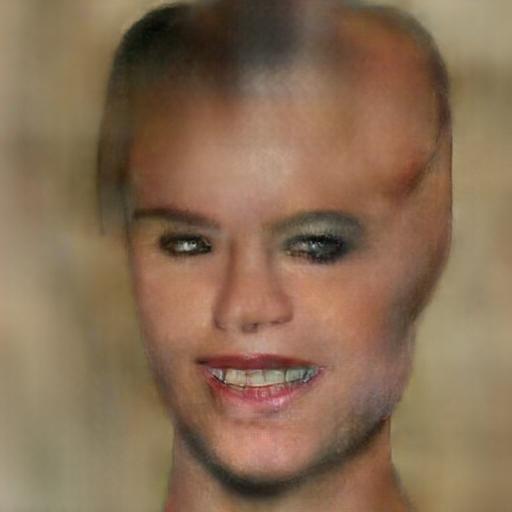}}
        \framebox{\includegraphics[width=0.16\linewidth]{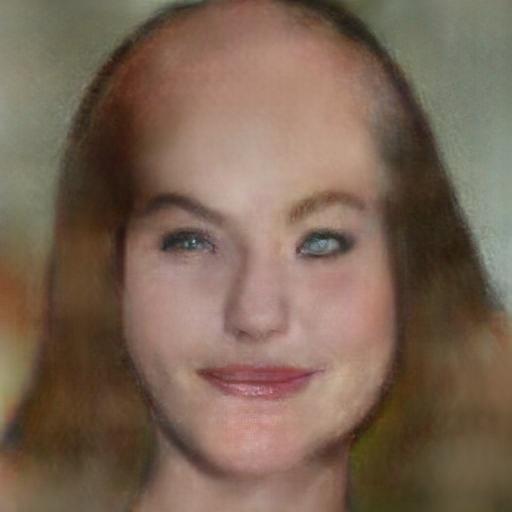}}
        \framebox{\includegraphics[width=0.16\linewidth]{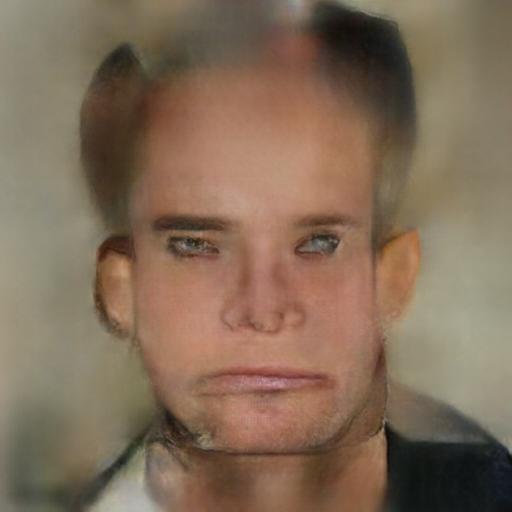}}
        \framebox{\includegraphics[width=0.16\linewidth]{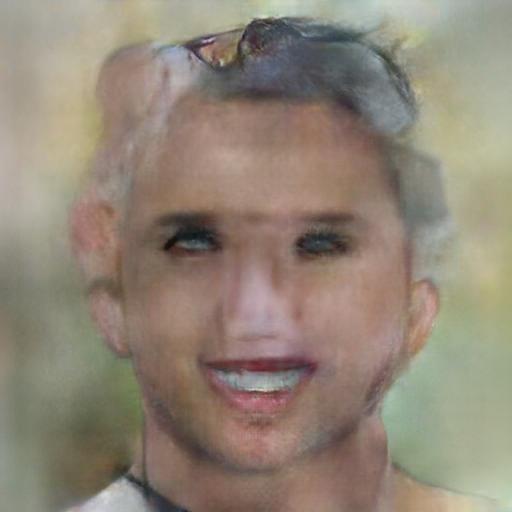}}\\

         \hspace{-3mm}
        \rotatebox{90}{\qquad \large{iSketchNFill}}
         &
         \hspace{-3mm}
        \framebox{\includegraphics[width=0.16\linewidth]{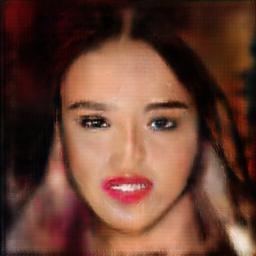}}
        \framebox{\includegraphics[width=0.16\linewidth]{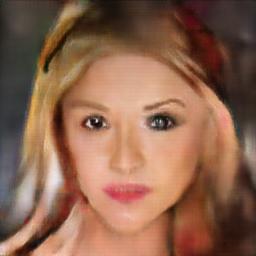}}
        \framebox{\includegraphics[width=0.16\linewidth]{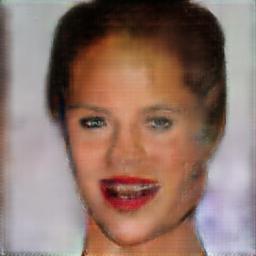}}
        \framebox{\includegraphics[width=0.16\linewidth]{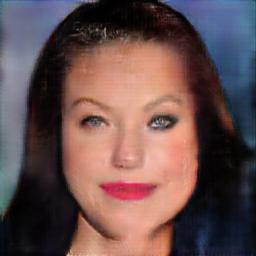}}
        \framebox{\includegraphics[width=0.16\linewidth]{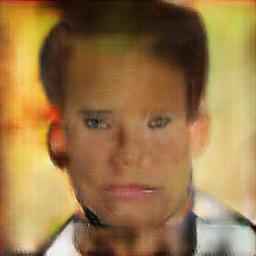}}
        \framebox{\includegraphics[width=0.16\linewidth]{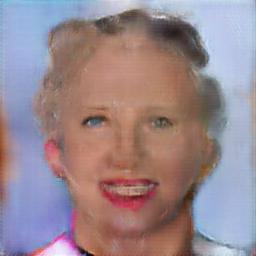}}\\
         \hspace{-3mm}
        \rotatebox{90}{\qquad\qquad \large{Ours}}
         &
         \hspace{-3mm}
        \framebox{\includegraphics[width=0.16\linewidth]{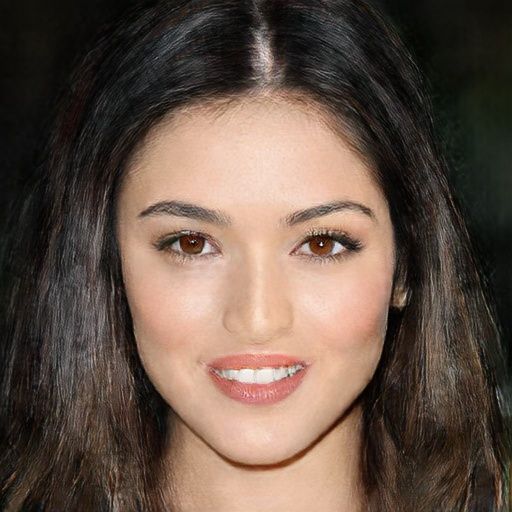}}
        \framebox{\includegraphics[width=0.16\linewidth]{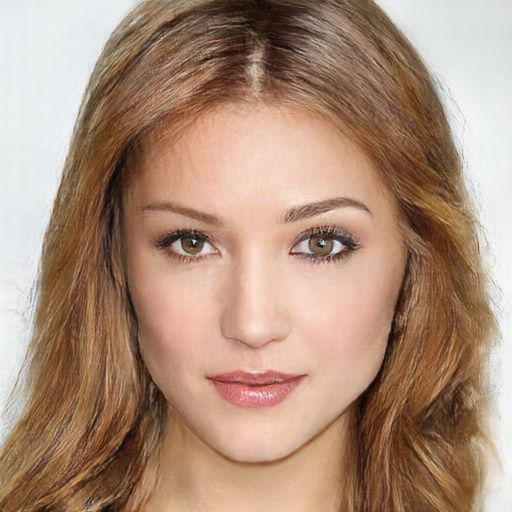}}
        \framebox{\includegraphics[width=0.16\linewidth]{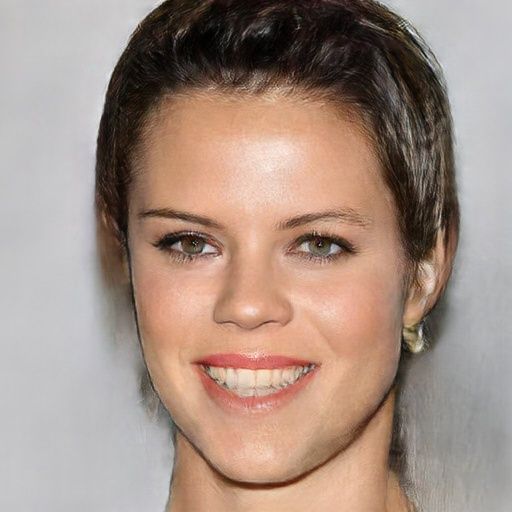}}
        \framebox{\includegraphics[width=0.16\linewidth]{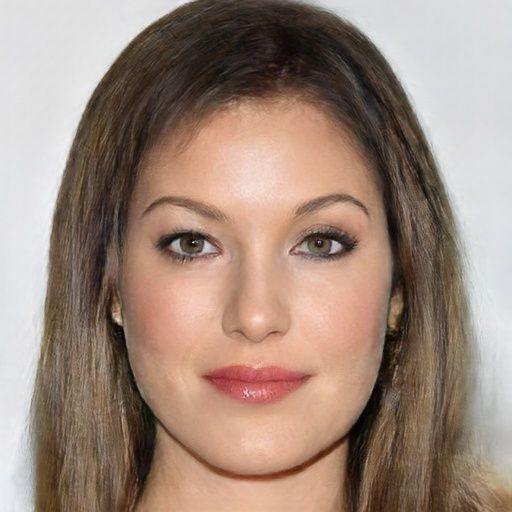}}
        \framebox{\includegraphics[width=0.16\linewidth]{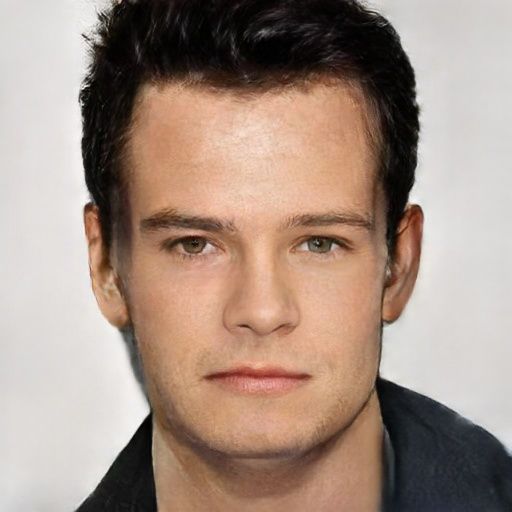}}
        \framebox{\includegraphics[width=0.16\linewidth]{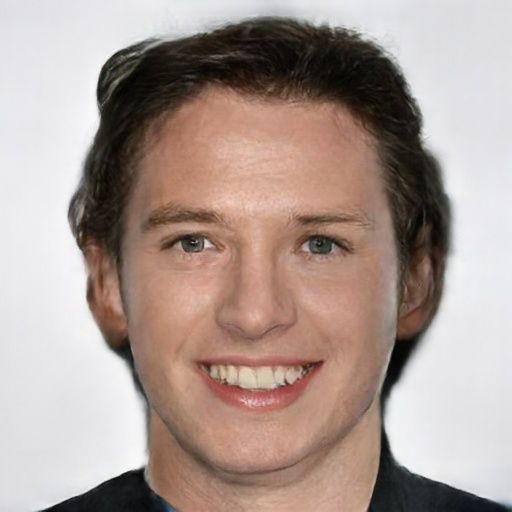}}

    \end{tabular}\\

    \caption{Comparisons with the state-of-the-art methods given the same input sketches (Top Row).
    }
    \label{fig:visual_comparison}
\end{figure*}

\subsection{Comparison with Existing Solutions}\label{sec: visual_comp}

We compare our method with the state-of-the-art methods for image synthesis conditioned on sketches, including \emph{pix2pix} \cite{isola2017image}, \emph{pix2pixHD} \cite{wang2018high} and \emph{Lines2FacePhoto} \cite{li2019linestofacephoto} and \emph{iSketchNFill} \cite{ghosh2019interactive} in terms of visual quality of generated faces.
We use their released source code but for fair comparisons we train all the networks on our training dataset (Section \ref{sec:data_preparation}).
The (input and output) resolution for our method and \emph{pix2pixHD} is $512\times512$, while we have $256\times256$ for \emph{pix2pix} and \emph{Lines2FacePhoto} according to their default setting. In addition, for \emph{Lines2FacePhoto}, following their original paper, we also convert each sketch to a distance map as input for both training and testing. For \emph{iSketchNFill}, we train their shape completion module before feeding it to \emph{pix2pix} \cite{isola2017image} (acting as the appearance synthesis module). The input and output resolutions in their method are $256\times256$ and $128\times128$, respectively.

Figure \ref{fig:visual_comparison} shows representative testing results given the same sketches as input. It can be easily seen that our method produces more realistic synthesized results. Since the input sketches are rough and/or incomplete, they are generally different from the training data, making the compared methods fail to produce realistic faces. Although \emph{Lines2FacePhoto} generates a relatively plausible result given an incomplete sketch, its ability to handle data imperfections is rather limited.
We attempted to perform quantitative evaluation as well. However, none of the assessment metrics we tried, including Fr\'{e}chet Inception Distance ~\cite{heusel2017gans} and Inception Score ~\cite{salimans2016improved}, could  faithfully reflect visual perception.
For example, the averaged values of the Inception Score were 2.59 and 1.82 (the higher, the better) for pix2pixHD and ours, respectively. However, it is easily noticeable that our results are visually better than those by pix2pixHD.

\section{Applications}
Our system can be adapted for various applications. In this section we present two applications: face morphing and face copy-paste.

\subsection{Face Morphing}
\begin{figure*}
    \centering
    {\includegraphics[width=0.99\linewidth]{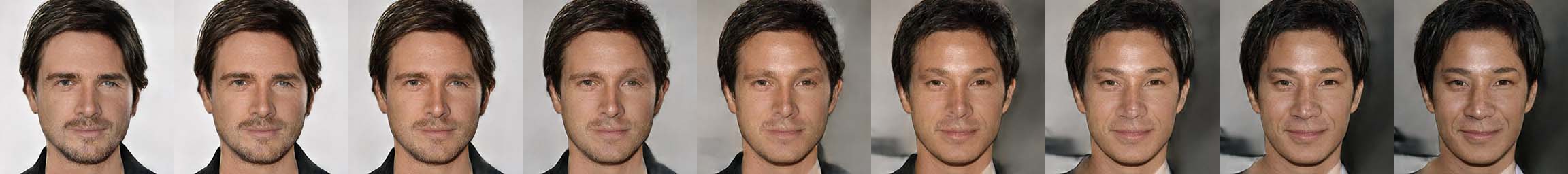}}
    {\includegraphics[width=0.99\linewidth]{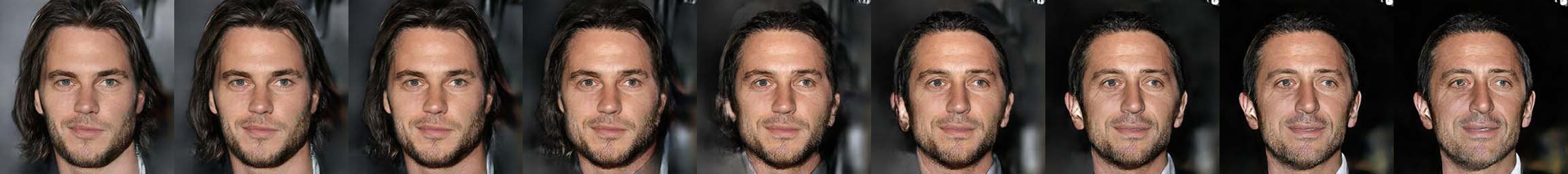}}
    {\includegraphics[width=0.99\linewidth]{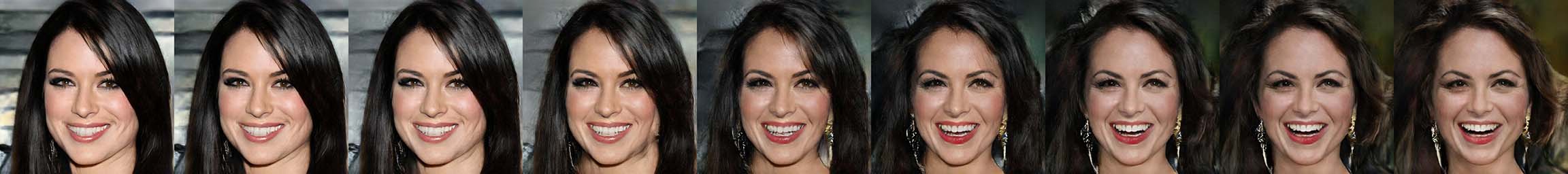}}
    {\includegraphics[width=0.99\linewidth]{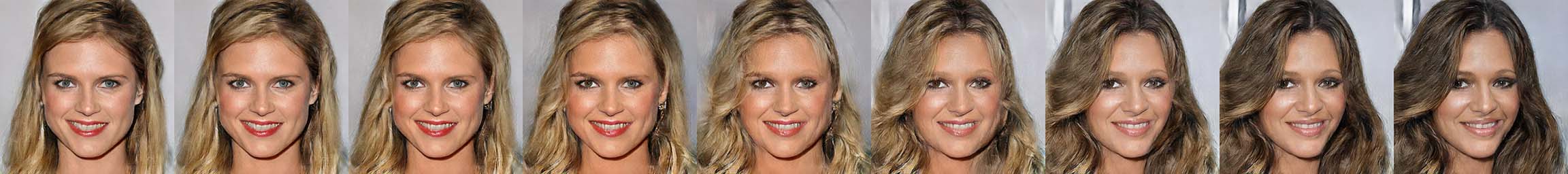}}

    \caption{
    Examples of face morphing by interpolating the component-level feature vectors of two given face {sketches (Leftmost and Rightmost are corresponding synthesized images)}.
    }
    \label{fig:interFace}
\end{figure*}

Traditional face morphing algorithms {~\cite{bichsel1996automatic}} often require a set of keypoint-level correspondence between two face images to guide semantic interpolation. We show a simple but effective morphing approach by 1) decomposing a pair of source and target face sketches in the training dataset into five components (Section \ref{para: module1}); 2) encoding the component sketches as feature vectors in the corresponding feature spaces; 3) performing linear interpolation between the source and target feature vectors for the corresponding components; 4) finally feeding the interpolated feature vectors to the \moduleTwo~and \moduleThree~module to get intermediate face images.
Figure \ref{fig:interFace} shows examples of face morphing using our method. It can be seen that our method leads to smoothly transforming results in identity, expression, and even highlight effects.

\subsection{Face Copy-Paste}
\begin{figure*}
    \centering
    \setlength{\fboxrule}{1pt}
    \setlength{\fboxsep}{-0.01cm}

    \includegraphics[width=0.33\linewidth]{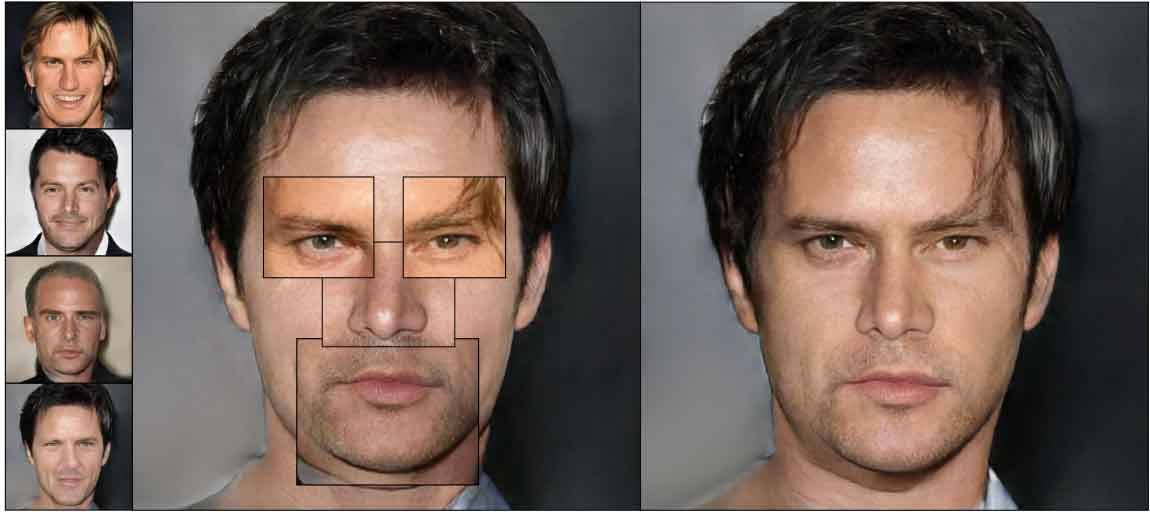}
    \includegraphics[width=0.33\linewidth]{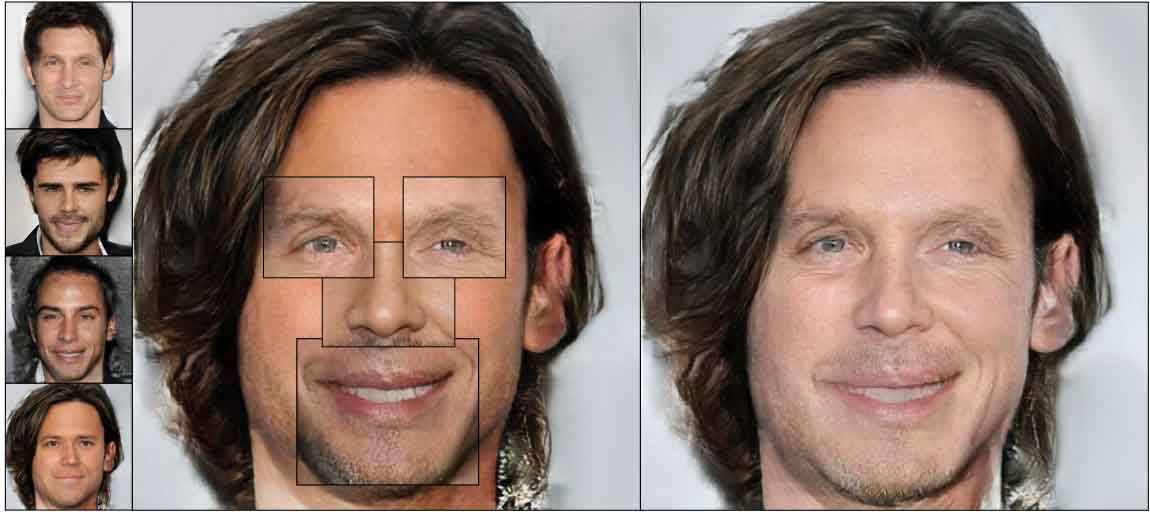}
    \includegraphics[width=0.33\linewidth]{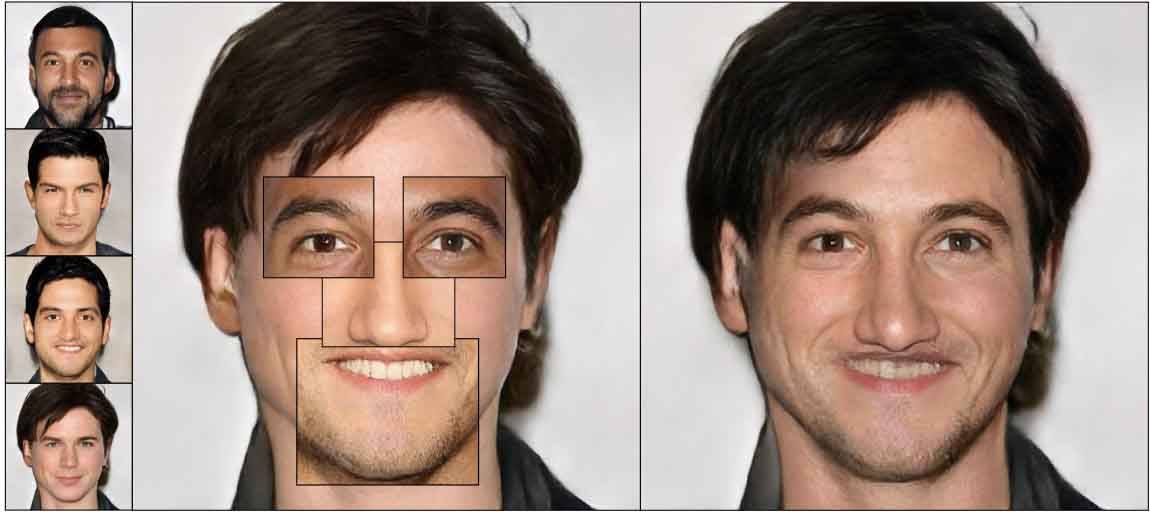}\\

    \includegraphics[width=0.33\linewidth]{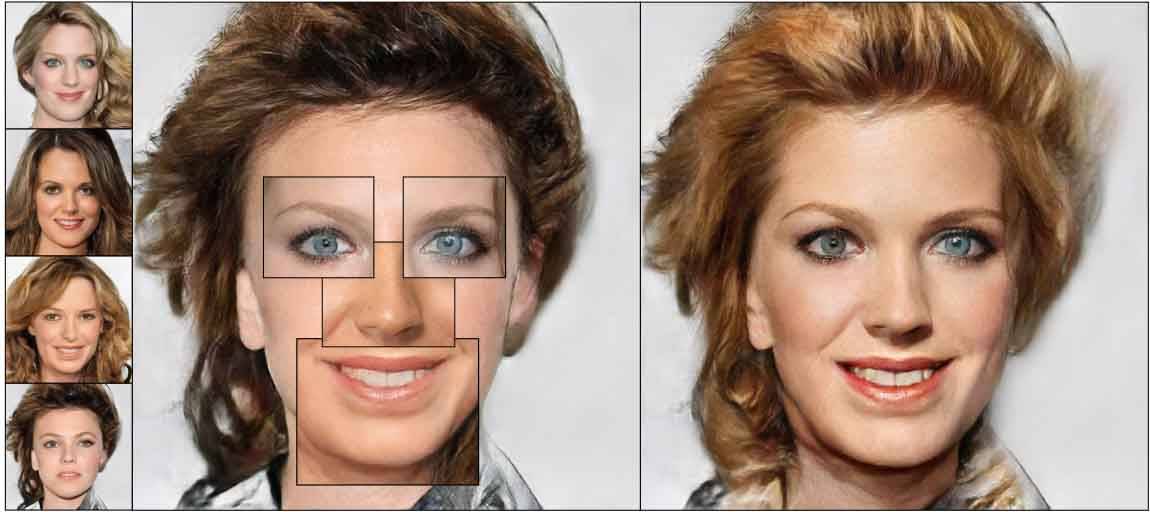}
    \includegraphics[width=0.33\linewidth]{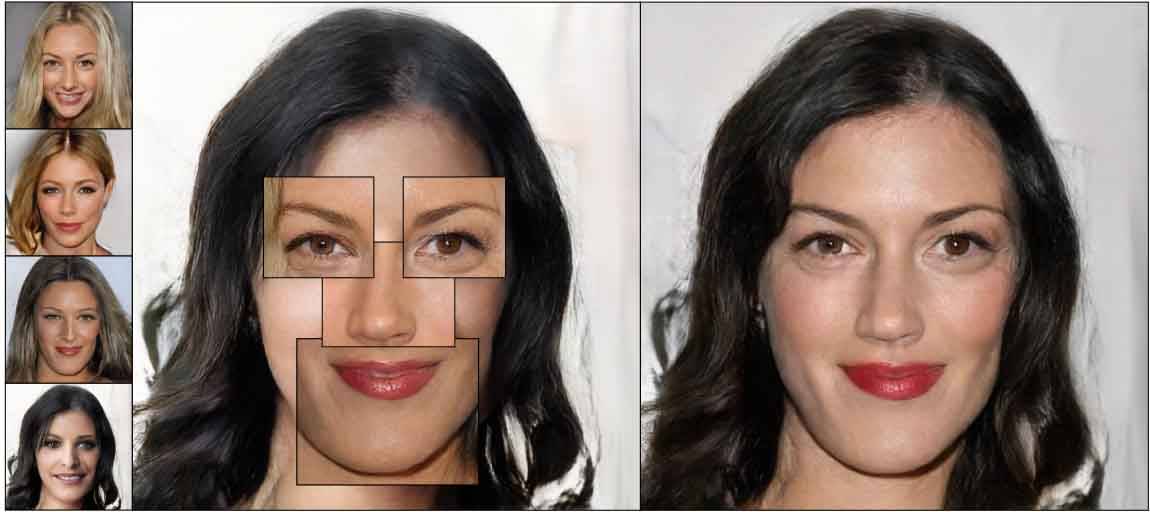}
    \includegraphics[width=0.33\linewidth]{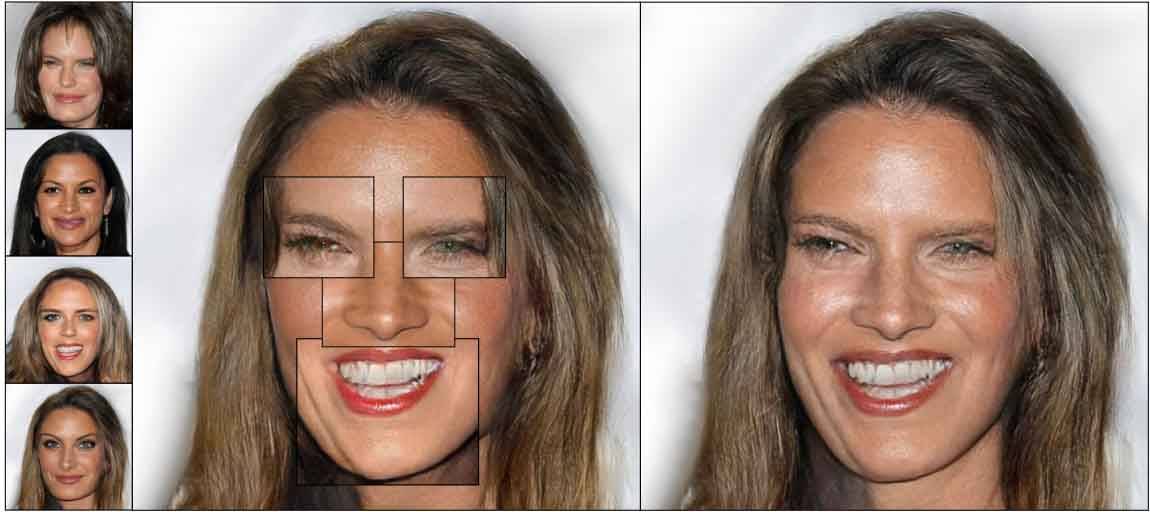}

    \caption{
    {In each set, we show color image (Left) of the source sketches (not shown here), a new face sketch (Middle) by directly recombining the cropped source sketches in the image domain, and a new face (Right) synthesized by using our method with the recombined sketches of the cropped components (eyes, nose, mouth, and remainder) as input.}
    }
    \label{fig:results_combine}
\end{figure*}

Traditional copy-paste methods (e.g., \cite{Shiming2018}) use seamless stitching methods on colored images. However, there will be situations where the hue of local areas is irrelevant. To address this issue, we recombine face components for composing new
faces, which can maintain the consistency of the overall color and lighting.
Specifically, it can be achieved by first encoding face component sketches (possibly from different subjects) as feature vectors and then combining them as new faces by using the \moduleTwo~and \moduleThree~modules.
This can be used to either replace components of existing faces with corresponding components from another source, or combining components from multiple persons.
Figure \ref{fig:results_combine} presents several synthesized new faces by re-combining eyes, nose, mouth and the remainder region from four source sketches. Our image synthesis sub-network is able to resolve the inconsistencies between face components from different sources in terms of both lighting and shape.

\section{Conclusion and Discussions}\label{sec:discussion}
In this paper we have presented a novel deep learning framework for synthesizing realistic face images from rough and/or incomplete freehand sketches. We take a local-to-global approach by first decomposing a sketched face into components, refining its individual components by projecting them to component manifolds defined by the existing component samples in the feature spaces, mapping the refined feature vectors to the feature maps for spatial combination, and finally translating the combined feature maps to realistic images. 
This approach naturally supports local editing and makes the  involved network easy to train from a training dataset of not very large scale. 
Our approach outperforms existing sketch-to-image synthesis approaches, which often require edge maps or sketches with similar quality as input. Our user study confirmed the usability of our system. We also adapted our system for two applications: face morphing and face copy-paste.

\begin{figure}
    \centering
    \setlength{\fboxrule}{0.5pt}
    \setlength{\fboxsep}{-0.01cm}
    \framebox{\includegraphics[width=0.45\linewidth]{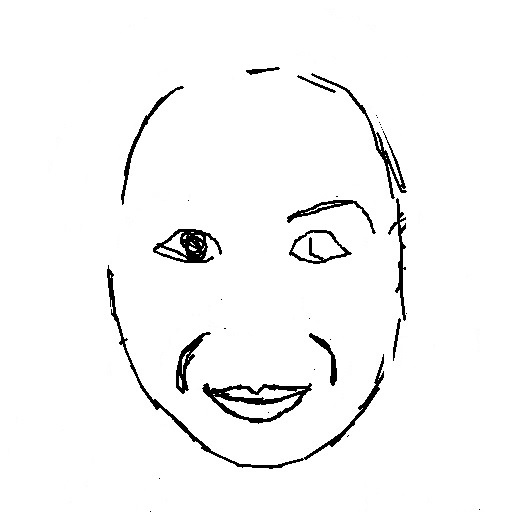}}
    \framebox{\includegraphics[width=0.45\linewidth]{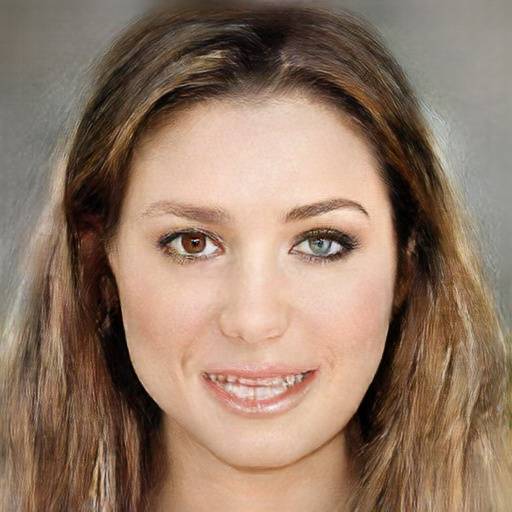}}
    \caption{A less successful example. The eyes in the generated image are of different colors.
    For the sketched mouth, it is slightly below an expected position, leading to a blurry result for this component.
    }
    \label{fig:falure}
\end{figure}

Our current implementation considers individual components rather independently. This provides flexibility (Figure \ref{fig:flexible}) but also introduces possible incompatibility problems. This issue is more obvious for the eyes (Figure \ref{fig:falure}), which are often symmetric. This might be addressed by introducing a symmetry loss ~\cite{huang2017beyond} or explicitly requiring two eyes from the same samples (similar to Figure \ref{fig:results_combine}).

Our work has focused on refining an input sketch component-by-component. In other words our system is generally able to handle errors within individual components, but is not designed to fix the errors in the layouts of components (Figure \ref{fig:falure}). To achieve proper layouts, we resort to a shadow-guided interface. In the future, we are interested in modeling spatial relations between facial components and fixing input layout errors. 

Our system takes black-and-white rasterized sketches as input and currently does not provide any control of color or texture in synthesized results. 
In a continuous drawing session, small changes in sketches sometimes might cause abrupt color changes.
This might surprise users and is thus not desirable for usability. We believe this can be potentially addressed by introducing a color control mechanism in generation. For example, we might introduce color constraints by either adding them in the input as additional hints or appending them to the latent space as additional guidance. In addition, adding color control is also beneficial for applications such as face morphing and face copy-and-paste. 

Like other learning-based approaches, the performance of our system is also dependent on the amount of training data. Although component-level manifolds of faces might be low dimensional, due to the relatively high-dimensional space of our feature vectors, our limited data only provides very sparse sampling of the manifolds. In the future we are interested in increasing the scale of our training data, and aim to model underlying component manifolds more accurately. This will also help our system to handle non-frontal faces, faces with accessories. 
It is also interesting to increase the diversity of results by adding random noise to the input.
Explicitly learning such manifolds and providing intuitive exploration tools in a 2D space would be also interesting to explore. 

Our current system is specially designed for faces by making use of the fixed structure of faces. How to adapt our idea to support the synthesis of objects of other categories is an interesting but challenging problem.

\begin{acks}

This work was supported by Beijing Program for International S\&T Cooperation Project~(No. Z191100001619003), Royal Society Newton Advanced Fellowship (No. NAF$\backslash$R2$\backslash$192151), Youth Innovation Promotion Association CAS, CCF-Tencent Open Fund and Open Project Program of the National Laboratory of Pattern Recognition (NLPR) (No. 201900055).  
Hongbo Fu was supported by an unrestricted gift from Adobe and grants from the Research Grants Council of the Hong Kong Special Administrative Region, China (No. CityU 11212119, 11237116), City University of Hong Kong (No. SRG 7005176), and the Centre for Applied Computing and Interactive Media (ACIM) of School of Creative Media, CityU.

\end{acks}


\bibliography{ref_new}
\bibliographystyle{ACM-Reference-Format}

\end{document}


\title{Supplemental Materials\\DeepFaceDrawing:  Deep Generation of Face Images from Sketches}
\maketitle

\section{Overview}
In this supplemental material, we have listed the conducted experiments, additional gallery of the user study results, and more details of the implementation.

\section{Ablation Study of K Selection}
\begin{figure*}[h!]
    \centering
    \includegraphics[width=0.15\linewidth]{images/K-choose/sample1/hand-draw.jpg}
    \includegraphics[width=0.15\linewidth]{images/K-choose/sample1/3.jpg}
    \includegraphics[width=0.15\linewidth]{images/K-choose/sample1/6.jpg}
    \includegraphics[width=0.15\linewidth]{images/K-choose/sample1/10.jpg}
    \includegraphics[width=0.15\linewidth]{images/K-choose/sample1/15.jpg}
    \includegraphics[width=0.15\linewidth]{images/K-choose/sample1/20.jpg}\\
    Input Sketch \qquad\quad $K=3$ \qquad  \quad$K=6$\qquad\quad$K=10$  \qquad \quad  $K=15$  \quad \qquad $K=20$\qquad\qquad
    \caption{The effect of selection of K in manifold projection. We set $K = 10$ in our implementation.
    }
    \label{fig:K-choose}
\end{figure*}

\section{Gallery of the User Study Results}
We have listed more results obtained from the usability study in Figures \ref{fig:results_more} and  \ref{fig:results_more1}.

\begin{figure*}[h!]
    \centering
    \setlength{\fboxrule}{0.5pt}
    \setlength{\fboxsep}{-0.03cm}
    
    \begin{minipage}{\linewidth}
    \framebox{\includegraphics[width=0.19\linewidth]{images/results/man14/hand-draw.jpg}}
    \framebox{\includegraphics[width=0.19\linewidth]{images/results/man11/hand-draw.jpg}}
    \framebox{\includegraphics[width=0.19\linewidth]{images/results/man8/hand-draw.jpg}}
    \framebox{\includegraphics[width=0.19\linewidth]{images/results/man9/hand-draw.jpg}}
    \framebox{\includegraphics[width=0.19\linewidth]{images/results/man17/hand-draw.jpg}}
    \end{minipage}
    \begin{minipage}{\linewidth}
    \framebox{\includegraphics[width=0.19\linewidth]{images/results/man14/colorized.jpg}}
    \framebox{\includegraphics[width=0.19\linewidth]{images/results/man11/colorized.jpg}}
    \framebox{\includegraphics[width=0.19\linewidth]{images/results/man8/colorized.jpg}}
    \framebox{\includegraphics[width=0.19\linewidth]{images/results/man9/colorized.jpg}}
    \framebox{\includegraphics[width=0.19\linewidth]{images/results/man17/colorized.jpg}}
    \end{minipage}
    \begin{minipage}{\linewidth}
    \framebox{\includegraphics[width=0.19\linewidth]{images/results/sample5/hand-draw.jpg}}
    \framebox{\includegraphics[width=0.19\linewidth]{images/results/sample38/hand-draw.jpg}}
    \framebox{\includegraphics[width=0.19\linewidth]{images/results/sample33/hand-draw.jpg}}
    \framebox{\includegraphics[width=0.19\linewidth]{images/results/sample32/hand-draw.jpg}}
    \framebox{\includegraphics[width=0.19\linewidth]{images/results/sample36/hand-draw.jpg}}
    \end{minipage}
    \begin{minipage}{\linewidth}
    \framebox{\includegraphics[width=0.19\linewidth]{images/results/sample5/colorized.jpg}}
    \framebox{\includegraphics[width=0.19\linewidth]{images/results/sample38/colorized.jpg}}
    \framebox{\includegraphics[width=0.19\linewidth]{images/results/sample33/colorized.jpg}}
    \framebox{\includegraphics[width=0.19\linewidth]{images/results/sample32/colorized.jpg}}
    \framebox{\includegraphics[width=0.19\linewidth]{images/results/sample36/colorized.jpg}}
    \end{minipage}
    \caption{More results: Part I.}
    \label{fig:results_more}
\end{figure*}

\begin{figure*}[h!]
    \centering
    \setlength{\fboxrule}{0.5pt}
    \setlength{\fboxsep}{-0.03cm}
    \begin{minipage}{\linewidth}
    \framebox{\includegraphics[width=0.19\linewidth]{images/results/man15/hand-draw.jpg}}
    \framebox{\includegraphics[width=0.19\linewidth]{images/results/man16/hand-draw.jpg}}
    \framebox{\includegraphics[width=0.19\linewidth]{images/results/man18/hand-draw.jpg}}
    \framebox{\includegraphics[width=0.19\linewidth]{images/results/man10/hand-draw.jpg}}
    \framebox{\includegraphics[width=0.19\linewidth]{images/results/man19/hand-draw.jpg}}
    \end{minipage}
    \begin{minipage}{\linewidth}
    \framebox{\includegraphics[width=0.19\linewidth]{images/results/man15/colorized.jpg}}
    \framebox{\includegraphics[width=0.19\linewidth]{images/results/man16/colorized.jpg}}
    \framebox{\includegraphics[width=0.19\linewidth]{images/results/man18/colorized.jpg}}
    \framebox{\includegraphics[width=0.19\linewidth]{images/results/man10/colorized.jpg}}
    \framebox{\includegraphics[width=0.19\linewidth]{images/results/man19/colorized.jpg}}
    \end{minipage}
    \begin{minipage}{\linewidth}
    \framebox{\includegraphics[width=0.19\linewidth]{images/results/sample31/hand-draw.jpg}}
    \framebox{\includegraphics[width=0.19\linewidth]{images/results/sample34/hand-draw.jpg}}
    \framebox{\includegraphics[width=0.19\linewidth]{images/results/sample35/hand-draw.jpg}}
    \framebox{\includegraphics[width=0.19\linewidth]{images/results/sample37/hand-draw.jpg}}
    \framebox{\includegraphics[width=0.19\linewidth]{images/results/sample3/hand-draw.jpg}}
    \end{minipage}
    \begin{minipage}{\linewidth}
    \framebox{\includegraphics[width=0.19\linewidth]{images/results/sample31/colorized.jpg}}
    \framebox{\includegraphics[width=0.19\linewidth]{images/results/sample34/colorized.jpg}}
    \framebox{\includegraphics[width=0.19\linewidth]{images/results/sample35/colorized.jpg}}
    \framebox{\includegraphics[width=0.19\linewidth]{images/results/sample37/colorized.jpg}}
    \framebox{\includegraphics[width=0.19\linewidth]{images/results/sample3/colorized.jpg}}
    \end{minipage}
    \begin{minipage}{\linewidth}
    \framebox{\includegraphics[width=0.19\linewidth]{images/results/man20/hand-draw.jpg}}
    \framebox{\includegraphics[width=0.19\linewidth]{images/results/man21/hand-draw.jpg}}
    \framebox{\includegraphics[width=0.19\linewidth]{images/results/man22/hand-draw.jpg}}
    \framebox{\includegraphics[width=0.19\linewidth]{images/results/man23/hand-draw.jpg}}
    \framebox{\includegraphics[width=0.19\linewidth]{images/results/man24/hand-draw.jpg}}
    \end{minipage}
    \begin{minipage}{\linewidth}
    \framebox{\includegraphics[width=0.19\linewidth]{images/results/man20/colorized.jpg}}
    \framebox{\includegraphics[width=0.19\linewidth]{images/results/man21/colorized.jpg}}
    \framebox{\includegraphics[width=0.19\linewidth]{images/results/man22/colorized.jpg}}
    \framebox{\includegraphics[width=0.19\linewidth]{images/results/man23/colorized.jpg}}
    \framebox{\includegraphics[width=0.19\linewidth]{images/results/man24/colorized.jpg}}
    \end{minipage}
    \caption{More results: Part II.}
    \label{fig:results_more1}
\end{figure*}

\section{Sketches Used in the Perceptive Evaluation Study}
In this section, we demonstrate the input sketches used in the perceptive evaluation in Figure \ref{fig:user_study_sketches}.
\begin{figure*}[h!]
    \centering
    \setlength{\fboxrule}{1pt}
    \setlength{\fboxsep}{-0.03cm}
    \framebox{\includegraphics[width=0.15\linewidth]{images/user_study_sketches/0.jpg}}
    \framebox{\includegraphics[width=0.15\linewidth]{images/user_study_sketches/1.jpg}}
    \framebox{\includegraphics[width=0.15\linewidth]{images/user_study_sketches/2.jpg}}
    \framebox{\includegraphics[width=0.15\linewidth]{images/user_study_sketches/3.jpg}}
    \framebox{\includegraphics[width=0.15\linewidth]{images/user_study_sketches/4.jpg}}
    \framebox{\includegraphics[width=0.15\linewidth]{images/user_study_sketches/5.jpg}}
    \framebox{\includegraphics[width=0.15\linewidth]{images/user_study_sketches/6.jpg}}
    \framebox{\includegraphics[width=0.15\linewidth]{images/user_study_sketches/7.jpg}}
    \framebox{\includegraphics[width=0.15\linewidth]{images/user_study_sketches/8.jpg}}
    \framebox{\includegraphics[width=0.15\linewidth]{images/user_study_sketches/9.jpg}}
    \framebox{\includegraphics[width=0.15\linewidth]{images/user_study_sketches/10.jpg}}
    \framebox{\includegraphics[width=0.15\linewidth]{images/user_study_sketches/11.jpg}}
    \framebox{\includegraphics[width=0.15\linewidth]{images/user_study_sketches/12.jpg}}
    \framebox{\includegraphics[width=0.15\linewidth]{images/user_study_sketches/13.jpg}}
    \framebox{\includegraphics[width=0.15\linewidth]{images/user_study_sketches/14.jpg}}
    \framebox{\includegraphics[width=0.15\linewidth]{images/user_study_sketches/15.jpg}}
    \framebox{\includegraphics[width=0.15\linewidth]{images/user_study_sketches/16.jpg}}
    \framebox{\includegraphics[width=0.15\linewidth]{images/user_study_sketches/17.jpg}}
    \framebox{\includegraphics[width=0.15\linewidth]{images/user_study_sketches/18.jpg}}
    \framebox{\includegraphics[width=0.15\linewidth]{images/user_study_sketches/19.jpg}}
    \framebox{\includegraphics[width=0.15\linewidth]{images/user_study_sketches/20.jpg}}
    \framebox{\includegraphics[width=0.15\linewidth]{images/user_study_sketches/21.jpg}}
    \caption{Sketch inputs used in the perceptive evaluation.}
    \label{fig:user_study_sketches}
\end{figure*}
\shuyu{
\section{Comparison of Different Methods with Test Set Input}
We illustrate the visual comparison given the test set sketches (i.e., the edge maps) as input in Figure \ref{fig:visual_comparison}.}

\begin{figure*}[h!]
    \centering
    \setlength{\fboxsep}{-0.03cm}
    \framebox{\includegraphics[width=0.16\linewidth]{images/test_set/sketch/759.jpg}}
    \framebox{\includegraphics[width=0.16\linewidth]{images/test_set/pix/759.jpg}}
    \framebox{\includegraphics[width=0.16\linewidth]{images/test_set/line/759.jpg}}
    \framebox{\includegraphics[width=0.16\linewidth]{images/test_set/HD/759.jpg}}
    \framebox{\includegraphics[width=0.16\linewidth]{images/test_set/ifill/759.jpg}}
    \framebox{\includegraphics[width=0.16\linewidth]{images/test_set/our/759.jpg}}
    
    \framebox{\includegraphics[width=0.16\linewidth]{images/test_set/sketch/4007.jpg}}
    \framebox{\includegraphics[width=0.16\linewidth]{images/test_set/pix/4007.jpg}}
    \framebox{\includegraphics[width=0.16\linewidth]{images/test_set/line/4007.jpg}}
    \framebox{\includegraphics[width=0.16\linewidth]{images/test_set/HD/4007.jpg}}
    \framebox{\includegraphics[width=0.16\linewidth]{images/test_set/ifill/4007.jpg}}
    \framebox{\includegraphics[width=0.16\linewidth]{images/test_set/our/4007.jpg}}
    
    \framebox{\includegraphics[width=0.16\linewidth]{images/test_set/sketch/16078.jpg}}
    \framebox{\includegraphics[width=0.16\linewidth]{images/test_set/pix/16078.jpg}}
    \framebox{\includegraphics[width=0.16\linewidth]{images/test_set/line/16078.jpg}}
    \framebox{\includegraphics[width=0.16\linewidth]{images/test_set/HD/16078.jpg}}
    \framebox{\includegraphics[width=0.16\linewidth]{images/test_set/ifill/16078.jpg}}
    \framebox{\includegraphics[width=0.16\linewidth]{images/test_set/our/16078.jpg}}

    \caption{Generation results given test set sketches. For columns from left to right: Sketch input, pix2pix, Lines2FacePhoto, pix2pixHD, iSketchNFill and Ours.}
    \label{fig:visual_comparison}
\end{figure*}

\shuyu{
\section{Ablation Study of the Selection of the Feature Dimension}
We show the MSE metrics of the reconstructed sketch with different latent feature dimensions. The quantitative results are elaborated in Table \ref{tab:my_label}.}
\begin{table}[h!]
    \centering
    \begin{tabular}{ccccc}
        \hline
       Latent Dim  & left-eye & right-eye & nose & mouth 
        \\
        \hline
        128 & 0.370 & 0.323 & 0.423 & 0.359 
        \\
        256 & 0.247 & 0.228 & 0.288 & 0.287 
        \\
        {512} & \textbf{0.131} & \textbf{0.126} & \textbf{0.177} & \textbf{0.197} 
        \\
        \hline
    \end{tabular}
    \caption{An ablation study for the number of feature dimensions in Component Embendding. Comparing the value of the MSE loss for each component in different feature dimensions, the loss decreases with the increase of the number of feature dimensions. 
    }
    \label{tab:my_label}
\end{table}


\shuyu{
\section{Network Architectures}
Our network consists of Component Embedding, Feature Mapping Module and Image Synthesis Module.
\syc{In training, we use Adam optimizer [Kingma and Ba 2014] with $\beta1=0.5$ and $\beta2=0.999$ and the 
initial learning rate is 0.0002.}

\subsection{Component Embedding Architectures}

Our Component Embedding module contain\hb{s} five auto-encoders. Each auto-encoder consists of five encoding layers (Table~\ref{tab:conv2D}) and five decoding layers (Table~\ref{tab:convTrans2D}). The encoding layer is shown in Table~\ref{tab:conv2D}. We add a fully connected layer in the middle to ensure the latent descriptor is of 512 dimensions for all the five components. The details of Component Embedding module are 
shown in Table~\ref{tab:CE}.

\subsection{Feature Matching Architectures}
The Feature Matching module contains five decoding networks, which take as input the compact feature vectors obtained from the component manifolds and convert them to the corresponding size of feature maps for subsequent generation. See the details in Table~\ref{tab:FM}.

\subsection{Image Synthesis Architectures}

Our Image Synthesis module adopts a GAN architecture utilizing a generator and a discriminator to generate real face images from the fused feature maps. The details of the generator are 
illustrated in Table~\ref{tab:IS_G}.
The discriminator employs a multi-scale discriminating manner: scale the input feature maps and the generated images in three different levels and go through three different sub-discriminators. See Table ~\ref{tab:IS_D} for more details.

\begin{table}[h!]
    \begin{minipage}{0.45\linewidth}\centering
        \centering
        \begin{tabular}{c}
             \hline
            \textbf{F(x)}\\
            \hline
            Conv2d\\
            BatchNorm2d\\
            LeakyReLU\\
            \hline
        \end{tabular}
        \caption{\textbf{Conv2D-Block}}
        \label{tab:conv2D}
    \end{minipage}
    \begin{minipage}{0.45\linewidth}\centering
        \centering
        \begin{tabular}{c}
             \hline
            \textbf{F(x)}\\
            \hline
            ConvTranspose2d\\
            BatchNorm2d\\
            LeakyReLU\\
            \hline
        \end{tabular}
        \caption{\textbf{ConvTrans2D-Block}}
        \label{tab:convTrans2D}
    \end{minipage}
\end{table}

\begin{table}[h!]
    \centering
    \begin{tabular}{ccc}
        \hline
         \textbf{Layer} & \textbf{Output Size} & \textbf{Filter} \\
        \hline
         Input & $1\times H\times W$  \\
        \hline
        Conv2D-Block & $32\times \frac{H}{2}\times \frac{W}{2}$ & $1\to32$\\
        Resnet-Block & $32\times \frac{H}{2}\times \frac{W}{2}$ & $32\to32\to32$ \\
        \hline
        Conv2D-Block & $64\times \frac{H}{4}\times \frac{W}{4}$ & $32\to64$\\
        Resnet-Block & $64\times \frac{H}{4}\times \frac{W}{4}$ & $64\to64\to64$ \\
        \hline
        Conv2D-Block & $128\times \frac{H}{8}\times \frac{W}{8}$ & $64\to128$\\
        Resnet-Block & $128\times \frac{H}{8}\times \frac{W}{8}$ & $128\to128\to128$ \\
        \hline
        Conv2D-Block & $256\times \frac{H}{16}\times \frac{W}{16}$ & $128\to256$\\
        Resnet-Block & $256\times \frac{H}{16}\times \frac{W}{16}$ & $256\to256\to256$ \\
        \hline
        Conv2D-Block & $512\times \frac{H}{32}\times \frac{W}{32}$ & $256\to512$\\
        Resnet-Block & $512\times \frac{H}{32}\times \frac{W}{32}$ & $512\to512\to512$ \\
        \hline
        Fully Connected & 512 & $512\times \frac{H}{32}\times \frac{W}{32}\to512$\\
        \hline
        \hline
        Fully Connected & $512\times \frac{H}{32}\times \frac{W}{32}$ & $512\to 512 \times \frac{H}{32}\times \frac{W}{32}$\\
        \hline
        Resnet-Block & $512\times \frac{H}{32}\times \frac{W}{32}$ & $512\to512\to512$ \\
        ConvTrans2D-Block & $256\times \frac{H}{16}\times \frac{W}{16}$ & $512\to256$\\
        \hline
        Resnet-Block & $256\times \frac{H}{16}\times \frac{W}{16}$ & $256\to256\to256$ \\
        ConvTrans2D-Block & $128\times \frac{H}{8}\times \frac{W}{8}$ & $256\to128$\\
        \hline
        Resnet-Block & $128\times \frac{H}{8}\times \frac{W}{8}$ & $128\to128\to128$ \\
        ConvTrans2D-Block & $64\times \frac{H}{4}\times \frac{W}{4}$ & $128\to64$\\
        \hline
        Resnet-Block & $64\times \frac{H}{4}\times \frac{W}{4}$ & $64\to64\to64$ \\
        ConvTrans2D-Block & $32\times \frac{H}{2}\times \frac{W}{2}$ & $64\to32$\\
        \hline
        Resnet-Block & $32\times \frac{H}{2}\times \frac{W}{2}$ & $32\to32\to32$ \\
        ConvTrans2D-Block & $32\times H\times W$ & $32\to32$\\
        \hline
        Resnet-Block & $32\times H\times W$ & $32\to32\to32$ \\
        ReflectionPad2d\\
        Conv2D-Block & $1\times H\times W$ & $32\to 1$\\
        \hline
        Output & $1\times H\times W$ \\
        \hline
    \end{tabular}
    \caption{The architecture of the Component Embedding Module.}
    \label{tab:CE}
\end{table}

\begin{table}[h!]
    \centering
    \begin{tabular}{ccc}
         \textbf{Layer} & \textbf{Output Size} & \textbf{Filter} \\
         \hline
        Input & 512  \\
        \hline
        Fully Connected & $512\times \frac{H}{32}\times \frac{W}{32}$ & $512\to 512 \times \frac{H}{32}\times \frac{W}{32}$\\
        \hline
        Resnet-Block & $512\times \frac{H}{32}\times \frac{W}{32}$ & $512\to512\to512$ \\
        ConvTrans2D-Block & $256\times \frac{H}{16}\times \frac{W}{16}$ & $512\to256$\\
        \hline
        Resnet-Block & $256\times \frac{H}{16}\times \frac{W}{16}$ & $256\to256\to256$ \\
        ConvTrans2D-Block & $128\times \frac{H}{8}\times \frac{W}{8}$ & $256\to128$\\
        \hline
        Resnet-Block & $128\times \frac{H}{8}\times \frac{W}{8}$ & $128\to128\to128$ \\
        ConvTrans2D-Block & $64\times \frac{H}{4}\times \frac{W}{4}$ & $128\to64$\\
        \hline
        Resnet-Block & $64\times \frac{H}{4}\times \frac{W}{4}$ & $64\to64\to64$ \\
        ConvTrans2D-Block & $32\times \frac{H}{2}\times \frac{W}{2}$ & $64\to32$\\
        \hline
        Resnet-Block & $32\times \frac{H}{2}\times \frac{W}{2}$ & $32\to32\to32$ \\
        ConvTrans2D-Block & $32\times H\times W$ & $32\to32$\\
        \hline
        Resnet-Block & $32\times H\times W$ & $32\to32\to32$ \\
        ReflectionPad2d\\
        Conv2D-Block & $32\times H\times W$ & $32\to 32$\\
        \hline
        Output & $32\times H\times W$ \\
        \hline
    \end{tabular}
    \caption{The architecture of the Feature Matching Module.}
    \label{tab:FM}
\end{table}

\begin{table}[h!]
    \centering
    \begin{tabular}{ccc}
    \hline
        &\textbf{Generator} &\\
         \hline
         \textbf{Layer} & \textbf{Output Size} & \textbf{Filter} \\
        \hline
         Input & $32\times 512\times 512$  \\
        \hline
        Conv2D-Block & $56\times 512\times 512$ & $32\to56$\\
        Conv2D-Block & $112\times 256\times 256$ & $56\to112$ \\
        Conv2D-Block & $224\times 128\times 128$ & $112\to224$ \\
        Conv2D-Block & $448\times 64\times 64$ & $224\to448$ \\
        \hline
        Resnet-Block ($\times 9$) & $448\times 64\times 64$ & $448\to448\to448$ \\
        \hline
        ConvTrans2D-Block & $224\times 128\times 128$ & $448\to224$ \\
        ConvTrans2D-Block & $112\times 256\times 256$ & $224\to112$ \\
        ConvTrans2D-Block & $224\times 512\times 512$ & $448\to224$ \\
        ConvTrans2D-Block & $3\times 512\times 512$ & $224\to3$ \\
        \hline
        Output & $3\times 512\times 512$ \\
        \hline
    \end{tabular}
    
    \caption{The architecture of Generator in the Image Synthesis Module.}
    \label{tab:IS_G}
\end{table}

\begin{table}[h!]
    \centering
    \begin{tabular}{ccc}
        \hline
       &\textbf{Discriminating Unit (DisUnit)}&\\
         \hline
         \textbf{Layer} & \textbf{Output Size} & \textbf{Filter} \\
        \hline
         Input & $(32+3)\times H\times W$  \\
        \hline
        Conv2D-Block & $64\times \frac{H}{2}\times \frac{W}{2}$ & $(32+3)\to64$\\
        Conv2D-Block & $128\times \frac{H}{4}\times \frac{W}{4}$ & $64\to128$ \\
        Conv2D-Block & $256\times \frac{H}{8}\times \frac{W}{8}$ & $128\to256$ \\
        Conv2D-Block & $512\times \frac{H}{8}\times \frac{W}{8}$ & $256\to512$ \\
        Conv2D-Block & $512\times \frac{H}{8}\times \frac{W}{8}$ & $512\to512$ \\
        \hline
        Output & $512\times \frac{H}{8}\times \frac{W}{8}$ \\
        \hline
    \end{tabular}
    \begin{tabular}{cccc}
        \hline
        \textbf{Discriminator} \\
         \hline
         \textbf{Layer} & \textbf{D1 Output Size}& \textbf{D2 Output Size}& \textbf{D3 Output Size} \\
        \hline
         Input & $35\times 512\times 512$& $35\times 512\times 512$& $35\times 512\times 512$  \\
        \hline
        AvgPool & - &$35\times 256\times 256$ & $35\times 256\times 256$\\
        AvgPool & - & -& $35\times 128\times 128$ \\
        \hline
        DisUnit & $512\times 64\times 64$ & $512\times 32\times 32$ &$512\times 32\times 32$ \\
        \hline
        Output & $512\times 64\times 64$ & $512\times 32\times 32$ &$512\times 32\times 32$ \\
        \hline
    \end{tabular}
    
    \caption{The architecture of the Discriminator in Image Synthesis Module.}
    \label{tab:IS_D}
\end{table}

}